%% file: wrapper.tex
\documentclass{elsart}
\include{mysymbols}
\begin{document}
\pdfgraphics
\begin{frontmatter}
\title{Path integration over closed loops and Gutzwiller's trace formula}
\author[Helsinki]{P.~Muratore-Ginanneschi}
\address[Helsinki]{Department of Mathematics, University of Helsinki, P.O. Box 4,  
00014, Helsinki, Finland\\
paolo.muratore-ginanneschi@helsinki.fi}
\begin{abstract}
In $1967$ M.C.~Gutzwiller succeeded to derive the semiclassical expression of the 
quantum energy density of systems exhibiting a chaotic Hamiltonian 
dynamics in the classical limit. The result is known as the Gutzwiller trace 
formula. 

The scope of this review is to present in a self-contained way 
recent developments in functional determinant theory allowing to revisit the 
Gutzwiller trace formula in the spirit of field theory.

The field theoretic setup permits to work explicitly at every step of the 
derivation of the trace formula with invariant quantities of classical 
periodic orbits. R.~Forman's theory of functional determinants of linear, non singular 
elliptic operators yields the expression of quantum quadratic fluctuations around 
classical periodic orbits directly in terms of the monodromy matrix of the 
periodic orbits.

The phase factor associated to quadratic fluctuations, the Maslov phase,  
is shown to be specified by the Morse index for closed extremals, 
also known as Conley and Zehnder index.    
\end{abstract}
\end{frontmatter}
\newpage
    \pagenumbering{roman}
    \tableofcontents
    \include{introduction}
    \pagenumbering{arabic}
    \include{density} 
    \include{measure}
    \include{Morse}
    \include{moduli}    
    \include{conclusions}
    \appendix
    \include{measure-appendix}
    \include{hamiltonian}
    \include{Lie} 
    \include{fresnel}

\include{bibliography}
\end{document}

%% file: mysymbols.tex
\newif\ifpdf\ifx\pdfoutput\undefined\pdffalse\else\pdfoutput=1\pdftrue\fi
\newcommand{\pdfgraphics}{\ifpdf\DeclareGraphicsExtensions{.pdf,.jpg}\else\fi}
\usepackage{graphicx,amssymb,amsmath,epsfig,times}
\journal{Physics Reports}

\newcommand{\de}{\partial}
\newcommand{\cl}{\mathit{c}\ell}

\newcommand{\rea}{\mathrm{Re}}
\newcommand{\ima}{\mathrm{Im}}
\newcommand{\spe}{\mathrm{Sp}}
\newcommand{\tr}{\mathrm{Tr}}
\newcommand{\Tr}{\mathbf{Tr}}
\newcommand{\ke}{\mathrm{Ker}}
\newcommand{\ind}{\mathrm{ind}^{-}}
\newcommand{\nul}{\mathrm{nul}}
\newcommand{\mo}{\mathfrak{t}}
\newcommand{\mos}{\mathfrak{s}}
\newcommand{\mov}{\mathfrak{v}}
\newcommand{\lift}{\Lambda \circ}
\newcommand{\liftd}{\Lambda^\dagger\circ}
\newcommand{\id}{\mathsf{I}}
\newcommand{\mass}{\mathsf{L}_{\dot{q}\,\dot{q}}}
\newcommand{\dmass}{\dot{\,\,\mathsf{L}}_{\dot{q}\,\dot{q}}}

\newcommand{\vpt}{\mathsf{L}_{\dot{q}\,q}}
\newcommand{\dvptau}{\dot{\,\,\mathsf{L}}_{\dot{q}\,q\,;\tau}}
\newcommand{\vptau}{\mathsf{L}_{\dot{q}\,q\,;\tau}}
\newcommand{\pot}{\mathsf{L}_{q\,q}}
\newcommand{\potau}{\mathsf{L}_{q\,q\,;\tau}}
\newcommand{\dU}{\mathsf{U}}
\newcommand{\Bott}{\mathsf{Z}}
\newcommand{\avh}{\bar{\mathsf{H}}}
\newcommand{\geo}{\mathrm{ge}}

\newcommand{\gph}{\mathrm{Gr}}
\newcommand{\flu}{\delta q}
\newcommand{\dflu}{\frac{d\delta q}{dt\,\,}}
\newcommand{\nflu}{\nabla \flu}
\newcommand{\entau}{\lambda_{n\,,\tau}}

\newcommand{\ellntau}{\ell_{n\,,\tau}}
\newcommand{\ha}{\mathcal{H}}
\newcommand{\la}{\mathcal{L}}
\newcommand{\qla}{\mathsf{D}^{2}_{q_{\cl}}\la}
\newcommand{\ac}{\mathcal{S}}
\newcommand{\qac}{\delta^{2}\mathcal{S}}
\newcommand{\rac}{\mathcal{W}}
\newcommand{\mes}{\mathcal{D}}

\newcommand{\Ho}{H}
\newcommand{\feo}{K}
\newcommand{\Gr}{G}
\newcommand{\fo}{L}
\newcommand{\fobc}{L_{\mathfrak{B}}}
\newcommand{\fobctau}{L_{\mathfrak{B}\,,\tau}}
\newcommand{\foCtau}{L_{\mathfrak{C}\,,\tau}}
\newcommand{\foD}{L_{\mathrm{Dir.}}}

\newcommand{\foP}{L_{\mathrm{Per.}}}
\newcommand{\foPtau}{L_{\mathrm{Per.}\,,\tau}}
\newcommand{\Grtau}{L^{-1}_{\mathfrak{B}\,,\tau}}
\newcommand{\Grbc}{L^{-1}_{\mathfrak{B}}}

\newcommand{\bcc}{\mathfrak{C}}

\newcommand{\GrCtau}{L^{-1}_{\mathfrak{C}\,,\tau}}

\newcommand{\mf}{\mathfrak{M}}
\newcommand{\lmf}{L\mathfrak{M}}

\newcommand{\bc}{\mathfrak{B}}

\newcommand{\rn}{\mathbb{R}}
\newcommand{\cn}{\mathbb{C}}

\newcommand{\si}{\mathbb{L}^{2}}
\newcommand{\sbc}{\mathbb{L}^{2}_\mathfrak{B}([T'\,,T],\mathbb{R}^{d})}

\newcommand{\sbcD}{\mathbb{L}^{2}_\mathrm{Dir.}([T'\,,T],\mathbb{R}^{d})}

\newcommand{\tgs}{\mathbf{T} \mathfrak{M}}
\newcommand{\qtgs}{\mathbf{T}_{q_{\cl}} \mathfrak{M}}
\newcommand{\qttgs}{\mathbf{T}\mathbf{T}_{q_{\cl}} \mathfrak{M}}
\newcommand{\po}{\mathcal{P}}
\newcommand{\pobc}{\mathcal{P}^{-1}_{\mathfrak{B}\,,\tau}}
\newcommand{\poC}{\mathcal{P}^{-1}_{\mathfrak{C}\,,\tau}}
\newcommand{\tfu}{\mathfrak{F}}
\newcommand{\ctgs}{\mathbf{T}^{*} \mathfrak{M}}

\newcommand{\qtctgs}{\mathbf{T}\mathbf{T}^{*}_{q_{\cl}} \mathfrak{M}}

\newcommand{\mon}{\mathsf{M}}
\newcommand{\fu}{\mathsf{F}}
\newcommand{\sy}{\Omega}
\newcommand{\sym}{\mathsf{J}}



%% file: introduction.tex
\section*{Introduction}
\label{sec:introduction}

The Gutzwiller trace formula is the semiclassical expression of the 
energy density of a quantum system 
\cite{Gutzwiller,Gutzwiller2,Gutzwiller3,Gutzwiller4,Gutzwiller7,Gutzwiller8}. 
Despite of the mathematical difficulties with which it is intertwined, 
it represents the cornerstone of the present understanding of the manifestations of 
the quasi-stochastic nature of the trajectories of a generic, chaotic, 
classical system in the properties of the quantum mechanics 
associated to it by Bohr's correspondence principle.

The trace formula states that the energy spectrum of a non relativistic quantum system 
is given by a series over all the periodic orbit of the classical system.
The discovery of the trace formula revived the intuition of Bohr 
and Sommerfeld \cite{Bohr,Sommerfeld} in the early days of quantum theory that 
the energy spectrum of a generic quantum system can be expressed in terms of the invariant 
properties of the periodic orbits of the corresponding classical system.
The derivation of the trace formula is highly non trivial due to the singular
nature of the classical limit of quantum mechanics.\\
Feynman's path integrals offer the physically most transparent route to
take the limit. Indeed path integrals were the starting point of 
Gutzwiller's analysis. At the same time, the very discovery in the late sixties of the 
trace formula corresponded also to a major progress in the understanding of real time 
path integration. Namely Gutzwiller realised that the known 
Van Vleck-Pauli-Morette \cite{Morette,Pauli,VanVleck} approximation of the 
semiclassical propagator had to be corrected by the 
introduction of a further phase factor in correspondence of conjugate points of 
the classical trajectory. The phase factor turned out to be specified by the Morse 
index for {\em open} extremals \cite{Morse}. The occurrence of the Morse index 
in the semiclassical propagator is experimentally observable. 
Namely, the phase of the propagator enters the Bohr-Sommerfeld quantisation 
condition and its generalisations as they emerge from the trace formula. \\
The role and significance of the propagator phase were further clarified 
by the ensuing fundamental investigations of the semiclassical limit carried out 
by Maslov \cite{MaslovFedoriuk}, Voros \cite{Voros} and Miller \cite{Miller}.
 
The works of Gutzwiller and Maslov laid also the basis for the development of 
functional semiclassical methods in field theory. Gutzwiller's idea to 
compute atomic spectra without resorting to the construction of wave functions
is of great advantage in field theory where state functionals are exceedingly complicated
objects. Dashen, Hasslacher and Neveu 
\cite{DashenHasslacherNeveu1,DashenHasslacherNeveu2,DashenHasslacherNeveu3} used the
trace formula as a paradigm in their ground-breaking investigations of particle spectra
of field theories. \\
Since the second half of the seventies the kinds of semiclassical
approximations underlying the trace formula became familiar tools
of theoretical physics. Among the early applications one can mention 
the monopole quantisation in the Georgi-Glashow model by 't Hooft \cite{tHooft} and 
Polyakov \cite{Polyakov} and the development the instanton formalism of 
Belavin, Polyakov, Schwartz and Tyupkin \cite{BelavinPolyakovSchwartzTyupkin} 
and 't Hooft \cite{tHooft2}.

The semiclassical approximation of path integrals is an infinite
dimensional implementation of stationary phase methods. 
Fluctuations around the field configuration dominating the path integral
in the semiclassical limit are, within leading order, generically quadratic.  
The resulting infinite dimensional Fresnel integral brings 
about the exigency of computing the {\em square root} of the determinant of the 
self-adjoint operator governing the quadratic fluctuations. 
Thus, a natural mathematical counterpart of path integral semiclassical methods is 
the theory of functional determinants.\\
Non-relativistic systems with {\em strictly positive} definite kinetic energy 
offer an intuitive framework to understand the mathematical issues interwoven 
with the evaluation of functional determinants. Since the quantum dynamics can be described 
by means of {\em configuration space} path integrals, the self-adjoint fluctuation operators 
emerging in the semiclassical approximation are of Sturm-Liouville type with 
spectrum bounded from below. The finite number of negative eigenvalues, 
the {\em Morse index}, by fixing the value of the phase of the quadratic 
path integral, provides a natural definition of the winding number
of the functional determinant in the complex plane.\\ 
An intrinsic interpretation of the result is achieved by embedding the fluctuation
operator in a family the elements of which are related by homotopy transformations. 
Thus, the phase of the functional determinant is specified by the winding number 
of the determinant bundle obtained by connecting the elements of the family
to a positive definite operator with zero Morse index. 
Homotopy transformations may clip the explicit form of the operator 
as well as the boundary conditions obeyed by its functional domain of definition.
A satisfactory theory should then be able to resolve the dependence 
of functional determinants versus the parameters entering the operator
and the boundary conditions. A powerful result was recently obtained 
by R.~Forman \cite{Forman} for functional determinants of a rather general class 
of elliptic operators. 
In the context of non-relativistic quantum mechanics Forman's result allows a 
particularly compact and intuitive expression of the semiclassical approximation 
of path integrals with {\em general} boundary conditions.
The result has the further merit to provide a topological characterisation of 
extremals of classical mechanical action functionals independently of the adoption 
of a Lagrangian or Hamiltonian formalism. In consequence of Forman's result,
it is possible to establish the equality of the functional determinants of the 
self-adjoint operators associated to the second variation around a classical trajectory 
in configuration and phase space. In this latter case quadratic fluctuations 
are governed by Dirac type operators the spectrum whereof is unbounded from below 
and from above.
The proof of index theorems for such Dirac operators is based on the construction of
an {\em infinite dimensional} Morse theory ultimately relying on spectral (Fredholm) flow 
theorems (see \cite{RobbinSalamon2,SalamonZehnder} and references therein; the particular 
case of open extremals was previously studied in the physical literature in 
\cite{LevitSmilansky,LevitMoehringSmilanskyDreyfus,MoehringLevitSmilansky} ).
In the periodic case, the resulting index is named in the mathematical literature 
after Conley and Zehnder who showed its role in the proof of existence statements 
for periodic solutions of time periodic and asymptotically linear Hamiltonians 
\cite{ConleyZehnder}.    
The equality of the functional determinants entails immediately that 
the Conley and Zehnder index coincides with the Morse index for closed extremals 
whenever both of them are defined. The result is proven in the mathematical literature 
by different, less intuitive methods \cite{AnLong}. In the physical literature
an equivalent result was obtained by Sugita in \cite{Sugita}. 

The modern theory of functional determinant provides considerable insight in the
Gutzwiller trace formula.\\
In the traditional derivation the resort to path integration is restricted to 
the construction of the quantum propagator. 
This is equivalent to applying functional determinant theory only to fluctuation
operators with Dirichlet boundary conditions exploiting a classical result of 
Gel'fand and Yaglom \cite{GelfandYaglom}. The semiclassical trace operation 
is carried out in a separate step. The re-parametrisation invariance 
\cite{Polyakov} of closed quantum paths contributing to the density of states is 
broken in the process of the stationary phase approximation. 
In consequence, classical periodic orbits do not appear at intermediate steps. 
They are shown to fully characterise the final result only after nontrivial algebraic 
manipulations which require a good deal of a priori physical insight.
More seriously the canonical invariance of the trace, although physically expected, 
needs to be proven apart. The proof relies on an highly non trivial mathematical 
apparatus developed by Arnol'd \cite{Arnold2} and investigated in the physical literature 
by Littlejohn, Creagh and Robbins \cite{CreaghRobbinsLittlejohn,Littlejohn2,Robbins}.\\ 
The situation is much simpler if the stationary phase approximation is carried out 
directly on the loop space where trace path integrals have support. 
Periodic functional determinants completely specify the contribution of quadratic 
fluctuations. The topological invariance of the phase factors is guaranteed by the
Morse index theory for {\em closed} extremals \cite{Morse}. The topological
formulation of Morse index theory of Bott \cite{Bott} and Duistermaat 
\cite{Duistermaat,CushmanDuistermaat} renders then a priori available 
an arbitrary number of equivalent prescriptions for the explicit evaluation of the index.

The aim of the present report is to review at a level as elementary as possible 
the general mathematical methods needed for a pure path integral derivation of the 
Gutzwiller trace formula.
The systematic application of these methods to the energy density of a non relativistic quantum 
system in the semiclassical limit is itself a novelty.
The merits of the path integral formulation are particularly evident when the Lagrangian of the
classical system is invariant under extra continuous symmetries beyond time translations. 
In the path integral formalism, such situations present no conceptual difference from the
case when the energy is the only conserved quantity. In all cases trace path integrals are
handled in a canonically invariant way by means of the Faddeev-Popov method 
\cite{Faddeev,FaddeevPopov}.

The report is organised as follows. In chapter~\ref{density} the
definition of the quantum density of states is recalled. The expression of the 
semiclassical approximation found by Gutzwiller is then qualitatively discussed. 
Chapter~\ref{Forman} focuses on functional determinant theory and Forman's identity.
Since the motivation is the semiclassical approximation of path integrals, 
the probabilistic interpretation of these latter ones is recalled in the first section of the 
chapter. Further details together with an outline of the recent rigorous proof 
\cite{AnderssonDriver} of the existence of the covariant representation 
of the path integral measure are given in appendix~\ref{shorttime}.
Chapter~\ref{Morse} summarises basic results of Morse index theory \cite{Morse,Milnor}.
The goal is to motivate the use of the symplectic geometry and differential
topology concepts applied in the modern formulation of the theory 
\cite{Duistermaat,CushmanDuistermaat,SalamonZehnder,AnLong2}
and to emphasise their physical relevance. 
The theory is illustrated with examples. Finally, it is shown 
how to derive from the general formalism practical prescriptions to compute the 
phase factors intervening in the Gutzwiller trace formula.
Chapter~\ref{moduli}, the last, deals with implementation of the Faddeev-Popov method
for trace path integrals. The Gutzwiller trace formula is then proven to follow 
immediately from the application of the method.\\
The material is expounded in each chapter in an as far as possible self-consistent
way in order to allow independent reading. The main text is complemented by appendices
summarising basic concepts of stochastic calculus and classical mechanics. 

The author's wish is that this may serve the reader interested not only in the Gutzwiller
trace formula but also in understanding methods of path integration of general use in
theoretical physics.


%% file: density.tex
\section{The density of states in quantum mechanics}
\label{density}

The aim of the present chapter is to set the scene for the path integral methods
which will be illustrated in the rest of this work.
Some basic concepts of non relativistic quantum dynamics are shortly recalled in order
to derive the relation between the density of states and the propagator. 
The relation is the starting point for the semiclassical approximations 
which led Gutzwiller to write his trace formula.
Finally, the qualitative features and the significance for quantum mechanics of the 
trace formula are shortly discussed.

\subsection{From the Schr\"odinger equation to the energy density}
\label{density:qm}

Non relativistic quantum mechanics for a spin-less particle is governed by the
Schr\"odinger equation
\begin{eqnarray}
i\,\hbar\, \de_T \psi-\Ho\,\psi=0\nonumber\\
\label{density-qm:Schroedinger}
\end{eqnarray}
where $\psi$ is the wave function, $\Ho$ the Hamiltonian operator and $\hbar$
Planck's constant. This latter is a pure number if absolute units are adopted
\begin{eqnarray}
[\,P\,]\,=\,-\,[\,Q\,]\,=\,1
\label{density-qm:units}
\end{eqnarray}
$Q$ and $P$ being respectively the spatial position and the canonically 
conjugated momentum variables.
The wave function $\psi$ is in general a complex valued function. It 
specifies the probability amplitude to observe the system in the 
position $Q$ at time $T$: the modulus squared $|\psi|^2$ defines the probability 
density of the same event (see for example \cite{Dirac,FeynmanHibbs,LandauLifshitz,Sakurai}). \\
In typical situations a quantum particle of mass $m$ interacts with an
external potential $U$ and with a vector potential $A_{\alpha}$.
Most often the physical configuration space $\mf$ is modeled by the Euclidean space
$\mathbb{R}^{d}$, eventually represented in non Cartesian coordinates. A slight 
generalisation is to imagine $\mf$ to be a Riemann manifold which is either compact
or $\mathbb{R}^{d}$ and has time independent metric $g_{\alpha\,\beta}$.
In such a case the Hamilton operator is
\begin{eqnarray}
\Ho\,\psi\,:=\,\frac{1}{2} \,g^{\alpha\,\beta}\frac{(P_{\alpha}+A_{\alpha})\,
(P_{\beta}+A_{\beta})}{m}\,\psi+U\,\psi
\label{density-qm:Hamilton}
\end{eqnarray}
where $A_{\alpha}$ and $U$ are  functions of $Q$ and eventually $T$ and
$g^{\alpha\,\beta}$ is the inverse of $g_{\alpha\,\beta}$. 
In (\ref{density-qm:Hamilton}) the $P_\alpha$'s denote momentum 
operators in position representation. They act on all objects to 
their right as covariant derivatives 
\begin{eqnarray}
P_{\alpha}:=-i\,\hbar\,\nabla_\alpha
\label{density-qm:momentum}
\end{eqnarray}
and therefore they obey the Leibniz rule
\begin{eqnarray}
P_{\alpha}A^{\alpha}\psi=(P_{\alpha}A^{\alpha})\psi+A^{\alpha}P_{\alpha}\psi
\label{density-qm:Leibniz}
\end{eqnarray}
The explicit form of the covariant derivative is fixed by the compatibility 
condition with the metric assigned to $\mf$. 
These geometric concepts are recalled in appendix~\ref{geometry}.\\
The Hamilton operator acts on the space $\si_{\mf}$ of square integrable functions 
on $\mf$ such that the Schr\"odinger equation is self-adjoint with respect to 
the scalar product 
\begin{eqnarray}
&&\langle \varphi\,, \psi\rangle\,=\,
\int_{\mf}d^dQ\sqrt{g(Q)}\,(\varphi^{*}\psi)(Q,T)
\nonumber\\
&&g:=\det\{g_{\alpha\,\beta}\}
\label{density-qm:scalarproduct}
\end{eqnarray}
If the potentials  $A_{\alpha}$ and $U$ are time independent it is possible
to introduce the stationary Schr\"odinger equation
\begin{eqnarray}
E\,\psi_E -\Ho\,\psi_E=0
\label{density-qm:stationary}
\end{eqnarray}
describing a quantum phenomenon occurring at a constant energy $E$. 
For bounded physical systems, the stationary equation admits solutions only 
if the energy assumes discrete, quantised, values. The energy levels are in such a 
case labelled by a set of integers referred to as quantum numbers.
 
The imaginary factor appearing in the Schr\"odinger equation (\ref{density-qm:Schroedinger})
ensures invariance under time reversal of the dynamics of isolated quantum systems. 
Nevertheless it is useful to introduce the forward time evolution operator or propagator. 
The kernel $\feo$ of the propagator is specified by the solution of
\begin{eqnarray}
\begin{array}{ll}
(i\,\hbar\,\de_T-\Ho)\feo(Q,T|Q',T')=-i\,\delta(T-T')\delta(Q-Q')\,,
&\quad T\,\geq\, T'
\\
\feo(Q,T|Q',T')=0\,,&\quad T\,<\, T'
\end{array}
\label{density-qm:propagator}
\end{eqnarray}
The propagator is the inverse of the Hamilton operator for
Cauchy boundary conditions in time. The propagator owns 
its name because it governs the propagation forward in time of
wave functions 
\begin{eqnarray}
\psi(Q,T)=\int_{\mf}d^{d}Q'\sqrt{g(Q')} \,\feo(Q,T|Q',T')\,\psi(Q',T') 
\label{density-qm:propagatoraction}
\end{eqnarray}
If the Hamiltonian is time independent and the energy spectrum discrete 
the propagator is amenable in the sense of $\si_{\mf}$ to a series over
the eigenfunctions of the stationary Hamilton equation
\begin{eqnarray}
\feo(Q,T|Q',T')=\theta(T-T')\,\sum_n\,\psi_n(Q)\,\psi_n^{*}(Q')\,
e^{-i\,\frac{E_n\,(T-T')}{\hbar}}
\label{density-qm:series}
\end{eqnarray}
where $\theta(T-T')$ is the Heaviside step function
\begin{eqnarray}
\theta(T-T')=\left\{
\begin{array}{lll}
1 \quad&\quad \mathrm{if} & T\,\geq\,T'
\\
0 \quad&\quad \mathrm{if} & T\,<\,T'
\end{array}
\right.
\label{density-qm:Heaviside}
\end{eqnarray}
Analogous representations exist for continuous and mixed spectra.

The energy spectrum of the quantum system can be computed from the Fourier-Laplace 
transform of the propagator. The latter is given by
the analytic continuation in the upper complex energy-plane 
\begin{eqnarray}
&&\int_{0}^{\infty} \frac{dT}{\hbar} \,e^{i\,\frac{Z\,T}{\hbar}}\,\feo(Q,T|Q',0)
=i\,\sum_n\,\frac{\psi_n(Q)\,\psi_n^{+}(Q')}{Z-E_n}:=
i\,\Gr(Q|Q',E)
\nonumber\\
&& Z\in\cn\,,\quad \ima Z\,>\,0 
\label{density-qm:Greenkernel}
\end{eqnarray}
The result is proportional to the kernel of the Green function $\Gr$, 
the resolvent of the time independent problem
\begin{eqnarray}
&&(Z\,-\Ho)\Gr(Q|Q',Z)=\delta(Q-Q') \,, \quad \quad \ima \,Z\,>0
\nonumber\\
&&\Gr(Q|Q',Z)=\int_{0}^{\infty} \frac{dT}{i\,\hbar} \,e^{i\,\frac{Z\,T}{\hbar}}
\,\feo(Q,T|Q',0)
\label{density-qm:resolvent}
\end{eqnarray}
The poles of $\Gr$ are located on the real axis and coincide with the 
energy levels of the quantum system.
Using the Plemelj identity \cite{Wyld} 
\begin{eqnarray}
\left.\lim_{\ima\,Z\,\downarrow 0}\frac{1}{Z-E_n}\right|_{E=\rea\,Z}=\mathrm{P.V.}
\frac{1}{E-E_n}-i\,\pi\,\delta(E-E_n)
\label{density-qm:Plemelj}
\end{eqnarray}
where $\mathrm{P.V.}$ denotes the integral principal value, 
the announced relation between the energy density of the system and 
the trace of $\Gr$ is found
\begin{eqnarray}
\rho(E)=-\left.\lim_{\ima\,Z\,\downarrow 0}\,\ima \frac{1}{\pi}\int_{\mf}d^dQ\,
\sqrt{g(Q)}\Gr(Q|Q,Z)\right|_{E=\rea\,Z}
\label{density-qm:trace}
\end{eqnarray}
Physically (\ref{density-qm:trace}) means that the energy density is a scalar 
quantity in non-relativistic Quantum Mechanics, independent of the representation of 
the Hilbert space. 

Finally, in terms of the propagator the energy density reads
\begin{eqnarray}
\rho(E)=-\left.
\lim_{\ima\,Z\,\downarrow 0}\,
\ima\,\int_{0}^{\infty}\frac{dT}{i\,\pi\,\hbar} \,e^{i\,\frac{Z\,T}{\hbar}}\,
\int_{\mf}d^{d}Q\sqrt{g(Q)}\,\feo(Q,T|Q,0)\right|_{E=\rea\,Z}
\label{density-qm:density}
\end{eqnarray}

\subsection{The semiclassical limit and the Gutzwiller trace formula}
\label{density:Gutzwiller}

The relation between the trace of the propagator and the energy density 
provides a way to extract information on the spectrum of a quantum system
without solving the stationary Schr\"odinger equation (\ref{density-qm:stationary}).
Indeed the propagator can be regarded as a more fundamental object than the
Schr\"odinger equation. According to Feynman's formulation of quantum mechanics 
\cite{Feynman,FeynmanHibbs} the propagator is the ``sum''
\begin{eqnarray}
\feo(Q,T|Q',T')\,\thicksim\, \mbox{``}\sum_{\mathrm{paths}}\,e^{i\,\frac{\ac(\mathrm{path})}
{\hbar}}\,\mbox{''}
\label{density-Gutzwiller:pi}
\end{eqnarray}
extended over all paths in configuration space  connecting $Q$ to $Q'$ 
in the time interval $[T',T]$. Each path contributes with a phase specified by
the action functional $\ac$. 
The precise meaning of the sum will be recalled later.
Here it is interesting to observe that classical mechanics is recovered in 
the singular limit of $\hbar$ tending to zero. The limit 
is directly meaningful in absolute units. In general units it 
will correspond to the vanishing of an overall adimensional parameter 
obtained from the Planck constant times other invariant quantities of the 
physical system.\\ 
Bohr's correspondence principle requires the recovery of classical
mechanics when $\hbar$ tends to zero while all the other parameters are held fixed.
The existence of the limit entails the possibility to investigate 
a range of phenomena for which quantum effects are weak by means of
an asymptotic expansion around the classical limit, the semiclassical 
expansion. 

Broadly speaking the semiclassical approximation is expected to apply 
to the description of phenomena occurring at energy scales large in comparison
to the mean energy level spacing.
Typical examples are encountered in the context of mesoscopic physics where the 
investigation of highly excited states of atoms as well as the transport 
properties of solid-state devices are amenable to semiclassical methods.
However, the domain of applicability extends even to the ground state 
of certain classes of quantum systems. The rescaling of the action functional $\ac$ 
often evinces that the small $\hbar$ limit is equivalent to a small coupling 
r\'egime of some nonlinear term \cite{Coleman}.
This observation has proven to have far reaching consequences in the
investigation  of tunneling phenomena where analytic perturbation theory is not available, 
both in systems with a finite number (Quantum Mechanics) and infinite (Field Theory)
number of degree of freedoms. Finally loop expansions around 
a ``vacuum'' state can be ordered in powers of $\hbar$ \cite{Zinn}.

The main result of semiclassical methods is to express quantum observables 
in terms of classical objects.
One of the striking differences between classical and quantum dynamics is that 
the latter gives a linear evolution law for the probability {\em amplitudes}. 
In Classical Mechanics a generic, non-linear, Hamiltonian exhibits a chaotic dynamics: 
exponential sensitivity to initial conditions in bounded regions leading to stretching
and folding in classical phase space. 
In Quantum Mechanics there is sound evidence that the evolution of time dependent 
observables, broadly speaking quantities related to the squared absolute value 
of probability amplitudes, exhibits no chaotic behavior \cite{Berry}.
However, there are many experimental evidences (see \cite{BrackBhaduri} for review)
supporting the conjecture that the energy spectrum may reflect the properties of
a classically chaotic motion. The inference is motivated by the correspondence
principle. The quantum energy levels are determined by the existence of eigenfunctions 
of the Hamilton operator. 
The correspondence principle leads to associating them to the invariant sets of 
the classical dynamics. For a generic autonomous system invariant sets are the 
energy surface, the tori produced by eventual extra symmetry of the theory and 
classical periodic orbits. \\
The role of classical invariant sets in quantisation is understood
in the particular case of classically integrable systems. In many cases integrability 
follows from the possibility to separate variables\footnote{A famous counter example 
is the Toda lattice \cite{Henon,Flaschka} (see also \cite{GreeneTaborCarnevale} for review). 
The quantisation of the Toda lattice is investigated in \cite{Gutzwiller5,Gutzwiller6}.}.
A classical Hamiltonian system is separable in $d$ dimensions if, 
including the Hamiltonian function $\ha$, there are $d$ independent integrals 
of the motion $(\ha,\ha_1,...,\ha_{d-1})$ Poisson commuting with each other. 
The set of first integrals $(\ha,\ha_1,...,\ha_{d-1})$ is then said to be in 
involution.  
Using these $d$ first integrals one can introduce action-angles coordinates 
$(A_1,...,A_d,\theta_1,...,\theta_d)$ which are canonical coordinates such
that the Hamiltonian $\ha=\ha(A_1,...,A_d)$ and the other first
integrals become functions of the action variables alone.
If a physical system is classically separable, 
semiclassical quantisation gives a simple rule for the spectrum
of a complete set of observables. Essentially this is a generalisation
of the Bohr and Sommerfeld rule (see for example \cite{Bohr,Sommerfeld,LandauLifshitz,BrackBhaduri})
and it is known as the Einstein-Brillouin-Keller or EBK quantisation rule
\cite{Einstein,Keller}. 
The EKB quantisation predicts the quantisation of the action variables
\begin{eqnarray}
A_j=\left(n_j+\frac{\mu_j}{4}\right)\, \hbar 
\label{EKB}
\end{eqnarray}
with the $n_j$'s and $\mu_j$'s specifying the set of quantum numbers of the 
system \cite{LandauLifshitz}.
The relevance of (\ref{EKB}) is mainly conceptual. It establishes a remarkable
connection between the occurrence of classical periodic orbits with quantum 
spectra. 
Namely, the quantised values of the action variables in (\ref{EKB}) 
correspond to closed curves on the tori defined by the constants of the motion.
From the practical point of view, the Schr\"odinger equation always separates when 
the classical problem is separable \cite{Gutzwiller7}
rendering thus available the exact expression of the energy spectrum.\\
In generic classical systems there are no constants of the motion other
than the energy. Moreover it is known that in a chaotic system the number
of primitive periodic orbits proliferates exponentially with the period $T$. 
The phenomenon has no correspondence in the observed densities of
energy levels.\\
A further difficulty, pointed out by M.~Berry, towards the understanding of 
how classic chaotic behavior translates into quantum spectra arises from the infinite 
time limit which is intertwined with both concepts \cite{Berry,Berry3}.
The very definition of typical indicators of chaotic behavior like of Ljapunov exponents, 
R\'enyi entropies (see for example \cite{ArrowsmithPlace,BeckSchlogl,webbook}) 
and algorithmic complexity \cite{Ford} requires infinite sequences of data. Such data
are provided by the same infinite time limit which defines invariant states.
However, the correspondence principle does not guarantee that the classical 
limit will be interchangeable with the infinite time limit and counterexamples 
are known \cite{Berry3}.

The difficulties listed above stress the relevance of the breakthrough 
achieved in a series of impressive papers \cite{Gutzwiller,Gutzwiller2,Gutzwiller3,Gutzwiller4} 
by M.~Gutzwiller with the discovery of the trace formula which now bears his name.
Gutzwiller's trace formula yields a general semiclassical expression for the quantum 
energy density. In the case when the energy is the only 
conserved quantity, the trace formula takes the form 
\begin{eqnarray}
&&\rho(E)=\sum_{n}\, \delta(E-E_n)\cong
\nonumber\\
&&\int\!\frac{d^{d}Qd^{d}P}{( 2\,\pi \,\hbar)^{d}}\, 
\delta\left(E-{\ha}(P,Q)\right)+\ima\!
\sum_{o\in \mathrm{p.p.o.}}\frac{i\,T_o}{\pi\,\hbar}
\sum_{r=1}^{\infty}\,\frac{e^{i\,r\,\frac{\rac_o(E)}{\hbar}-i\,\frac{\pi}{2}\,\aleph_{o,r}}}
{\sqrt{\left|\det_{\perp}[\id_{2 d}-M_{o}^{r}]\right|}}
\label{density-Gutzwiller:formula}
\end{eqnarray}
The semiclassical density of states is seen to comprise two terms of different nature.\\
The first term is a microcanonical average of the classical Hamiltonian, 
associated by Bohr's correspondence principle to the Hamilton operator the 
spectrum whereof is sought. The existence of a similar contribution has been known 
for long time in atomic physics from the Thomas-Fermi approximation. 
The microcanonical average brings in a smooth background dependence of the energy 
density on $E$.\\
The second term, Gutzwiller's genuine achievement, consists of a formal series 
ranging over all classical primitive periodic orbits of finite period $T_o$ and 
their repetitions $r$. 
Each orbit is represented in the series by a complex function of the energy.
The phase has an essential singularity in $\hbar$ proportional to the 
reduced action $\rac$ of the periodic orbit \cite{LandauLifshitz}.  
The reduced action depends continuously on the energy and therefore brings in 
strong oscillations in the energy density. The phase also receives contribution 
from $\aleph_{o}^{r}(E)$ a topological invariant of classical orbits usually 
referred to as the Maslov index. It carries information about the structural stability 
of the dynamics linearised around the orbit.\\
The amplitudes of the orbit contributions depend upon the monodromy
matrix $\mon$ of each primitive orbit. The monodromy matrix governs the linear 
stability in phase space of the periodic orbit over one period. 
It enters the trace formula through the inverse square root of the absolute value 
of the determinant of $\id_{2 d}-\mon$ restricted to the eigendirections transversal 
to the orbit.  Thus, the contribution of the more unstable orbits is exponentially
damped
\begin{eqnarray}
\frac{1}{\sqrt{\left|\det_{\perp}\left(\id_{2 d}-M(T)\right)\right|}}\sim 
\exp\left\{-\frac{h_{KS}\,T}{2}\right\}
\label{density-Gutzwiller:damping}
\end{eqnarray}
the decay rate being specified by the Kolmogorov-Sinai entropy 
\cite{Sinai,ArrowsmithPlace,BeckSchlogl,PaladinVulpiani}: the sum of the
positive Ljapunov exponents of the orbit.\\
Periodic orbits are therefore seen from the Gutzwiller trace formula to affect 
the spectrum both individually, through the essential singularity in $\hbar$, 
and collectively. 
The collective contribution gives rise to major physical and mathematical 
difficulties. For fixed values of the energy the number of periodic orbit is 
infinite. Moreover, in a chaotic system the number of periodic orbits  
proliferates exponentially with the period $T$ and growth rate given by the 
topological entropy $h_T$:
\begin{eqnarray}
\#(\mathrm{periodic\,\,orbits})\,\sim \exp\{h_T\,T\}\,,\qquad T\,\uparrow\,\infty
\label{density-Gutzwiller:proliferation}
\end{eqnarray}
The topological entropy is the R\'enyi entropy of order zero  
\cite{Sinai,ArrowsmithPlace,BeckSchlogl,PaladinVulpiani}.
If one assumes that on average the topological and Kolmogorv-Sinai entropies are 
equal, the diminishing amplitude of orbits of period $T$ is dominated by their
proliferation. Thus, the series consists effectively of terms with exponentially 
growing amplitudes.
The estimate indicates that a literal interpretation of the Gutzwiller trace formula 
is problematic. Nevertheless experimental and numerical evidences extensively 
reviewed in \cite{BrackBhaduri} support the existence in some mathematical sense 
of a semiclassical approximation to the density of states related 
to the trace formula above. 
In particular, it has been conjectured that convergence may arise on the 
basis of the topological organisation which is often observed in the
occurrence of periodic orbits. Longer orbits should be ``shadowed'' by shorter ones 
begetting in this way mutual compensations \cite{Gutzwiller8,Cvitanovic,ArtusoAurellCvitanovic}. 
An up-to-date survey of the current research in these directions can be found 
in ref.~\cite{webbook}. \\
The difficulties listed above make the series over periodic orbit conditionally
convergent at best. Nevertheless the existence of a number of special examples 
where the trace formula has been successfully applied encourages to take very 
seriously the insight it offers in the quantum behaviour of classically chaotic systems.  
In the words of Gutzwiller, ``as physicists, we have to make a compromise between 
logic and intuition'' \cite{Gutzwiller7}.


%% file: measure.tex
\section{Quadratic path integrals and functional determinants}
\label{Forman}

Most systems of physical relevance are described by Lagrangians with 
{\it strictly positive definite} kinetic energy. Under such an assumption
the quantum propagator can be written as a configuration space path integral.
The semiclassical approximation of quantum observables requires the explicit 
evaluation of quadratic path integrals.
In configuration space the task is equivalent to computing the determinant and 
the index of an elliptic second order linear differential operator. The index
is here defined as the number of negative eigenvalues. 
Forman's theorem reduces the computation of the functional determinant to that 
of the determinant of the fundamental solution (Poisson map) of the linear 
homogeneous problem associated with the nullspace of the elliptic operator.
The theorem is based on the construction of homotopy transformations between
elliptic operators. Using the same positive definiteness assumption, Morse theory  
permits to classify different homotopy classes and in this way to 
compute the index.  

\subsection{Path integrals and Lagrangians}
\label{Forman:measure}

Feynman \cite{Feynman,FeynmanHibbs} introduced path integrals as fundamental 
objects governing quantum dynamics. In non relativistic quantum mechanics 
Feynman's intuition has now evolved into a rigorous mathematical theory
\cite{AlbeverioHoeghKrohn,DeWittMorette1,DeWittMorette2,DeWittMoretteElworthy,DeWittMoretteElworthyNelsonSammelman,DeWittMoretteMaheshwariNelson,NelsonSheeks,CartierDeWittMorette}.\\
In the present work quantum mechanical path integrals are derived from 
an analytic continuation of the Wiener measure. The use of the analytic 
continuation has the advantage to unravel the probabilistic interpretation of 
Feynman path integrals. Furthermore it provides a unified formalism for quantum 
and statistical mechanics.

The connection between the Feynman path integral and the Wiener measure can be
established starting form partial differential equations.
Consider a continuous, forward-in-time stochastic process with support
on the same configuration space $\mf$ where the Schr\"odinger equation 
(\ref{density-qm:Hamilton}) was defined. 
The fundamental object describing the stochastic process is the 
conditional probability also called transition probability density $\feo_z$ 
to find the process in a point at time $T$ given the position at a 
previous time $T'$. 
By definition $\feo_z$ transforms as a scalar with respect to the 
invariant measure of $\mf$ under a change of variables 
$Q\,\rightarrow \,\tilde{Q}$
\begin{eqnarray}
\int_{\mf}d^dQ \sqrt{g(Q)}\feo_z(Q,T\,|\,Q',T')=
\int_{\tilde{\mf}}d^d\tilde{Q} \sqrt{\tilde{g}(\tilde{Q})} 
\tilde{\feo}_z(\tilde{Q},T\,|\,\tilde{Q}',T')
\label{Forman-measure:scalar}
\end{eqnarray}
In the presence of a drift field $\upsilon^\alpha$ and of a damping potential $\phi$, 
the transition probability is governed by the covariant Fokker-Planck equation          
\begin{eqnarray}
&&\frac{\de\,\feo_z}{\de T}=-\frac{z}{\hbar}\left[\frac{1}{2\,m} \,g^{\alpha\,\beta}
P_{\alpha}\,P_{\beta}+\frac{i}{z}\,
P_{\alpha}\,\upsilon^{\alpha}+\phi\,\right]\feo_z\,,\quad \forall\,\, 
T\,\geq \,T'\,>\,\infty
\nonumber\\
&& \lim_{T\,\downarrow\,T'}\feo_z(Q,T\,|\,Q',T')\,=\,\frac{\delta^{(d)}(Q-Q')}{\sqrt{g(Q')}}
\label{Forman-measure:covariant}
\end{eqnarray}
The link between the Fokker-Planck equation and the Schr\"odinger equation
is established by an analytical continuation in the parameter $z$ usually
referred to as Wick rotation. If $z$ is rotated along the unit circle of the complex
plane from the real to the imaginary positive semi-axis, the identifications
\begin{eqnarray}
&&\upsilon^{\alpha}\,=\,\frac{A^{\alpha}}{m}
\nonumber\\
&&\phi\,=\,U+\frac{A_{\alpha}A^{\alpha}}{m}+
\frac{i\,\hbar}{2\,m} \nabla_\alpha A^{\alpha}
\label{Forman-measure:identifications}
\end{eqnarray}
recover the Hamilton operator (\ref{density-qm:Hamilton}).
In the second of the (\ref{Forman-measure:identifications}) the momentum 
operator has been replaced with a covariant derivative in order to emphasise 
that it acts only on the vector potential $A^{\alpha}$.

A physically convenient picture of the Fokker-Planck equation 
is achieved by applying to it with the method of the characteristics. 
The second order spatial derivatives in  (\ref{Forman-measure:covariant}) impose 
the characteristics to be solutions of stochastic differential 
equations \cite{DeWittMoretteElworthy,IkedaWatanabe,Oksendal,KaratzasShreve}.
If the metric is time independent, it is possible to write {\em covariant} equations 
for the characteristics of (\ref{Forman-measure:covariant}) in the guise of the system of 
{\em Stratonovich} stochastic differential equations:
\begin{eqnarray}
\begin{array}{lc}
dq^{\alpha}(t)\,=\,\upsilon^{\alpha}(q(t),t)\,dt+\sqrt{\frac{\hbar\,z}{m}}\,
\sigma^{\alpha}_{k}(t)\diamond dw_{k}(t)\,,
\,\, 
&\,\, q^{\alpha}(T')=Q^{\prime\,\alpha}
\\
d\sigma^{\alpha}_{k}(t)\,=\,-\Gamma^{\alpha}_{\mu\,\nu}(q(t))\,\sigma^{\mu}_{k}(t)
\diamond dq^{\nu}(t)\,,\,\, 
&\,\, g^{\alpha\,\beta}(Q')=(\sigma_k^\alpha\sigma^\beta_k)(T')
\\
d\varsigma(t)\,=-\,\varsigma(t)\,\frac{z\,\phi(q(t),t)}{\hbar}\,dt\,,\,\, & \,\,\varsigma(T')=1
\end{array}
\nonumber\\
\label{Forman-measure:SDE}
\end{eqnarray}
where $q^{\alpha}$ are the coordinates of the position process while $\varsigma$ 
describes the damping.
The $\{\sigma^{\alpha}_{k}\}_{k=1}^{d}$ form a set of vielbeins parallel 
transported along a path $q^{\alpha}(t)$ by the Christoffel symbols $\Gamma^{\alpha}_{\mu\,\nu}$
specified by the metric $g_{\alpha\,\beta}$ (appendix~\ref{geometry}). 
The vielbeins project on $\mf$ the increments $dw^{k}$ of a Wiener process 
(Brownian motion) based on $\mathbb{R}^{d}$:
\begin{eqnarray}
&&\langle w_{k}(t)\rangle\,=\,0\,\quad\quad\quad \forall\,t\,\leq\,0
\nonumber\\
&&\langle dw_{k}(t)\,dw_{l}(t')\rangle\,=\,\delta_{k\,l}\,\delta(t-t')\,dt
\label{Forman-measure:noise}
\end{eqnarray}
Finally the symbol $\diamond$ highlights the Stratonovich's mid-point rule:
\begin{eqnarray}
\sigma^{\alpha}_{k}(t)\,\diamond dw_{k}:=\lim_{dt\,\downarrow\, 0}
\frac{\sigma^{\alpha}_{k}(t+dt)+\sigma^{\alpha}_{k}(t)}{2}\,
\left(w_{k}(t+dt)-w_{k}(t)\right)
\label{Forman-measure:Stratonovich}
\end{eqnarray}
The relevant feature of the mid-point discretisation is to cancel terms 
of the order $O(dw_k^2)\,\sim\,O(dt)$ in time differentials 
\cite{DeWittMoretteElworthy,KaratzasShreve,IkedaWatanabe,Oksendal,Stratonovich}. 
Ordinary differential calculus therefore applies to Stratonovich 
stochastic differential equations. More details on the geometric meaning 
of (\ref{Forman-measure:SDE}) can be found in \cite{IkedaWatanabe} and appendix~\ref{SDE}.
Here it is enough to stress that on a Riemann manifold $\mf$ covariant characteristics
for the Fokker-Planck equation (\ref{Forman-measure:scalar}) can be written only by means
of {\em path-dependent} vielbeins.

Thinking in terms of characteristic curves evinces the intuitive content of the
Feynman-Kac formula \cite{DeWittMoretteElworthy,KaratzasShreve,IkedaWatanabe,Oksendal}.
The transition probability density is the average over the Wiener measure of
the solutions of (\ref{Forman-measure:SDE}) connecting $Q'$ to $Q$ in the time
interval $[T'\,,T]$: 
\begin{eqnarray}
&&\sqrt{g(Q)}\, \feo_z(Q,T|\,Q',T')\,=\int\mes\mu(w(t))\, \varsigma(T)
\,\delta^{(d)}\left(q(T)-Q\right) 
\nonumber\\
&&\varsigma(T)\,=\,e^{-\frac{z}{\hbar}\int_{T'}^{T}dt\, \phi\left(q(t),t\right)}
\nonumber\\
&&q^{\alpha}(t)\equiv q^{\alpha}(t;T',Q',w(t))\,\quad\quad T\,\geq\,t\,\geq\,T'
\label{Forman-measure:solution}
\end{eqnarray}
The Wiener measure can be thought to have support on continuous paths $w$ in 
$[T'\,,T]$ with square integrable absolute value. A rigorous and compact 
discussion can be found in \cite{DeWittMoretteElworthy}.
The exact result
\begin{eqnarray}
\int_{w(T')=W'}^{w(T)=W}\mes\mu(w(t))\,=\,
\frac{e^{-\frac{(W-W')^{2}}{2\,(T-T')}}}{[2\,\pi\,(T-T')]^{\frac{d}{2}}}
\label{Forman-measure:Wienershorttimes}
\end{eqnarray}
motivates the representation of the Wiener measure usually encountered in the 
physical literature \cite{Zinn}
\begin{eqnarray}
\mes\mu(w(t))\,=\,\mes[w(t)]e^{-\int_{T'}^{T}\frac{\dot{w}^{2}(t)}{2}}
\label{Forman-measure:Wienerphysical}
\end{eqnarray}
This latter suggests to represent the Feynman-Kac formula directly as a path 
integral over the realisations of the position process. 
This ``change of measure'' is defined by means of the asymptotic expression of 
$\feo_z(Q,T|\,Q',T')$ for short displacements of the position process from
its initial state. Under reasonable smoothness assumptions on the drift
and the metric the asymptotics is obtained by substituting the short time solution
of the stochastic system (\ref{Forman-measure:SDE}) into the Feynman-Kac equation
and yields
\begin{eqnarray} 
\feo_z(Q,T'\!\!+\!dt|\,Q',T')=\!
\left(\frac{m}{2\,\pi\,z\,\hbar\,dt}\right)^{\frac{d}{2}}\!
e^{-\frac{1}{z\,\hbar}\!\int_{T'}^{T'+dt}\!\!\!dt\,\la_z(q_t,\dot{q}_t)}\!\!+o(dt)
\label{Forman-measure:shorttime}
\end{eqnarray}
The Lagrangian appearing in the exponential is \cite{AnderssonDriver}:
\begin{eqnarray}
\la_z(q,\dot{q})=\frac{m}{2}\,||\dot{q}- \upsilon||^{2}+z^2\,\phi+\frac{z\,\hbar}{2}\,
\nabla_\alpha \upsilon^{\alpha}-\frac{(z\,\hbar)^2\,R}{6\,m}
\label{Forman-measure:onesixth}
\end{eqnarray}
and it is evaluated along the {\em geodesic}, supposed unique for $dt$ 
small enough, connecting $Q$ to $Q'$. 
The notation in (\ref{Forman-measure:onesixth}) means 
\begin{eqnarray}
&&||\dot{q}-\upsilon||^{2}\,=\,g_{\alpha\,\beta}
(\dot{q}-\upsilon)^{\alpha}\,(\dot{q}-\upsilon)^{\beta}
\nonumber\\
&&\nabla_\alpha \upsilon^{\alpha}\,=\,\de_\alpha \upsilon^{\alpha}
+\Gamma^{\alpha}_{\alpha\,\beta}\upsilon^{\beta}
\end{eqnarray}
with $R$ the curvature scalar defined by the metric $g_{\alpha\,\beta}$.
The derivation of (\ref{Forman-measure:shorttime}) is summarised in 
appendix~\ref{SDE}.

The path integral for finite time differences follows by iterating $N$ times
the convolution integral of short-time kernels 
\begin{eqnarray}
&&\feo_z(Q,T'+2\,dt\,|\,Q',T')=(\feo_z\star\feo_z)(Q,T'+2\,dt\,|\,Q',T')
\nonumber\\
&&:=\int_{\mf}d^dQ''\sqrt{g(Q'')}\feo_z(Q,T'\!+\!2 dt|Q'',T'\!+\!dt)
\feo_z(Q'',T'\!+\!dt|Q',T')
\label{Forman-measure:semigroup}
\end{eqnarray}
and then taking the limit:
\begin{eqnarray}
\feo_z(Q,T\,|\,Q',T')&=&
\lim_{\substack{N\,\uparrow\,\infty\,\\ N\,dt=T-T'}}
(\feo_z\star...\star\feo_z)(Q,T|Q',T')
\nonumber\\ 
&:=&\int_{q(T')=Q'}^{q(T)=Q} \mes [\sqrt{g}q(t)]\,
e^{-\frac{1}{z\,\hbar}\,\int_{T'}^{T}dt\,\la_z}
\label{Forman-measure:pathintegral}
\end{eqnarray}
The procedure outlined above can be made completely rigorous \cite{AnderssonDriver}.

The quantum mechanical propagator corresponds to the analytical continuation
of (\ref{Forman-measure:pathintegral}) obtained by setting 
$z=\exp\{-i\,\theta\}$ and  rotating $\theta$ from zero
to $\theta$ equal $\pi/2$ \cite{Parisi}. 
The Feynman path integral for $z$ equal to $i$ is 
\begin{eqnarray}
\feo(Q,T\,|\,Q',T')=\int_{q(T')=Q'}^{q(T)=Q} \mes[\sqrt{g} q(t)]\, 
e^{\frac{i}{\hbar}\,\int_{T'}^{T}dt\,\la}
\label{Forman-measure:quantumpropagator}
\end{eqnarray}
where the Lagrangian
\begin{eqnarray}
\la(q,\dot{q})=\frac{m}{2}||\dot{q}-\upsilon||^2-\phi+\frac{i\,\hbar}{2}\,\nabla_\alpha 
\upsilon^{\alpha}+\frac{\hbar^2\,R}{6\,m}
\label{Forman-measure:Lagrangian}
\end{eqnarray}
with the identifications (\ref{Forman-measure:identifications}) assumes the form
\begin{eqnarray}
\la(q,\dot{q})=\frac{m}{2}\,g_{\alpha\,\beta}\dot{q}^{\alpha}\,\dot{q}^{\beta}
+\dot{q}_{\alpha}\,A^{\alpha}-U+\frac{\hbar^2\,R}{6\,m}
\label{Forman-measure:DeWitt}
\end{eqnarray}
In the last two formulae and from now on any reference to $z$ is dropped. 
The covariant path integral (\ref{Forman-measure:quantumpropagator}),
(\ref{Forman-measure:DeWitt}) was originally derived by B.S.~DeWitt in ref.
\cite{DeWitt} starting from the short time solution of the Schro\"dinger equation. 
Note that on a curved manifold the covariant Hamilton operator (\ref{density-qm:Hamilton}) 
corresponds to a Lagrangian comprising a curvature term vanishing in the classical 
limit (see also discussion in \cite{Schulman}).

For analytical and numerical purposes, it is often convenient to interpret
the limit of the iteration procedure (\ref{Forman-measure:pathintegral}) as the 
continuum limit of a single $N\,\times\,d$ lattice integral 
\cite{LangoucheRoekaertsTirapegui} with time mesh
\begin{eqnarray}
dt=\frac{T-T'}{N}
\label{Forman-measure:mesh}
\end{eqnarray}
A direct lattice construction of the quantum propagator is also possible. 
The Lagrangian in the exponential is recovered if $\upsilon^{\alpha}$ and 
$g_{\alpha\,\beta}$ are discretised according to the {\em mid-point rule} 
\cite{Schulman}. The discretisation rule reflects the interpolation with a 
differentiable curve, a geodesic, which was used to derive the short time 
transition probability density (\ref{Forman-measure:shorttime}).\\
A mid-point discretisation permits to import the rules of 
ordinary calculus for formal manipulations under path integral sign.
The circumstance was early realised by Feynman \cite{Feynman} by requiring the 
gauge invariance of the propagator of a quantum particle interacting with an 
electromagnetic field.

From the knowledge of the propagator it is possible to reconstruct the
dynamics of all relevant observables in quantum mechanics. This is generally
done by averaging over the propagator the kernel of a self-adjoint operator
\begin{eqnarray}
\langle \mathcal{O} \rangle&=&\int_{\mf\times\mf}d^dQ d^dQ'
\sqrt{g(Q)\,g(Q')}\,\mathcal{O}(Q,Q')\feo(Q,T\,|\,Q',T')
\label{Forman-measure:average}
\end{eqnarray}
If the average operation is reabsorbed in the definition of the functional 
measure, more general path integrals are obtained
\begin{eqnarray}
\langle \mathcal{O} \rangle=
\int_{\mathfrak{P}} \mes[\sqrt{g} q(t)]\, 
e^{\frac{i}{\hbar}\,\int_{T'}^{T}dt\,(\la+\la_{\mathcal{O}})}
\label{Forman-measure:unified}
\end{eqnarray}
To wit, the kernel $\mathcal{O}(Q'\,,Q)$ will not only modify the potential 
in the original Lagrangian (\ref{Forman-measure:DeWitt}) but it will also impose 
new boundary conditions on its lattice discretisation. The support of the path integral 
is then identified with the space of continuous paths $\mathfrak{P}$ satisfying 
such boundary conditions.
An example of (\ref{Forman-measure:unified}) is the trace of the propagator: 
setting $\mathcal{O}(Q'\,,Q)=\delta^{(d)}(Q'-Q)/\sqrt{g(Q)}$ yields
\begin{eqnarray}
\Tr K:=
\int_{\lmf} \mes[\sqrt{g} q(t)]\, 
e^{\frac{i}{\hbar}\,\int_{T'}^{T}dt\,\la}
\label{Forman-measure:trace}
\end{eqnarray}
where $\lmf$ is the loop space, the space of quantum paths closed in $[T',T]$.

\subsection{Semiclassical approximation and quadratic path integrals}
\label{Forman:quadratic}

The path integral (\ref{Forman-measure:trace}) suggests a physically intuitive 
picture of the semiclassical approximation. The leading order should correspond 
to quantum paths exploring the configuration space $\mf$ only in a neighborhood 
of classical trajectories satisfying the boundary conditions $\mathfrak{P}$.
Namely, had the path integral support on the set of at least once differentiable 
vector fields over $\mf$,
\begin{eqnarray}
\mathcal{C}_{\mathfrak{P}}=\{q(t)\in C^{(1)}([T'\,,T],\mf)\,|\,(q(T'),q(T))\in \mathfrak{P}\}
\label{Forman-quadratic:curves}
\end{eqnarray}
it would be possible to give a literal meaning to the time derivatives 
in the path integral Lagrangian. Once restricted to $\mathcal{C}_{\mathfrak{P}}$,
the path integral concentrates for vanishing $\hbar$ around those curves 
$q_{\cl}$ for which the action is stationary
\begin{eqnarray}
0=\left.\frac{\delta \ac}{\delta q^{\alpha}(t)}\right|_{q_{cl}(t)}= \left[
\frac{\delta}{\delta q^{\alpha}(t)}\int_{T'}^{T}dt \la\,\,\right]_{q_{cl}(t)}
\label{Forman-quadratic:extremal}
\end{eqnarray}
The latter condition is equivalent to require $q_{\cl}$ to satisfy
\begin{eqnarray}
&&\frac{\de \la}{\de q^\alpha}-\frac{d}{dt}\frac{\de \la}
{\de \dot{q}^{\alpha}}=0
\nonumber\\
&&\left.\flu^\alpha \frac{\de \la}{\de\dot{q}^{\alpha}}\right|_{T}-
\left.\flu^\alpha \frac{\de \la}{\de\dot{q}^{\alpha}}\right|_{T'}=0
\label{Forman-quadratic:actionextremum}
\end{eqnarray} 
for all fluctuations $\flu$ in the tangent space to $q_{\cl}$ 
\begin{eqnarray}
&&\mathbf{T}_{q_{\cl}}\mathcal{C}=\{\flu(t)\in C^{(1)}([T',T],\qtgs) 
\,|\, (\delta q(T'),\delta q(T))\in\bc \}
\nonumber\\&&
\bc\,:=\,\mathbf{T}_{(q_{\cl}(T'),q_{\cl}(T))}\mathfrak{P} 
\label{Forman-quadratic:conditions}
\end{eqnarray}
Thus the representation of quantum paths  
\begin{eqnarray}
q^{\alpha}(t)=q_{\cl}^{\alpha}(t)+\sqrt{\hbar}\,\flu^{\alpha}(t)
\label{Forman-quadratic:decomposition}
\end{eqnarray}
gives the asymptotic ``naive'' approximation of the path integral on 
$\mathcal{C}_{\mathfrak{P}}$
\begin{eqnarray} 
\langle \hat{\mathcal{O}} \rangle\,\sim\,\sum_{\{q_{\cl}\in\mathcal{C}\}}
e^{\frac{i}{\hbar}\,\int_{T'}^{T}\la_{\cl}}
\int_{\mathbf{T}_{q_{\cl}}\mathcal{C}} \mes[\sqrt{g}_{\cl} \flu(t)]\, 
e^{i\,\int_{T'}^{T}dt\,\qla}+O(\sqrt{\hbar})
\label{Forman-quadratic:naive}
\end{eqnarray}
the sum being extended to all extremal curves in $\mathcal{C}$.
According to (\ref{Forman-quadratic:naive}) the path integral
to evaluate is governed by the second variation Lagrangian
\begin{eqnarray}
\qla(\flu,\dot{\flu})=\frac{1}{2}\flu^{\alpha}\left[\overleftarrow{\frac{d}{dt}}\,\mass\,
\frac{d}{dt}+\overleftarrow{\frac{d}{dt}}\,\vpt+\vpt^{\dagger}\frac{d}{dt}+
\pot\,\right]_{\alpha\,\beta}\flu^{\beta}
\label{Forman-quadratic:Lagrangian}
\end{eqnarray}
with $\mass$, $\vpt$, $\pot$, $d\,\times\,d$ time dependent real 
matrices obtained from the second derivatives of the original Lagrangian
evaluated along the stationary trajectory $q_{\cl}$. The arrow over the 
derivatives indicates that they act to their left.
In particular the ``mass tensor'' $\mass$ is equal to the metric tensor 
evaluated along the extremal trajectory
\begin{eqnarray}
(\mass)_{\alpha\,\beta}(t):=m\,g_{\alpha\,\beta}(q_{\cl}(t))\,,\qquad\qquad 
\mass^{\alpha\,\beta}(t):=(\mass^{-1})_{\alpha\,\beta}(t)
\label{Forman-quadratic:masstensor}
\end{eqnarray}
and it is therefore symmetric and {\it strictly positive definite}. The measure terms
$\sqrt{g}_{\cl}$ are also evaluated along $q_{\cl}$.

Unfortunately the situation is more complicated. Feynman path 
integrals as well as the Wiener measure concentrate on nowhere 
differentiable paths. A discussion of intuitive appeal of such issue is given in 
appendix~3 of \cite{Coleman}. 
A rigorous justification of the semiclassical approximation requires more mathematical 
work  (see \cite{DeWittMoretteMaheshwariNelson,CartierDeWittMorette} and 
references therein). However, it turns out that formal rules of path integral 
calculus leading to the correct quantum theory can be inferred from the 
lattice representation of path integrals.

On a finite lattice the stationary phase approximation can be applied 
to the Lagrangian (\ref{Forman-measure:DeWitt}) or its 
generalisations discretised according to the mid-point rule. The stationary point
is seen to correspond to the discrete version of (\ref{Forman-quadratic:actionextremum}).
The integral over quadratic fluctuations reduces then to a multidimensional 
Fresnel integral. Definition and basic properties of Fresnel integrals are 
recalled in appendix~\ref{Fresnel}. The lattice quadratic action is
specified by an $N\,d\,\times\,N\,d$ dimensional symmetric matrix $\mathsf{L}_{N}$. 
The entries of $\mathsf{L}_{N}$ are read off the mid point discretisation of 
the path integral action and from the boundary conditions.
If the matrix $\mathsf{L}_{N}$ is non singular a straightforward computation 
performed in appendix~\ref{Fresnel} yields
\begin{eqnarray}
\iota^{(N)}(\bc)\,=\,\varkappa_\bc\,
\frac{e^{-i\frac{\pi}{2}\,\ind\mathsf{L}_{N}}}
{\sqrt{|\mathrm{Det}_{N}\mathsf{L}_{N}|}}
\label{Forman-quadratic:Fresnel}
\end{eqnarray}
The result requires some explanations.
The prefactor $\varkappa_\bc$ is a complex number depending only on 
the boundary conditions $\bc$ imposed on the lattice.\\
The symbol $\mathrm{Det}_{N}$ indicates a redefinition of the determinant. 
Namely, on a finite lattice with mesh (\ref{Forman-measure:mesh}) 
the propagator path integral gets an overall normalisation 
constant:
\begin{eqnarray}
\mathcal{N}_{N}=\left[\frac{m\,e^{-\,i\,\frac{\pi}{2}}}{2\,\pi\,dt\,\,\hbar}\right]
^{\frac{N\,d}{2}}
\label{Forman-quadratic:prefactor}
\end{eqnarray}
The $\mathrm{Det}_{N}$ operation is defined by setting
\begin{eqnarray}
\mathrm{Det}_{N}\mathsf{L}_{N}\,\propto\,\frac{\mathrm{det}\mathsf{L}_{N}}{\mathcal{N}^2_{N}}
\label{Forman-quadratic:functionaldeterminant}
\end{eqnarray}
the proportionality factor being $N$ independent and being fixed by $\varkappa_\bc$.
The re-definition of the determinant operation is immaterial on a finite lattice 
but leads to great simplifications in the continuum limit.
Finally the Fresnel integral is seen to produce the phase factor
$\ind\mathsf{L}_{N}$. Due to the normalisation (\ref{Forman-quadratic:prefactor}) 
the phase factor coincides with the {\em Morse index} of $\mathsf{L}_{N}$: 
the number of {\em negative eigenvalues} of $\mathsf{L}_{N}$. 

The lattice computation can formally repeated in the continuum by writing
the second variation of the action in terms of a non-singular, self-adjoint 
Sturm-Liouville operator. In general, evaluated on any two smooth vector
fields $\xi$, $\chi$, the second variation defines an infinite dimensional 
quadratic form 
\begin{eqnarray}
\qac(\xi,\chi)\,=\,\int_{T'}^{T}dt\,
\xi^{\alpha}\left[\overleftarrow{\frac{d}{dt}}\,\mass\,
\frac{d}{dt}+\overleftarrow{\frac{d}{dt}}\,\vpt+\vpt^{\dagger}\frac{d}{dt}+
\pot\,\right]_{\alpha\,\beta}\chi^{\beta}
\label{Forman-quadratic:qf}
\end{eqnarray}
Integration by parts yields
\begin{eqnarray}
\qac(\xi,\chi)=\xi^{\alpha}\,(\nabla \chi)_{\alpha}|_{T'}^{T}
+\int_{T'}^{T}dt\,\xi^{\alpha}\,\fo_{\alpha\,\beta}\,\chi^{\beta}
\label{Forman-quadratic:byparts}
\end{eqnarray}
where 
\begin{eqnarray}
\fo_{\alpha\,\beta}\,\chi^{\beta}:=-\,\frac{d}{dt}(\nabla \chi)_{\alpha}
+(\vpt^\dagger)_{\alpha\,\beta}\,\frac{d\,\chi^{\beta}}{dt}
+(\pot)_{\alpha\,\beta}\,\chi^{\beta}
\label{Forman-quadratic:differential}
\end{eqnarray}
and 
\begin{eqnarray}
(\nabla \chi)_{\alpha} \,:=\,(\mass)_{\alpha\,\beta}\,\frac{d \chi}{dt}^{\beta}
+(\vpt)_{\alpha\,\beta}\,\chi^{\beta}
\label{Forman-quadratic:momentum}
\end{eqnarray}
is the momentum canonically conjugated to $\chi$.
The notation is motivated in appendix~\ref{Hamilton}.
If the boundary conditions $\bc$ satisfied by the vector fields are such that
\begin{eqnarray}
\xi^{\alpha}\,(\nabla \chi)_{\alpha}|_{T'}^{T}=0
\label{Forman-quadratic:bc}
\end{eqnarray}
the differential operation $\fo$ specifies an  operator 
$\fobc$ in $\sbc$ self-adjoint with respect to the scalar product
\begin{eqnarray}
\langle \xi\,,\chi\rangle :=\,\frac{1}{T-T'}\int_{T'}^{T}dt\,
\xi^{\alpha}(\mass)_{\alpha\,\beta}\chi^{\beta}
\label{Forman-quadratic:scalarproduct}
\end{eqnarray}
The associated eigenvalue problem is
\begin{eqnarray}
\begin{array}{ll}
(\fo^{\alpha}_{\beta}\,\lambda_{n}^{\beta})(t)\,=\,
\ell_n\,\lambda_{n}^{\alpha}(t)\,,\quad&\quad 
\fo^{\alpha}_{\beta}=(\mass)^{\alpha\,\gamma}\,\fo_{\gamma\,\beta}
\\
 (\,\lambda_{n}(T'),\lambda_n(T)\,)\,\in\,\bc\,,\quad&\quad \forall\,n
\end{array}
\label{Forman-quadratic:eigenvalues}
\end{eqnarray}
Eigenvalues and eigenvectors in (\ref{Forman-quadratic:eigenvalues}) are 
labelled by quantum numbers collectively denoted by $n$.
The operator $\fobc$ inherits from the kinetic energy of the Lagrangian
a strictly positive prefactor of the highest derivative. 
The fact together with physically reasonable 
smoothness assumptions, permits to apply known results in functional analysis 
\cite{Dieudonne,Duistermaat} insuring that the eigenvalue problem
admits a countable number of solutions orthonormal with respect to 
(\ref{Forman-quadratic:scalarproduct}) with eigenvalues bounded from below and 
accumulating to infinity. The Morse index $\ind \fobc$ is therefore well defined.
\\ 
The condition (\ref{Forman-quadratic:bc}) is satisfied by the boundary 
conditions encountered in classical variational problems encompassed by 
Morse theory. Examples are Dirichlet, periodic, anti-periodic or focal boundary 
conditions. The unitary evolution laws of quantum mechanics renders these examples 
the ones typically encountered in semiclassical asymptotics.

The continuum limit of a quadratic path integral can be thought to have support 
on the space of square integrable fluctuations:
\begin{eqnarray}
&&\sbc:=
\nonumber\\&&\qquad
\{\flu(t)\in [T',T]\times\mathbb{R}^{d}\,|\, 
\langle \flu,\flu \rangle<\infty\,,\, (\,\flu(T'),\flu(T)\,)\in\bc\,\}
\label{Forman-quadratic:squareintegrable}
\end{eqnarray}
Any element of $\sbc$ can be represented as a series over the eigenvectors
\begin{eqnarray}
\flu^{\alpha}(t)\,=\,\sum_n\,\langle \lambda_n\,,\flu\rangle\, \lambda_n^{\alpha}(t)
\label{Forman-quadratic:element}
\end{eqnarray}
and therefore 
\begin{eqnarray}
\qac(\flu,\flu)=\langle \flu\,,\fobc\flu\rangle= \langle \fobc \flu\,,\flu\rangle
\label{Forman-quadratic:l2action}
\end{eqnarray}
The path integral measure consists then of the countable product of Fresnel integrals
over the amplitudes  $\langle \lambda_n\,,\flu\rangle$. The normalisation of each 
integral can be inferred from the lattice regularisation, compare with 
appendix~\ref{Fresnel} and \ref{formulae}.\\
The above picture of path integration requires $\fobc$ to be non singular:
\begin{eqnarray}
\ke\fobc\,=\,\emptyset
\label{Forman-quadratic:kernel}
\end{eqnarray}
The hypothesis is too restrictive for most physical applications and in chapter~\ref{moduli}
it will be shown how it can be loosened. For the rest of the present chapter, however, 
it is supposed to hold. Self-adjoint Sturm-Liouville operators governing the evolution 
of quantum paths in the neighborhood of a classical trajectory are often referred 
to as {\em fluctuation operators}.

In conclusion the leading order of the semiclassical approximation reads
\begin{eqnarray} 
\langle \hat{\mathcal{O}} \rangle\,\cong\,\sum_{\{q_{\cl}\in\mathcal{C}\}}
e^{\frac{i}{\hbar}\,\int_{T'}^{T}\la_{\cl}}
\int_{\sbc} \mes[\sqrt{g}_{\cl} \flu(t)]\, 
e^{i\,\int_{T'}^{T}dt\,\qla}
\label{Forman-quadratic:semiclassical}
\end{eqnarray}
The result is exact in the case of quadratic theories.
It remains therefore the problem to evaluate explicitly
\begin{eqnarray}
\int_{\sbc} \mes[\sqrt{g}_{\cl} \flu(t)]\, 
e^{i\,\int_{T'}^{T}dt\,\qla}=\varkappa_\bc\,
\frac{e^{-i\frac{\pi}{2}\,\ind\fobc}}{\sqrt{|\mathrm{Det}\fobc|}}
\label{Forman-quadratic:problem}
\end{eqnarray}
This can be done by combining classical Morse theory with Forman's theory
of functional determinants which will be expounded below.

\subsection{Functional determinants: from physics to mathematics}
\label{Forman:determinants}

The functional determinant of an infinite dimensional operator cannot be defined 
as the product of the eigenvalues. These latter are unbounded and the product would
in general diverge. Fortunately the physical motivation to introduce
functional determinants helps to overcome the difficulty. To streamline the 
notation, tensor indices will be from now on omitted whenever no ambiguity can 
arise.

Functional determinants come about as a result of the infinite dimensional Gaussian or 
Fresnel integrals entailed by quadratic path integrals. 
Equation (\ref{Forman-quadratic:Fresnel}) above, shows that $\mathrm{Det}_{N}$ is the 
quantity the continuum limit of which is really needed. 
$\mathrm{Det}_{N}$ has chance to converge in the continuum limit as the result of the mutual 
cancellation of two infinities. This is indeed the case each time it is possible to make sense 
of path integrals according to the definition given in (\ref{Forman-measure:pathintegral}). 
It is therefore appropriate to identify the functional determinant $\mathrm{Det}$ 
with the continuum limit of $\mathrm{Det}_{N}$.\\
It is a remarkable fact that in all cases when functional determinants 
exist according to the definition just given their expression can be as well
recovered if another, mathematically more satisfactory, definition is adopted. 
Such definition is supplied by the $\zeta$-function determinant of Ray and Singer 
\cite{RaySinger}.\\
Consider a non singular linear (partial) differential operator $O$ which is elliptic 
on a closed domain of definition. The attribute elliptic basically means here that the 
scalar or tensor coefficient of the highest derivative is strictly different from zero.
The zeta function of $O$ is the series  
\begin{eqnarray}
\zeta(s)=\sum_{\ell_j\,\in\,\spe O}\, \ell_j^{-\,s}:=\Tr\,O^{-s}
\,, \quad \quad \rea\,s\,>\,0
\label{detemninants:zeta}
\end{eqnarray}
summing the eigenvalues of $O$.
The symbol $\Tr$ in bold characters betokens the abstract trace operation
on $O$. The series is convergent for $\rea\,s$ large enough. 
Since a finite dimensional square matrix $\mathsf{O}$ with non-zero 
eigenvalues satisfies the equality
\begin{eqnarray}
\ln \mathrm{det} \mathsf{O}\,=\,- \left.\frac{d\,}{ds}\tr\,\mathsf{O}^{-s}\right|_{s=0}
\label{Forman-determinants:matrix}
\end{eqnarray}
with now $\tr$ the finite dimensional trace operation, it is tempting to surmise
\begin{eqnarray}
\ln \mathrm{Det} O\,:=\,-\,\left.\frac{d\,}{ds}\Tr\,O^{-s}\right|_{s=0} 
\label{Forman-determinants:definition}
\end{eqnarray}
Although $\Tr\,O^{-s}$ does not converge for $\rea\,s$ near zero,
the definition is consistent. Theorems of Seely \cite{Seely} insure that, under general conditions,
$\Tr\,O^{-s}$ can be analytically continued to a meromorphic function of the entire 
complex $s$-plane which is holomorphic in the origin.\\
The analytic continuation is most conveniently achieved by means of the Mellin 
transform of the fundamental solution of 
\begin{eqnarray}
&&(\de_u+O)\,f=0
\nonumber\\
&&f|_{u=0}=I
\end{eqnarray}
If $O$ is reasonably smooth
\begin{eqnarray} 
O^{-\,s}:=\frac{1}{\Gamma(s)}\int_{0}^{\infty}du\,u^{s-1}\,e^{-u\,O}
\label{Forman-determinants:Mellin}
\end{eqnarray}
is well defined and therefore
\begin{eqnarray} 
\Tr\,O^{-\,s}=\frac{1}{\Gamma(s)}\int_{0}^{\infty}du\,u^{s-1}\,
\Tr\,e^{-\,u\,O}
\label{Forman-determinants:trace}
\end{eqnarray}
can be used to define functional determinants.

The definition through the $\zeta$-function has the advantage to provide with a 
precise mathematical meaning the manipulations necessary to compute functional 
determinants without diagonalising the operator.
The idea is to write functional determinants in terms of quantities of lower 
functional dimensionality.
This is possible if the elliptic operator $O$ is thought as a member
of a one parameter, here denoted by  $\tau$, family of elliptic operators 
$O_\tau$ acting on the same functional domain. 
As $\tau$ varies from, say, $\tau_0$ to $\tau_1$ the elements of the 
family are smoothly deformed from $O_{\tau_0}$ to $O_{\tau_1}$. 
Along the deformation, functional determinants satisfy the differential equation
\begin{eqnarray}
\de_\tau \,\mathrm{ln}\mathrm{Det}\,O_\tau\,=\,\Tr
\left\{(\de_\tau O_\tau)\,O_{\tau}^{-1}\right\}
\label{Forman-determinants:homotopy}
\end{eqnarray}
introduced by Feynman in \cite{Feynman2}.
On the right hand side the differential operator $\de_\tau O_\tau$ acts
on the Green function $O_{\tau}^{-1}$. 
The differential equation translates into a useful prescription to compute $\mathrm{Det}\,O$
if the initial condition $\mathrm{Det}\,O_0$ is known. 
This is never a problem because there are examples of exactly solvable quadratic path 
integrals from which an initial condition can be read off. 
There are instead two other difficulties with the functional determinant flow 
equation (\ref{Forman-determinants:homotopy}).\\
The first problem is related to the exact meaning of the trace operation. 
The discussion can be made more concrete for the family of Sturm-Liouville operators 
acting as
\begin{eqnarray}
(&\fo&_{\tau})^{\alpha}_{\beta}:=
\nonumber\\
&&-\,\delta_{\beta}^{\alpha}\,\frac{d^{2}\,}{dt^{2}}-
\mass^{\alpha\,\gamma}(\dmass+\vptau-\vptau^\dagger)_{\gamma\,\beta}
\frac{d\,}{dt}+\mass^{\alpha\,\gamma}\,(\potau-\dvptau)_{\gamma\,\beta}
\label{Forman-determinants:differential}
\end{eqnarray}
with
\begin{eqnarray}
\spe\{\mass\}\,>\,0\,\qquad\qquad\forall\,t\,\in\,[T',T]
\label{Forman-determinants:spectrum}
\end{eqnarray}
on square integrable vector fields satisfying some boundary conditions $\bc$ in $[T',T]$.
Note that the tensor prefactor of the second order derivative in 
(\ref{Forman-determinants:differential}) is $\tau$ independent. 
At variance with the previous section, it is convenient to loosen for a while the 
hypothesis of self-adjointness. A broader class of Sturm-Liouville operators $\fobc$ 
is encompassed if the boundary conditions are described by a pair, 
{\em not} necessarily unique of $2\,d\,\times\,2\,d$ 
matrices $(\mathcal{Y}_{1}\,,\mathcal{Y}_{2})$ such that the Green functions $\Grbc$ 
satisfy
\begin{eqnarray}
&&(\fo_{\tau})^{\alpha}_{\gamma}(t)\,(\Grtau)^{\gamma}_{\beta}(t,t')\,=\,
\delta(t-t')\,\delta^{\alpha}_{\beta}
\nonumber\\
&&\mathcal{Y}_{1}\,\left[\begin{array}{c} \Grtau(T',t')\\
(\de_t\Grtau)(T',t')\end{array}\right]+
\mathcal{Y}_{2}\,\left[\begin{array}{c} \Grtau(T,t')\\
(\de_t\Grtau)(T,t')\end{array}\right]\,=\,0
\label{Forman-determinants:Green}
\end{eqnarray}
The simplest example is represented by Cauchy boundary conditions $\bcc$:
\begin{eqnarray}
&&\mathcal{Y}_{1}\,=\,\mathsf{I}_{2\,d}
\nonumber\\
&&\mathcal{Y}_{2}\,=\,0
\label{Forman-determinants:Cauchybc}
\end{eqnarray}
Cauchy boundary conditions will prove useful in what follows. \\
By inspection of (\ref{Forman-determinants:Green}) the derivative of the Green 
function displays a discontinuity on the diagonal at $t$ equal $t'$. 
In Cauchy's case in particular 
\begin{eqnarray}
\lim_{t\,\downarrow \,t'}\,\frac{d\,}{dt}\GrCtau(t,t')\,=\,-\,\id_{d}
\label{Forman-determinants:discontinuity}
\end{eqnarray}
An elliptic operator is said to be trace class when the associated singularity does 
not affect the determinant flow. Concretely  Sturm-Liouville operators $\fobctau$ are 
trace class when $\dot{\mathsf{\,\,L}}_{\dot{q}\,\dot{q}}$, 
$\mathsf{L}_{\dot{q}\,q\,;\tau}$ are equal to zero. 
Physically this is the case of quadratic fluctuations of a kinetic plus potential 
Lagrangian in flat space. 
However, if the metric becomes nontrivial as in non Cartesian coordinates or if
a vector potential sets in, Feynman's equation in the above form is ill-defined.
In the physical literature the problem is usually solved by coming back to the 
the lattice regularisation of the path integral and showing that the singular terms 
dash out in the continuum limit, see for example \cite{Barvinsky}.
The cancellation can be seen directly on the continuum in a more general, 
boundary condition independent framework due to R.~Forman \cite{Forman}. 
He proved that for a family of elliptic {\em ordinary differential} operators 
of $n$-th order the correct definition of the trace from the analytical 
continuation (\ref{Forman-determinants:Mellin}) is
\begin{eqnarray}
&&\de_\tau \ln \mathrm{Det}O_\tau=-\lim_{s\,\rightarrow\, 0}\Tr
\,\de_s\, \de_\tau\, O^{-\,s}=
\nonumber\\
&&\int_{T'}^{T}\!\!\!dt\,\tr\left\{\gimel(t)\lim_{t \downarrow t'}\de_\tau O_\tau(t)
O^{-1}_{\bc\,,\tau}(t,t')+\left[1-\gimel(t)\right]
\lim_{t\uparrow t'}\de_\tau O_\tau(t)
O^{-1}_{\bc\,,\tau}(t,t')\right\}
\nonumber\\
&&\gimel(t)=\left\{ \begin{array}{ll} \frac{1}{2}\, \mathsf{I}_d & 
\mbox{if $n$ is even}\\ \hat{\gimel}(t) & 
\mbox{if $n$ is odd} \end{array} \right. 
\label{Forman-determinants:traceclass}
\end{eqnarray}
The projector $\gimel$ weights the contribution to the trace of the limits
from above and below. 
For configuration space fluctuation operators $\fo_\tau$, $n$ is equal to
two. 
For general odd order differential operators, the projector $\gimel$
has a complicated expression. Examples of physical relevance of odd order 
differential operators are Dirac ones with $n$ equal one describing 
quadratic fluctuations in phase space \cite{Roepstorff,Zinn}. 
Fortunately, simplifications occur in that case since $\gimel$ reduces to zero 
or the identity when the matrix prefactor of the highest derivative in $O$ 
is the identity.
The proof of (\ref{Forman-determinants:traceclass}) is rather technical, 
the interested reader is referred to \cite{Forman} for details.\\
Cauchy boundary conditions provide the simplest application of Feynman's
determinant flow equation (\ref{Forman-determinants:differential}).
The family of Sturm-Liouville operators (\ref{Forman-determinants:differential})
is seen to share the same functional determinant if Cauchy boundary conditions
are imposed
\begin{eqnarray}
\ln \frac{\mathrm{Det}\fo_{\bcc\,,\tau_1}}{\mathrm{Det}\fo_{\bcc\,,\tau_0}}=\frac{1}{2}
\int_{\tau_{0}}^{\tau_{1}}d\tau \,\int_{T'}^{T}dt
\tr\,[\left(\de_\tau\,\vptau\,-\,\de_\tau\,\vptau^\dagger\,\right)\mass^{-1}]=0
\label{Forman-determinants:Cauchy}
\end{eqnarray}
The result will be exploited to construct explicitly functional determinants with 
general boundary conditions.

The second difficulty arising from (\ref{Forman-determinants:homotopy}) concerns the
classification of the admissible smooth deformations or, in the mathematical jargon, 
homotopy transformations.\\
Although the problem is general the discussion will focus on the 
Sturm-Liouville family (\ref{Forman-determinants:differential}) with {\em self-adjoint}
boundary conditions $\bc$. 
The smooth dependence of the spectrum of $\fobctau$ on $\tau$ may produce a divergence 
of the Green function if a zero eigenvalue is crossed during the deformation. 
Divergences are avoided if the homotopy connects operators having the same Morse index
$\ind\,\fobc$. The loosening of the condition leads to classify the homotopy 
transformations by means of the Morse indices of the operators they connect. 
Namely, the Morse index can be interpreted as the {\em winding number} of the phase 
of the functional determinant 
$\mathrm{Det}\fobctau$. Two functional determinants are then said to pertain to the same 
homotopy class if they have the same winding number.
On the other hand, if in principle the knowledge of the index is required in order to
construct functional determinants from (\ref{Forman-determinants:homotopy}) it is also 
possible to proceed in the opposite direction. 
In chapter~\ref{Morse} it will be shown that if a self-adjoint Sturm-Liouville 
operator $\fobc$ is smoothly deformed by the introduction of a positive definite,  
self-adjoint perturbation, its Morse index can only decrease or remain constant. 
Hence the number of negative eigenvalues is also specified by number of zeroes, counted 
with their degeneration, encountered by the determinant as the coupling constant of the 
positive definite perturbation is increased until the overall Sturm-Liouville operator 
becomes positive definite.\\
For the moment the knowledge of the index will be assumed and the attention focused
on the derivation of the explicit expression of functional determinants of 
Sturm-Liouville operators.

\subsection{Forman's identity}
\label{Forman:identity}

Forman's identity \cite{Forman} relates the ratio of the functional determinants 
of an elliptic operator $O$ with boundary conditions $\bc_1$ and $\bc_2$ to the 
determinant of the Poisson map of $O$. The Poisson map is the fundamental solution 
of the homogeneous problem associated to $O$.
Explicitly Forman's identity reads
\begin{eqnarray}
\de_\tau\mathrm{ln}\frac{\mathrm{Det}O_{\mathfrak{B}_2\,,\tau}}
{\mathrm{Det}O_{\mathfrak{B}_1\,,\tau}}
=\de_\tau\mathrm{ln}\det\mathcal{G}_{\tau}(\bar{\bc_2},\bar{\bc_1})
\label{Forman-identity:general}
\end{eqnarray}
$\mathcal{G}_{\tau}(\bar{\bc_2},\bar{\bc_1})$ is the Poisson map applied 
to boundary conditions $\bar{\bc}_{1}$ given by the linear complement 
of $\bc_1$ and projected on the linear complement $\bar{\bc}_{2}$ of 
the boundary conditions $\bc_2$.  \\
Although far more general, Forman's identity will be 
now derived for Sturm-Liouville operators. The proof goes along the same lines 
of the general case with the advantage that all operations can be given an 
explicit expression.

The functional determinant for Cauchy boundary condition 
permits to write a more general determinant flow equation
\begin{eqnarray}
&&\de_\tau\,\mathrm{ln}\frac{\mathrm{Det}\fobctau}{\mathrm{Det}\foCtau}\,=\,
\nonumber\\
&&\qquad\int_{T'}^{T}\int_{T'}^{T}dtdt'\delta(t-t')\,
\,\mathrm{Tr}\left\{(\de_{\tau}\fo_{\tau})(t)\,\left[
\Grtau(t,t')-\GrCtau(t,t')\right]\right\}
\label{Forman-identity:homotopy}
\end{eqnarray}
The above equation is well defined. The difference of two Green functions
\begin{eqnarray}
\fo_{\tau}(t)\,\left[\Grtau(t,t')-\GrCtau(t,t')\right]=0
\label{Forman-identity:difference}
\end{eqnarray}
is solution of the Poisson homogeneous problem $\fo\,f=0$ and therefore
does not have singularities on the diagonal. There is a further advantage 
in writing (\ref{Forman-identity:homotopy}): the integral on the right hand side can be 
performed exactly. The proof requires though to reformulate the
Green function problem as a first order one in $2\,d$-dimensions.
This is possible in consequence of the identity
\begin{eqnarray}
&&\int_{T'}^{T}\int_{T'}^{T}dtdt'\,\delta(t-t')\,
\mathrm{Tr}\left\{(\de_{\tau}\fo_{\tau})(t)\,\left[
\Grtau(t,t')-\GrCtau(t,t')
\right]\right\}=
\nonumber\\
&&\int_{T'}^{T}\int_{T'}^{T}dtdt'\,\delta(t-t')
\mathrm{Tr}\left\{(\de_{\tau}\po_{\tau})(t)
\left[\pobc(t,t')-\poC(t,t')\right]\right\}
\label{Forman-identity:passagetots}
\end{eqnarray}
where
\begin{eqnarray}
\po_{\tau}:=
\left[
\begin{array}{cc}
\id_d\,\frac{d\,}{dt}\quad&\quad-\, \id_d\\
\mass^{-1}(\potau-\dvptau) \quad &\quad -\,\frac{d\,}{dt}
-\mass^{-1}(\dmass+\vptau-\vptau^{\dagger})
\end{array}
\right]
\label{Forman-identity:Poincare}
\end{eqnarray}
and $\pobc(t,t')$, $\poC(t,t')$ are the Green functions in tangent space:
\begin{eqnarray}
&&\po_\tau(t)\,\pobc(t,t')\,=\,\delta(t-t')\,\id_{2\,d}\,
\qquad\qquad t, t'\,\in\,[T'\,,T]
\nonumber\\
&&\mathcal{Y}_{1}\,\pobc(T',t')+
\mathcal{Y}_{2}\,\pobc(T,t')\,=\,0
\label{Forman-identity:Green}
\end{eqnarray} 
The statement follows by a direct computation.\\
Note that according to (\ref{Forman-identity:Green}) the boundary conditions
$\bc$ are ``transversal'' to the matrices $\mathcal{Y}_{1}^{\dagger}$, 
$\mathcal{Y}_{2}^{\dagger}$. These latter describe the linear complement
$\bar{\bc}$ of $\bc$ in a sense which will be enunciated more precisely in 
section~\ref{Morse:geometry}.\\ 
The advantage of working in the tangent space is that the equations of the
motion and the boundary conditions can be treated on the same footing.
To wit, the homogeneous problem associated to (\ref{Forman-identity:Poincare}) is
\begin{eqnarray}
&&\po_{\tau}(t)\,\tfu_{\tau}(t\,,t')\,=\,0
\nonumber\\
&&\tfu_{\tau}(t'\,,t')\,=\,\id_{2\,d}
\label{Forman-identity:tangentspace}
\end{eqnarray}
Hence, the Green function fulfilling Cauchy boundary condition $\bcc$ 
(\ref{Forman-determinants:Cauchybc}) is
\begin{eqnarray}
&&\poC(t,t')\,=-\,\tfu(t\,,t')\,\theta(t-t')
\nonumber\\
&&\theta(t-t')\,=\,\left\{\begin{array}{cc} 1\quad & t\,>\,t' \\ 0 \quad & t\,\leq\,t'\end{array}
\right.
\label{Forman-identity:Cauchy}
\end{eqnarray}
By (\ref{Forman-identity:tangentspace}), (\ref{Forman-identity:Cauchy}) the Poisson
map $\tfu_{\tau}(t\,,t')$ of $\po_\tau$ is seen to obey boundary conditions
linearly complementary to Cauchy's.\\
Green functions satisfying boundary conditions $\bc$ are reducible to a 
linear combination of the Cauchy Green function with a particular
solution of the homogeneous equation
\begin{eqnarray} 
\pobc(t,t')\,=\,\poC(t,t')+\tfu_{\tau}(t,T')\,\Psi(T',t')
\label{Forman-identity:Greensolution}
\end{eqnarray}
since the compatibility condition
\begin{eqnarray}
\left[\mathcal{Y}_{1}\,+\mathcal{Y}_{2}\,\tfu_{\tau}(T\,,T')\right]\,\Psi(T',t')\,=\,
-\,\mathcal{Y}_{2}\,\poC(T,t')
\label{Forman-identity:boundary}
\end{eqnarray}
admits a unique solution for $\Psi(T',t')$ each time $\po_{\bc,\tau}$ does not 
have zero modes.
Thus the integrand in (\ref{Forman-identity:passagetots}) can be rewritten as
\begin{eqnarray}
&&\mathrm{Tr}\left\{(\de_{\tau}\po_{\tau})(t)\,\tfu_{\tau}(t,T')\Psi(T',t')
\right\}
\nonumber\\
&&=\mathrm{Tr}\left\{(\de_{\tau}\po_{\tau})(t)\,\tfu_{\tau}(t,T')
\frac{-\,1}{\mathcal{Y}_{1}\,+\mathcal{Y}_2\,\tfu_{\tau}(T\,,T')}
\mathcal{Y}_{2}\,\poC(T,t')\right\}
\label{Forman-identity:innerintegral}
\end{eqnarray}
The derivative of $\po_\tau$ along the homotopy can be eliminated using the identity
\begin{eqnarray}
\de_{\tau}\left[\po_\tau(t)\,\tfu_{\tau}(t\,,T')\right]\,=\,0
\label{Forman-identity:flow}
\end{eqnarray}
Furthermore, the circular symmetry of the trace allows to write 
\begin{eqnarray}
&&\mathrm{Tr}\left\{\po_{\tau}(t)\,(\de_{\tau}\tfu_{\tau})(t,T')\,
\frac{1}{\mathcal{Y}_{1}\,+\mathcal{Y}_{2}\,\tfu_{\tau}(T\,,T')}
\mathcal{Y}_{2}\,\poC(T,t')\right\}=
\nonumber\\
&&\mathrm{Tr}\left\{\poC(T,t')\,\po_{\tau}(t)
(\de_{\tau}\tfu_{\tau})(t,T')\,\frac{1}{\mathcal{Y}_{1}\,+\mathcal{Y}_{2}\,
\tfu_{\tau}(T\,,T')}\mathcal{Y}_{2}\right\}
\label{Forman-identity:argument}
\end{eqnarray}
An integration by parts and the $\delta$-function in the integrand of 
(\ref{Forman-identity:passagetots}) are then used to carry over from $t$ to $t'$ 
the derivatives in $\po_{\tau}$.
The explicit form of the Cauchy Green function (\ref{Forman-identity:Cauchy}) 
substantiates the otherwise formally obvious equality
\begin{eqnarray}
\poC(T,t')\overleftarrow{\po_{\tau}}(t')\,=\,\delta(T-t')\,\id_{2\,d}
\label{Forman-identity:transpose}
\end{eqnarray}
Substituting the last equation in (\ref{Forman-identity:passagetots}) one 
obtains
\begin{eqnarray}
&&\de_\tau\,\mathrm{ln}\frac{\mathrm{Det}\fobctau}{\mathrm{Det}\foCtau}\,=\,
\nonumber\\
&&\int_{T'}^{T}\!\int_{T'}^{T}\!\!dtdt'\,\delta(t-t') \delta(T-t')
\mathrm{Tr}\left\{
\frac{1}{\mathcal{Y}_{1}\,+\mathcal{Y}_2\,\tfu_{\tau}(T\,,T')}
\mathcal{Y}_{2}\,(\de_{\tau}\tfu_{\tau})(t,T')\right\}
\label{Forman-identity:finalintegral}
\end{eqnarray}
Therefore, provided the homotopy transformation does not encounter zero modes, 
the {\em Forman's identity} (\ref{Forman-identity:general}) becomes:
\begin{eqnarray}
\frac{\mathrm{Det}\fo_{\bc\,,\tau_1}}{\mathrm{Det}
\fo_{\bc\,,\tau_0}}\,=\,
\frac{\mathrm{det}[\mathcal{Y}_{1}+\mathcal{Y}_{2}\,\tfu_{\tau_{1}}(T\,,T')]}
{\mathrm{det}[\mathcal{Y}_{1}+\mathcal{Y}_{2}\,\tfu_{\tau_{0}}(T\,,T')]}
\label{Forman-identity:identity}
\end{eqnarray}

\subsection{Applications of the Forman identity}
\label{Forman:examples}

The physical content of Forman's identity is to assert that the leading 
contribution of quantum fluctuations to the semiclassical expansion is determined
by the linearised classical dynamics around the stationary trajectory. 
Canonical covariance of classical mechanics is better displayed in phase
space. Thus, it is convenient to rephrase Forman's identity in terms of phase 
space quantities. 

The linearised tangent space flow $\tfu(T\,,T')$ is connected by a linear transformation
to the linearised {\em phase space} flow $\fu(T,T')$:
\begin{eqnarray}
\fu(t,t')\,=\,\mathfrak{T}(t)\,\tfu(t,t')\mathfrak{T}^{-1}(t')
\label{Hamilton-examples:linear}
\end{eqnarray}
where the matrix $\mathfrak{T}(t)$ is defined in appendix~\ref{Hamilton:elementary}.
The linear phase space flow $\fu(T,T')$ is specified by the Poisson brackets 
(appendix~\ref{Hamilton}) of the flow solution of the first variation 
\begin{eqnarray}
\fu(T\,,T')\,=\,\left[\begin{array}{cc} 
\{q_{\cl}(T)\,,p_{\cl}(T')\}_{\mathrm{P.b.}}\quad & \quad-\,\{q_{\cl}(T)\,,q_{\cl}(T')\}
_{\mathrm{P.b.}} \\
\{p_{\cl}(T)\,,p_{\cl}(T')\}_{\mathrm{P.b.}}\quad & \quad-\,\{p_{\cl}(T)\,,q_{\cl}(T')\}
_{\mathrm{P.b.}}
\end{array} \right]
\label{Forman-examples:fundamental}
\end{eqnarray}
where $q_{\cl}$ and $p_{\cl}$ are classical positions and momenta and
each bracket symbolically represents a $d\,\times\,d$ set of relations.

In terms of phase space quantities the Sturm-Liouville problem 
is restated as
\begin{eqnarray}
&&(L\,\lambda_n)(t)\,=\,\ell_n\,\lambda_n(t)\,,\qquad\qquad\qquad\qquad
T'\,\leq t\,\leq\,T
\nonumber\\
&&\mathsf{Y}_{1}\,\left[\begin{array}{c} \lambda_n(T')\\
(\nabla \lambda_n)(T')\end{array}\right]+
\mathsf{Y}_{2}\,\left[\begin{array}{c} \lambda_n(T)\\
(\nabla \lambda_n)(T)\end{array}\right]\,=\,0
\label{Forman-examples:SturmLiouville}
\end{eqnarray}
while Forman's identity becomes 
\begin{eqnarray}
\mathrm{Det}\fobc([T',T])=\,\varkappa_{\bc}
\,\mathrm{det}[\mathsf{Y}_{1}+\mathsf{Y}_{2}\,\fu(T\,,T')]
\label{Forman-examples:identity}
\end{eqnarray}
If the homotopy transformation takes place in the space of self-adjoint operators, 
the prefactor $\varkappa_{\bc}$ can be evaluated on 
any reference operator $\fo_{\bc\,,\tau_0}$
\begin{eqnarray}
\varkappa_{\bc}\,=\,\frac{\mathrm{Det}\fo_{\bc\,,\tau_0}([T',T])\,
e^{i\,\pi (\ind \fobc-\ind \fo_{\bc\,,\tau_0})}}
{\mathrm{det}[\mathsf{Y}_{1}+\mathsf{Y}_{2}\,\fu_{\tau_{0}}(T\,,T')]}
\label{Forman-examples:constant}
\end{eqnarray}
by keeping track of the change of the Morse index along the homotopy path.
The situation is illustrated by some examples.

\subsubsection{Dirichlet boundary conditions}

Dirichlet boundary conditions are the most common in applications. They are obeyed
by fluctuations along a classical trajectory connecting in $[T'\,,T]$ two assigned 
positions in space. 
Therefore they are associated to the semiclassical approximation of the propagator.
In the Dirichlet case the phase space matrix pair in (\ref{Forman-examples:SturmLiouville})
can be chosen as 
\begin{eqnarray}
\mathsf{Y}_{1}=\left[ 
\begin{array}{cc} \id_d \quad&\quad 0 \\ 
0\quad &\quad 0 \end{array} \right], \quad
\mathsf{Y}_{2}=\left[ 
\begin{array}{cc} 0\quad &\quad 0 
\\ \id_d\quad &\quad 0  \end{array} \right]
\label{Forman-examples:Dirichletmatrices}
\end{eqnarray}
The functional determinant is therefore 
\begin{eqnarray}
\mathrm{Det} \foD([T',T])\,=\,\varkappa_{\mathrm{Dir.}}\,\det 
[-\,\{q_{\cl}(T)\,,q_{\cl}(T')\}_{\mathrm{P.b.}}]
\label{Forman-examples:Dirichletdeterminant}
\end{eqnarray}
If the Lagrangian $\la$ describes a single particle with mass $m$ in flat space
the proportionality constant is most conveniently evaluated from free particle 
motion.
The formulae given in appendix~\ref{formulae} give in one dimension
\begin{eqnarray}
&&\left.\mbox{Det}\left[-m\,\frac{d^{2}}{dt^{2}}\right]\right|_{\mathrm{Dir.}}\!\!([T',T])
=\left[\frac{2\,\pi\,\hbar\,(T-T')}{m}\right]e^{i\,\frac{\pi}{2}}\,,
\nonumber\\
&&-\,\{q_{\cl}^{\alpha}(T)\,,q_{\cl}^{\beta}(T')\}_{\mathrm{P.b.}}=\left(\frac{T-T'}{m}\right)\,
\delta^{\alpha\,\beta}
\label{Forman-examples:freemotion}
\end{eqnarray}
Quantum fluctuations for free particle motion are quadratic with positive definite 
spectrum. Therefore the Morse index is equal to zero. When such information is 
combined with (\ref{Forman-examples:freemotion}) one finally gets into
\begin{eqnarray}
&\mathrm{Det}&\!\foD([T',T])=
\nonumber\\
&&(2\,\pi\,\hbar)^{d}\,e^{\,i\,\pi\,
\left[\ind\foD([T',T])+\frac{d}{2}\right]}|\det 
\{q_{\cl}(T)\,,q_{\cl}(T')\}_{\mathrm{P.b.}}|
\label{Forman-examples:Dirichlet}
\end{eqnarray}
The result is usually presented in a slightly different form.
The action integral evaluated along $q_{\cl}$ becomes a function of the
initial and final point of the trajectory
\begin{eqnarray}
\ac(Q,T,Q',T')=\int_{T'}^{T}dt\,\la(q_{\cl}\,,\dot{q}_{\cl}\,,t)
\label{Forman-examples:Hamiltonprincipal}
\end{eqnarray}
The action with the above functional dependence is often referred to 
as Hamilton principal function.
The partial derivatives of $\ac(Q,T;Q',T')$ are proven to fulfill 
\cite{Arnold,LandauLifshitz,deAlmeida}
\begin{eqnarray}
&&P_\alpha\,:=\,p_{\cl\,,\alpha}(T)\,=\,\frac{\de \ac}{\de Q^{\alpha}}\nonumber\\
&&P_\alpha^{\prime}:=\,p_{\cl\,,\alpha}(T')=-\frac{\de \ac}{\de Q^{\prime\,\alpha}} 
\label{Forman-examples:momenta}
\end{eqnarray}
The total differential of the initial momenta with respect to the positions at
time interval ends is obtained from the second derivatives of the Hamilton 
principal function
\begin{eqnarray}
dp_{\cl\,,\alpha}(T')\,=-\,d Q^{\beta}\frac{\de^{2}\,\ac}{\de Q^{\beta}\de Q^{\prime\,\alpha}}-
\,d Q^{\prime\,\beta}\frac{\de^{2}\,\ac}{\de Q^{\prime\,\beta}\de Q^{\prime\,\alpha}}
\label{Forman-examples:variation}
\end{eqnarray}
The Poisson brackets $\{q_{\cl}^{\alpha}(T)\,,q_{\cl}^{\beta}(T')\}_{\mathrm{P.b.}}$
yield the displacement of the final position $q_{\cl}^{\alpha}(T)$ versus a change
of the initial momentum $p_{\cl\,\beta}(T')$ while all other initial momenta and positions 
are kept fixed. 
This is exactly the same as the inverse of the differential 
(\ref{Forman-examples:variation}) when the $Q^{\prime\,\beta}$'s are held
constant. It follows
\begin{eqnarray}
\left|\mathrm{det}\frac{\de^{2}\,\ac}{\de Q^{\beta}\de Q^{\prime\,\alpha}}\right|\,=\,
\frac{1}{|\mathrm{det}[\{q_{\cl}(T)\,,q_{\cl}(T')\}_{\mathrm{P.b.}}]|}
\label{Forman-examples:determinants}
\end{eqnarray} 
Thus the familiar form of the semiclassical propagator is recovered
\begin{eqnarray}
&&\feo(Q,T|Q',T')\,\cong\,
\nonumber\\
&&\qquad\sqrt{\left|\det\frac{\de^{2}\,\ac}{\de Q^{\beta}\de Q^{\prime\,\alpha}}\right|}
\frac{e^{i\,\frac{\ac(Q,T,Q',T')}{\hbar}-i\,\frac{\pi}{2}\,\left[
\ind\foD([T',T])+\frac{d}{2}\right]}}{[2\,\hbar\,\pi]^{d/2}}
\label{Forman-examples:propagator}
\end{eqnarray}
The semiclassical propagator is named after Van Vleck \cite{VanVleck}, Pauli 
\cite{Pauli} and DeWitt-Morette \cite{Morette} who rediscovered it in independent 
contributions. The relation with functional determinants with Dirichlet boundary conditions
was first proven by Gel'fand and Yaglom \cite{GelfandYaglom}.
Finally the complete form of the semiclassical approximation featuring the Morse 
index for open extremals is due to Gutzwiller. 

The expression of the semiclassical propagator in curved spaces is also known
\cite{DeWitt,LevitSmilansky2}. In order to recover it from Forman's theorem one 
needs to evaluate the functional determinant in a reference case. The free 
particle supplies again the needed information although more work is demanded
in comparison to the flat space counterpart. The interested reader can find the 
details of the computation in appendix~9.2 of ref. \cite{LangoucheRoekaertsTirapegui}.
It turns out that the resulting functional determinant predicts the semiclassical 
asymptotics 
\begin{eqnarray}
&&\feo(Q,T|Q',T')\,\cong\,
\nonumber\\
&&\qquad\sqrt{\left|\det\frac{\de^{2}\,\ac}{\de Q^{\beta}\de Q^{\prime\,\alpha}}\right|}
\,\frac{e^{i\,\frac{\ac(Q,T,Q',T')}{\hbar}
-i\,\frac{\pi}{2}\left[\ind\foD([T',T])+
\frac{d}{2}\right]}}{|g(Q)|^{\frac{1}{4}}[2\,\hbar\,\pi]^{d/2}|g(Q')|^{\frac{1}{4}}}
\label{Forman-examples:DeWitt}
\end{eqnarray}
The result coincides with De Witt's computation of the quantum propagator 
\cite{DeWitt} holding for arbitrary time separations. 
For small times the Morse index is zero as will be discussed in 
chapter~\ref{Morse}. Moreover for small times and small spatial separations
 \cite{DeWittMoretteElworthyNelsonSammelman,Schulman} the equality holds
\begin{eqnarray}
&&\left(\frac{T-T'}{m}\right)^{\frac{d}{2}}\,\frac{1}{|g(Q)|^{\frac{1}{4}}}
\sqrt{\left|\det\frac{\de^{2}\,\ac}{\de Q^{\beta}\de Q^{\prime\,\alpha}}\right|}
\frac{1}{|g(Q')|^{\frac{1}{4}}}
\nonumber\\
&&\qquad= 1+\frac{R_{\mu\,\nu}(Q')}{6}(Q-Q')^{\mu}(Q-Q')^{\nu}+o(\Delta Q^3)
\label{Forman-examples:approximations}
\end{eqnarray}
in the limit 
\begin{eqnarray}
\qquad O(dT)\sim O(Q-Q')^2\,\downarrow\, 0
\label{Forman-examples:limit}
\end{eqnarray}
Thus the small time expansion of section~\ref{Forman:measure} is recovered if 
$\ac$ is identified with the action specified by the classical part, 
$\hbar$ set to zero, of the Lagrangian (\ref{Forman-measure:DeWitt}).

\subsubsection{Periodic Boundary conditions}

Periodic boundary conditions are those of interest for trace formulae. 
They are enforced by the choice 
\begin{eqnarray}
\mathsf{Y}_{1}\,=\,-\,\mathsf{Y}_{2}\,=\,\id_{2\,d}
\label{Forman-examples:periodicmatrices}
\end{eqnarray}
Forman's identity yields
\begin{eqnarray}
\mathrm{Det}\foP([T',T])=\varkappa_{\mathrm{Per.}}\,
\mathrm{det}\left[\id_{2\,d}-\fu(T,T')\right]
\label{Forman-examples:periodic}
\end{eqnarray}
If the Sturm-Liouville operator arises from the second variation evaluated on
a prime periodic trajectory, the time difference $T_{\cl}=T-T'$ is the period 
of the orbit. The attribute prime means that the trajectory has wended its path 
along the orbit only once.
The matrix 
\begin{eqnarray}
\mon\,:=\,\fu(T,T')
\end{eqnarray}
is referred to as the monodromy matrix. The stability of any trajectory 
on the orbit is governed by the same $\mon$ 
\cite{Krein,GelfandLidskii,Moser,Arnold,deAlmeida}.  
A short summary of the properties of the monodromy is provided in 
appendix~\ref{Hamilton:periodic}. As it will be shown in chapter~\ref{moduli},
the monodromy matrix acquires an eigenvalue and a generalised eigenvalue equal to one 
in correspondence of any one parameter continuous symmetry of the classical trajectory.  
The functional determinant (\ref{Forman-examples:periodic}) is therefore zero. 
Forman's identity can be applied generically only to Sturm-Liouville
operators stemming from non-autonomous Lagrangian. 

The argument does not apply to oscillators. In a generic time interval $[T',T]$, 
stable oscillators do not admit classical periodic orbits. 
In consequence the functional determinant is non zero for it receives contribution
only by quantum fluctuations. The Lagrangian of a one dimensional unstable oscillator 
is also a positive definite quadratic form
\begin{eqnarray}
\la\,=\,\frac {m\,\dot{q}^2}{2}+\frac{m\,\omega^2\,q^2}{2}
\label{Forman-examples:unstable}
\end{eqnarray}
so the Morse index is equal to zero. The unstable oscillator provides the reference
case in flat spaces wherefrom $\varkappa_{\mathrm{Per.}}$ can be determined.
By comparison with the path integral formulae of appendix~\ref{formulae}:
\begin{eqnarray}
&&\mathrm{Det}\left[-\,\de_t^2+\omega^2\right]_{\mathrm{Per.}}([T',T])=
4\,\sinh^{2}\left(\omega\,\frac{T-T'}{2}\right)
\nonumber\\
&&\mathrm{det}\left[\id_{2}-\fu(T,T')\right]=-\,4\,\sinh^{2}\left(\omega\,\frac{T-T'}{2}\right)
\nonumber
\end{eqnarray}
it is found
\begin{eqnarray}
\mathrm{Det}\,\foP(T,T')=e^{i\,\pi\,\ind\foP([T',T])}\,
|\mathrm{det}\left[\id_{2\,d}-\fu(T,T')\right]|
\label{fluctuationperiodics}
\end{eqnarray}

\subsubsection{Smooth deformation of the boundary conditions}

A final example of applications of the Forman identity is the variation 
of functional determinants under rotations of the boundary conditions 
\cite{Forman}.

Under a rotation of phase space such that
\begin{eqnarray}
x^{(\mathsf{R})}(t)=\mathsf{R}(t)\,x(t)
\label{Formansexamples:rotation}
\end{eqnarray}
the linear flow transforms as
\begin{eqnarray}
\fu^{(\mathsf{R})}(T,T')=\mathsf{R}^{-1}(T)\,\fu(T,T')\,\,\mathsf{R}(T')
\nonumber
\end{eqnarray}
Accordingly, the functional determinant of (\ref{Forman-examples:SturmLiouville}) 
obeys
\begin{eqnarray}
\frac{\mathrm{Det} \fobc^{(\mathsf{R})}([T',T])}{\mathrm{Det} \fobc([T',T])}=\frac{\det 
\left[\mathsf{Y}_{1}+\mathsf{Y}_{2}\,\fu^{(\mathsf{R})}(T,T')\right]}
{\det\left[\mathsf{Y}_{1}+\mathsf{Y}_{2}\,\fu(T,T')\right]}
\label{Formansexamples:rotated}
\end{eqnarray}


%% file: Morse.tex
\section{Intersection forms and Morse index theorems}
\label{Morse}

The construction of functional determinants by means of Feynman's equation
(\ref{Forman-determinants:homotopy}) requires an homotopy classification
of the determinant bundle. In the case of self-adjoint Sturm-Liouville operators 
the Morse index provides the classification. 

Morse indices are topological invariants. The circumstance gives great 
freedom in their evaluation. The classical index theory
devised by Morse \cite{Morse,Milnor} provides the explicit expressions of 
the indices associated to {\em general} self-adjoint boundary conditions. 
The ``Sturm intersection theory'' introduced by Bott \cite{Bott} and further 
refined by Duistermaat \cite{Duistermaat} and Salamon and Robbin \cite{RobbinSalamon} 
evinces the equivalence of Morse theory with the topological characterisation 
(Maslov index theory) of the structural stability of solutions of classical variational 
problems introduced by Maslov \cite{MaslovFedoriuk} and Arnol'd \cite{Arnold2}. 

The ``Sturm intersection theory'' focuses on the properties 
of the eigenvalue (Fredholm) flow \cite{RobbinSalamon2} of self-adjoint 
Sturm-Liouville operators. The scope of this chapter is to provide an 
elementary review of the intersection theory. The point of view
is complementary to one adopted in the literature on the Gutzwiller 
trace formula to (see for example \cite{Littlejohn2,Littlejohn,CreaghRobbinsLittlejohn,Robbins})
where Maslov indices are deduced from the properties of metaplectic operators.

\subsection{Morse index and symplectic geometry}
\label{Morse:geometry}

Let $\sbc$ be the space of square integrable vector fields in $\rn^{d}\times[T'\,,T]$ 
satisfying some assigned boundary conditions $\bc$. The differential operation
\begin{eqnarray}
\fo:=-\,\id_{d}\frac{d^{2}\,}{dt^{2}}-
\mass^{-1}(\dmass+\vptau-\vptau^\dagger)
\frac{d\,}{dt}+\mass^{-1}\,(\potau-\dvptau)
\label{Morse-geometry:SL}
\end{eqnarray}
acting on $\sbc$ defines a Sturm-Liouville operator $\fobc$. 
The operator is self-adjoint with respect to the scalar product 
(\ref{Forman-quadratic:scalarproduct}) if for all $\xi$, $\chi$ belonging 
to $\sbc$ the boundary form on the right hand side of the equality 
\begin{eqnarray}
\langle\chi,\fobc\, \xi\rangle=\langle\fobc^{\dagger} \chi\,, \xi\rangle-
(\chi^{\dagger}\nabla\xi)|_{T'}^{T}+(\xi^{\dagger}\nabla\chi)|_{T'}^{T}
\label{Morse-geometry:selfadjoint}
\end{eqnarray}
vanishes. All Sturm-Liouville operators $\fobc$ considered in this chapter 
will satisfy the condition. \\
In ref. \cite{Bott} R.~Bott restated the requirement of vanishing boundary form 
in a way explicitly invariant under linear canonical transformations. 
He observed that the boundary form can be recast as the difference of skew products 
of the phase space lifts of the vector fields $\xi$, $\chi$: 
\begin{eqnarray}
-(\chi^{\dagger}\nabla\xi)(t)+(\xi^{\dagger}\nabla\chi)(t)=
(\liftd\chi\sym\, \lift\xi)(t)\,,\qquad\qquad t=T'\,,T
\label{Morse-geometry:lift}
\end{eqnarray}
where $\sym$ is the symplectic pseudo-metric in Darboux coordinates 
(appendix~\ref{Hamilton}),
\begin{eqnarray}
\sym=\left[\begin{array}{cc} 0\,\,&\,\, -\,\id_d 
\\ \id_d \,\,&\,\, 0\end{array}\right]
\label{Morse-geometry:sym}
\end{eqnarray}
and $\Lambda$ denotes the lift operation. This latter associates to configuration space
vector fields their canonical momenta. For a linear theory it means 
\begin{eqnarray}
\lift\chi=\left[\begin{array}{c} \chi \\ \nabla\chi\end{array}\right]
\end{eqnarray}
Hence $\fobc$ is self-adjoint if for all $\xi$, $\chi$ in $\sbc$
\begin{eqnarray}
0=(\liftd \chi\sym\, \lift \xi)(T)-(\liftd \chi\sym\, \lift \xi)(T')
\label{Morse-geometry:skew}
\end{eqnarray}
The requirement is fulfilled if the boundary conditions $\bc$ admit 
the representation 
\begin{eqnarray}
\left\{\begin{array}{c}
\lift \chi(T')=\Bott_1\,y
\\
\lift \chi(T)=\Bott_2\,y
\end{array}\right.\,,
\quad\quad \Bott_1^{\dagger}\,\sym\,\Bott_1= \Bott_2^{\dagger}\,\sym\,\Bott_2\,, 
\quad\quad \Bott_1\,,\Bott_2\,\in\,\rn^{2 d}\times\rn^{2 d}
\label{Morse-geometry:Bott}
\end{eqnarray}
for some $y$ in $\rn^{2 d}$ and matrices $(\Bott_1\,,\Bott_2)$ independent
of the interval end times $T'$, $T$. In the literature 
the matrices $(\Bott_1\,,\Bott_2)$ are often referred to as Bott's 
Hermitian pair. 

Bott pairs characterise geometrically self-adjoint boundary conditions $\bc$.\\
A Sturm-Liouville problem in $d$ dimensions is completely specified by $2 d$ 
linearly independent boundary conditions. If they are self-adjoint, they define 
in $\rn^{4 d}$ the $2 d$-dimensional subspace 
\begin{eqnarray}
\bc=\{\,y\,\in\,\rn^{4 d}\,|\,y=
\left[\begin{array}{c}\Bott_1\,x\,\\\Bott_2\,x\end{array}\right]\!\!,\,\,
\Bott_1^{\dagger}\sym \Bott_1=\Bott_2^{\dagger}\sym \Bott_2\,,\,\,\,
\Bott_1,\Bott_2\in\rn^{2 d}\times\rn^{2 d}\,\}
\label{Morse-geometry:bc}
\end{eqnarray} 
Furthermore, in $\rn^{4 d}$ the condition (\ref{Morse-geometry:skew}) 
is equivalent to the vanishing of the non degenerate symplectic form $\beth$ obtained by 
evaluating $\sym\oplus(-\sym)$ on any two elements of $\bc$
\begin{eqnarray}
\beth(\chi,\xi):=
\left[\!\begin{array}{c} \lift\chi(T')\\ \lift\chi(T) \end{array}\!\right]^{\dagger}\!
\left[\begin{array}{cc} \sym\,\,&\,\, 0 
\\ 0 \,\,&\,\, -\,\sym\end{array}\right]\!\!
\left[\!\begin{array}{c} \lift\xi(T')\\ \lift\xi(T)\end{array}\! \right]
\label{Morse-geometry:beth}
\end{eqnarray}
The maximal number of linear independent vectors annihilating a non 
degenerate symplectic form like (\ref{Morse-geometry:beth}) is 
exactly $2 d$ in a $4 d$-dimensional space \cite{Arnold}. 
A manifold is called {\em Lagrangian} if it contains the maximal 
number of linear independent vectors pairwise annihilating a symplectic form.
Hence $\bc$ is a Lagrangian manifold, specifically an hyper-plane, of 
$\rn^{4 d}$ with respect to $\beth$.\\
In the enunciation of Forman's theorem, the boundary conditions associated to 
$\fobc$ were imposed in the form
\begin{eqnarray}
\mathsf{Y}_1\,\lift\lambda_n(T')+\mathsf{Y}_2\,\lift\lambda_n(T)=0
\label{Morse-geometry:Formanbc}
\end{eqnarray}
for any eigenvector $\lambda_n$ of $\fobc$.
The geometrical meaning is a transversality condition in $\rn^{4 d}$
\begin{eqnarray}
\left[\begin{array}{c}\mathsf{Y}_1^{\dagger}\, x\\ \mathsf{Y}_2^{\dagger}\,x
\end{array}\right]^{\dagger}
\left[\begin{array}{c}\Bott_1\, y\\ \Bott_2\, y
\end{array}\right]=0\,,\qquad\qquad \forall\, x,y\,\in\,\rn^{2 d} 
\label{Morse-geometry:transversality}
\end{eqnarray}
In other words, the matrix pair $(\mathsf{Y}_1^{\dagger},\mathsf{Y}_2^{\dagger})$ 
describes a linear complement of the boundary conditions $\bc$. \\
Also the zero modes of $\fobc$ admit a symplectic geometry characterisation.
The symplectic form $\beth$ vanishes on vectors framed by the graph 
\begin{eqnarray}
\gph\, \mathsf{S} \,y:=\left[\begin{array}{c} y \\ \mathsf{S} y\end{array}\right]\,
\quad\quad \forall y\,\in\,\rn^{2 d}
\label{Morse-geometry:graph}
\end{eqnarray}
of any symplectic matrix $\mathsf{S}$
\begin{eqnarray}
\mathsf{S}^{\dagger}\,\sym\,\mathsf{S}\,=\,\sym
\label{Morse-geometry:symplectic}
\end{eqnarray}
Any flow $\fu(t,T')$ solving a linear Hamiltonian system draws a curve in
the symplectic group $Sp(2 d)$ (appendix~\ref{Hamilton}). In particular
$\fu(t,T')$ can be identified as the phase space flow associated
to the $2 d$ linear independent solutions $\jmath(t)$ of the 
homogeneous problem
\begin{eqnarray}
&&(\fo\,\jmath)^{\alpha}(t)=0
\nonumber\\
&&\jmath^{\alpha}(t)=\fu^{\alpha}_{\beta}(t,T')\jmath^{\beta}(T')+
\fu^{\alpha}_{\beta}(t,T')(\nabla\jmath)^{\beta}(T')\,,\qquad\quad\alpha,\beta=1,...,d 
\label{Morse-geometry:Jacobi}
\end{eqnarray}
In variational problems the $\jmath(t)$'s are called Jacobi fields.
The existence of zero modes of $\fobc$ corresponds to a non empty intersection
between the Lagrangian manifolds $\bc$ and $\gph\fu(T,T')$.
By Forman's theorem and the transversality relation 
(\ref{Morse-geometry:transversality}), 
intersections are analytically characterised by the equivalence
\begin{eqnarray}
\bc\,\cap\,\gph\fu(T,T')\,\neq\,\emptyset \qquad \Leftrightarrow \qquad 
\det[\mathsf{Y}_1+\mathsf{Y}_2\,\fu(T,T')]=0
\label{Morse-geometry:intersection}
\end{eqnarray}
The description of zero modes as intersection of Lagrangian manifolds plays
a major role in the explicit construction of Morse indices.

The semiclassical approximation brings about Sturm-Liouville operators
such that
\begin{eqnarray}
\qac(\chi\,,\xi)\,=\,\langle \chi,\fobc \xi \rangle
\label{Morse-geometry:pairing}
\end{eqnarray}
holds true. The equality offers a direct way to compute the Morse index of $\fobc$ 
\cite{Bott,Duistermaat}. The bilinear form can be rewritten using 
canonical momenta 
\begin{eqnarray}
\qac(\chi\,,\xi)\,=\,\int_{T'}^{T}dt\,
\left[(\nabla\chi)^{\dagger}\,\mass^{-1} \nabla \xi+
\chi^{\dagger}(\pot-\vpt^{\dagger}\mass^{-1}\vpt)\xi\,\right]
\label{Morse-geometry:momenta}
\end{eqnarray}  
The replacement
\begin{eqnarray}
\pot\,\Rightarrow\,\pot+\tau\,\dU
\label{Morse-geometry:replacement}
\end{eqnarray}
renders the quadratic form
\begin{eqnarray}
\qac_{\tau}(\xi\,,\xi)\,=\,\int_{T'}^{T}dt\,
\left[(\nabla\xi)^{\dagger}\,\mass^{-1} \nabla \xi+
\xi^{\dagger}(\pot+\tau\dU-\vpt^{\dagger}\mass^{-1}\vpt)\xi\,\right]
\label{Morse-geometry:perturbed}
\end{eqnarray}  
positive definite for values of the coupling constant $\tau$ large enough 
provided $\dU$ and $\mass$ are {\em strictly positive definite}. 
If the boundary conditions $\bc$ are non-local $\dU$ should be chosen such 
that $\fobctau$ remains well defined. For example, if $\fobc$ is $T-T'$-periodic 
then any matrix time dependent matrix $\dU$ such that 
\begin{eqnarray}
\dU(t+T-T')=\dU(t) 
\end{eqnarray}
is admissible.\\
The Morse index receives contribution each time the deformation path encounters
a Sturm-Liouville $\fobctau$ with zero modes. 
Since an arbitrary vector field $\xi$ in $\sbc$ is independent of $\tau$
\begin{eqnarray}
\de_{\tau}\qac_{\tau}(\xi\,,\xi)\,=\,\langle \xi\,,\dU\,\xi\,\rangle\,>\,0
\label{Morse-geometry:positive}
\end{eqnarray}  
and the index decreases monotonically for increasing $\tau$.\\
It is instructive to see how (\ref{Morse-geometry:positive}) 
comes about by looking directly at the eigenvectors of $\fobc$.
For the sake of the notation, the parametric dependence on $\tau$
at $\tau$ equal to zero is from now on omitted.
The derivative of an eigenvalue of $\fobctau$ reads
\begin{eqnarray}
\de_{\tau}\ellntau=\langle \de_\tau\entau\,,\fobctau \entau\rangle+
\langle \entau\,,\dU \,\entau\rangle+
\langle \entau\,,\fobctau \de_{\tau}\entau\rangle
\label{Morse-geometry:eigenvalue1}
\end{eqnarray}
where $\entau$ denotes the eigenvector
\begin{eqnarray}
(\fobctau \entau)(t)=\ellntau \entau(t)\,,\quad\quad (\lift\entau(T')\,,\lift\entau(T))\,\in\,\bc
\label{Morse-geometry:eigenvalue2}
\end{eqnarray}
A double integration by parts in the last addend in (\ref{Morse-geometry:eigenvalue1}) 
gives
\begin{eqnarray}
\de_{\tau}\ellntau\,=\,\ellntau\de_\tau\langle \entau\,,\entau\rangle+
\langle \entau\,,\dU \entau\rangle+\beth(\de_{\tau}\entau\,,\entau)
\label{Morse-geometry:eigenvalue3}
\end{eqnarray}
The first term is zero because eigenvectors are normalised to one at every $\tau$,
the third term vanishes because for all $\tau$ the eigenvectors satisfy the boundary
conditions. The final result is familiar in quantum mechanics: 
\begin{eqnarray}
\de_{\tau}\ellntau=\langle \entau\,,\dU \entau\rangle
\label{Morse-geometry:eigenvalue}
\end{eqnarray}
Thus the Morse index of $\fobc$ coincides with the number of zero modes encountered 
during the deformation for $\tau$ larger than zero; the contribution at $\tau$ equal to 
zero would add to the Morse index the nullity of $\fobc$.\\
In practice the prescription requires to solve for arbitrary $\tau$ 
the linear Hamiltonian system  
\begin{eqnarray}
&&\sym\frac{d \fu_{\tau}}{dt}(t,T')=\mathsf{H}_{\tau}(t)\,\fu_{\tau}(t,T')
\nonumber\\
&&\fu_{\tau}(T',T')=\id_{2 d}
\label{Morse-geometry:Hamilton}
\end{eqnarray}
with
\begin{eqnarray}
\mathsf{H}_{\tau}(t)=\left[\begin{array}{cc} 
-\pot+ \vpt^{\dagger}\,\mass^{-1}\,\vpt-\tau\,\mathsf{U} 
\quad&\quad -\, \vpt^{\dagger}\mass^{-1}
\\
-\,\mass^{-1}\,\vpt \quad & \quad \mass^{-1}
\end{array}\right]\!(t)
\label{Morse-geometry:Hamiltonian}
\end{eqnarray}
The flow generates the family of linear Lagrangian manifolds 
$\gph\,\fu_{\tau}(T,T')$ of $\rn^{4 d}$. The intersections with 
$\bc\sim (\Bott_1\,,\Bott_2)$ yield the index: 
\begin{eqnarray}
\ind \fobc([T',T]) &=& \sum_{\{\tau>0\,|\,\bc\cap\,\gph\,\fu_{\tau}(T,T')\,\}}\!\!\!\!\!\!
\nul\left\{\Bott_2-\fu_{\tau}(T,T')\Bott_1\right\}
\nonumber\\ 
&=&\sum_{\{\tau>0\,|\,\bc\cap\,\gph\,\fu_{\tau}(T,T')\,\}}\!\!\!\!\!\!
\nul\left\{\mathsf{Y}_1+\mathsf{Y}_2\fu_{\tau}(T,T')\right\} 
\label{Morse-geometry:index}
\end{eqnarray}
In the last row the transversality relation 
(\ref{Morse-geometry:transversality}) has been used.\\
The prescription is illustrated by the following elementary example.

\subsubsection{Example: Morse index of the harmonic oscillator}
\label{Morse-geometry:oscar}

The Sturm-Liouville operator paired with a one dimensional 
harmonic oscillator of mass $m$ acts on functions in $\sbc$ as
\begin{eqnarray}
\fo_{\tau}=-m\,\frac{d^{2}}{dt^{2}}-m\,\omega^2\,(1-\tau)
\label{Morse-geometry-oscar:SL}
\end{eqnarray}
As above, $\tau$ equal zero corresponds to the Sturm-Liouville
operator the Morse index whereof is needed. At $\tau$ equal unity
a semi positive definite operator is certainly attained.
For any $\tau$ the classical equations of the motion admit the
phase space representation 
\begin{eqnarray}
\left[
\begin{array}{cc} 
0 \,\,&\,\, -1
\\
1 \,\,&\,\, 0
\end{array}\right]\,
\left[\begin{array}{c} \dot{q} \\ \dot{p}
\end{array}\right]=
\left[\begin{array}{cc} m\,\omega^2\,(1-\tau) \,\,&\,\, 0
\\
0 \,\,&\,\, 1
\end{array}\right]\left[\begin{array}{c} q \\ p
\end{array}\right]
\label{Morse-geometry-oscar:equation}
\end{eqnarray}
with fundamental solution
\begin{eqnarray}
\fu_{\tau}(t,0)=\left[
\begin{array}{cc} 
\cos(\omega\sqrt{1-\tau}\,t) \quad&\quad 
\frac{\sin(\omega\,\sqrt{1-\tau}\,t)}{m\,\omega\sqrt{1-\tau}}
\\
-\,m\,\omega\sqrt{1-\tau}\,\sin(\omega\sqrt{1-\tau}\, t) \quad&\quad 
\cos(\omega\sqrt{1-\tau}\, t)
\end{array}\right]
\label{Morse-geometry-oscar:fundamental}
\end{eqnarray}
The Morse index can be now computed using Forman's theorem.
Consider 
\begin{enumerate}
\item[i] Dirichlet boundary conditions in $[0,T]$. \\
By (\ref{Forman-examples:Dirichlet}) the absolute value of the 
functional determinant is
\begin{eqnarray}
|\mathrm{Det}\foD([0,T])|=2\,\pi\,\hbar\,
\frac{|\sin(\omega\,\sqrt{1-\tau}\,T)|}{m\,\omega\sqrt{1-\tau}}
\label{Morse-geometry-oscar:Dirichlet}
\end{eqnarray}
The index is given by the number of zeroes for $\tau$ ranging in
$]0,1]$. Zeroes occur for
\begin{eqnarray}
\omega\sqrt{1-\tau}\, T=\,k\,\pi>0 
\label{Morse-geometry-oscar:zeroes}
\end{eqnarray}
and are non degenerate. Hence the result of the path integral computation
of appendix~\ref{formulae:oscar} is recovered: 
\begin{eqnarray}
\ind\foD([0,T])=\mbox{int}\left[\frac{\omega T}{\pi}\right]
\label{Morse-geometry-oscar:Dirichletindex}
\end{eqnarray}
The integer part function ``int'' is defined as the largest integer less 
or equal to the argument.
\item[ii] Periodic boundary conditions in $[0,T]$.\\ 
The absolute value of the functional determinant is 
\begin{eqnarray}
|\mathrm{Det}\foP([0,T])|=4\,\sin^2\left(\frac{\omega\,\sqrt{1-\tau}\,T}{2}\right)
\label{Morse-geometry-oscar:periodic}
\end{eqnarray}
and vanishes for
\begin{eqnarray}
\omega\sqrt{1-\tau}\, T=2\,k\,\pi \geq 0 
\label{Morse-geometry-oscar:periodiczeroes}
\end{eqnarray}
The zeroes are doubly degenerate but at $\tau$ equal to one where the degeneration
is single. Hence, the result of the computation of appendix~\ref{formulae:oscar} 
is recovered: 
\begin{eqnarray}
\ind \foP([0,T])=1+2\,\mbox{int}\left[\frac{\omega T}{2\,\pi}\right]
\label{Morse-geometry-oscar:periodicindex}
\end{eqnarray}
\end{enumerate}

\subsection{Classical Morse index theory}
\label{Morse:Morse}

The relevance of the prescription given in the previous section is mainly conceptual.
Classical Morse theory \cite{Morse} permits a more direct evaluation of the index.\\
In chapter~IV of ref.~\cite{Morse} Morse devises a general formalism encompassing all 
self-adjoint Sturm-Liouville operators.
Here instead the attention is restricted to self-adjoint boundary conditions 
such that:
\begin{eqnarray}
\qac(\chi\,,\xi)\,=\,\langle \chi,\fobc \xi \rangle
\label{Morse-Morse:pairing}
\end{eqnarray}
The evaluation of the index is accomplished first for local boundary 
conditions. The paradigm is represented by Dirichlet boundary 
conditions. The Morse index for Dirichlet boundary conditions (open extremals)
is derived here by adapting the method of 
ref.'s~\cite{LevitMoehringSmilanskyDreyfus,MoehringLevitSmilansky}. 
The result is then exploited to construct the indices associated to non local
boundary conditions. The procedure is exemplified by the periodic case. 

\subsubsection{Dirichlet and local boundary conditions}
\label{Morse-Morse:local}

Dirichlet boundary conditions are local. Locality enforces
self-adjointness in consequence of the independent vanishing of the two skew 
products in (\ref{Morse-geometry:selfadjoint}).
Therefore any self-adjoint operator $\fobc$ acting on $\sbc$ can be seen as
member of a family of operators $\fobctau$ parametrised by the length of the
time interval $[T',\tau]$. Consequently, varying $\tau$ generates a spectral flow.

For times short enough, Dirichlet boundary conditions render the bilinear form
(\ref{Morse-Morse:pairing}) positive definite
\begin{eqnarray}
\langle\xi\,,\foD\,\xi\rangle|_{[T',T'+dt]}\,=\,\qac_{[T',T'+dt]}(\xi,\xi)\sim
\left(\frac{d \xi^\dagger}{dt}\,\mass\,\frac{d \xi}{dt}\right)(T')\,dt\,>\,0
\label{Morse-Morse-local:shorttime}
\end{eqnarray}
as $\mass$ is strictly positive definite by assumption. 

The spectral flow encounters a zero mode each time a Jacobi field 
(\ref{Morse-geometry:Jacobi}) fulfills Dirichlet boundary conditions 
in $[T',\tau]$. The direction of the crossing is given by
\begin{eqnarray}
\de_{\tau}\ellntau|_{\ellntau=0}=\left.\frac{1}{\tau-T'} \int_{T'}^{\tau}dt
(\entau^{\dagger}\foD \de_\tau\entau)(t)\right|_{\ellntau=0}
\label{Morse-Morse-local:time}
\end{eqnarray}
all other derivatives vanishing on a zero mode. The integral localises by means of a 
double integration by parts. The statement is proven by observing that  
on the boundary of the interval $[T',\tau]$ any eigenvector acquires, beside 
the parametric, a functional dependence on $\tau$:
\begin{eqnarray}
\entau(T')=0
\nonumber\\
\entau(\tau)=0
\end{eqnarray}
The total differential with respect to $\tau$ is zero: 
\begin{eqnarray}
&&\frac{d\entau}{d\tau\,}(T')\equiv(\de_\tau\entau)(T')=0  
\label{Morse-Morse-local:derone}\\
&&\frac{d\entau}{d\tau\,}(\tau)
=\frac{d \entau}{dt}(\tau)+(\de_{\tau}\entau)(\tau)=0
\label{Morse-Morse-local:dertwo}
\end{eqnarray}
Combined with an integration by parts of (\ref{Morse-Morse-local:time}), equation 
(\ref{Morse-Morse-local:dertwo}) yields
\begin{eqnarray}
\de_{\tau}\ellntau|_{\ellntau=0}
=-\left.\left(\frac{d\,\entau}{dt\,}^{\dagger}\mass\frac{d\entau}{dt\,}\right)
(\tau)\right|_{\ellntau=0}\,<\,0
\label{Morse-Morse-local:Dirichletflow}
\end{eqnarray}
since the mass tensor $\mass$ is strictly positive definite by hypothesis. 
The Morse index is monotonically increasing as a function of the time interval
$[T',\tau]$ length.

In variational calculus Dirichlet boundary conditions arise from the 
second variation around an open extremal $q_{\cl}$ of a classical 
action $\ac$: a trajectory connecting two assigned points in 
configuration space in the time interval $[T,T']$. 
Jacobi fields with the property
\begin{eqnarray}
\jmath(T')=\jmath(\tau)=0\,,\qquad\qquad  \mbox{for some}\,\, \tau\,\in [T',T]
\label{Morse-Morse-local:conjugate}
\end{eqnarray}
are said to have an $q_{\cl}(T')$-{\em conjugate point} in $\tau$.
The most famous of the Morse index theorems states that:

{\bf {\em Theorem}}{\em(Morse):  
Let $q_{cl}$ be an extremum of $\ac$ in $[T',T]$ and assume the
kinetic energy positive definite. The Morse index of $\qac_{[T',T]}$ is 
equal to the number of conjugate points to $q_{\cl}(T')$. Each conjugate point
is counted with its multiplicity.
}

The theorem can be rephrased using functional determinants.\\ 
Any linear Hamiltonian flow admits the square block representation
\begin{eqnarray}
\fu(t\,,T')\,=\,\left[
\begin{array}{cc} 
\mathsf{A}(t\,T') & \mathsf{B}(t\,,T') \\ \mathsf{C}(t\,,T') & \mathsf{D}(t\,,T')
\end{array}
\right]
\label{Morse-Morse-local:flow}
\end{eqnarray}
(see appendix~\ref{Hamilton:periodic} for details).
By Forman's theorem the Morse index with Dirichlet boundary conditions is
\begin{eqnarray}
\ind \foD([T',T])\,=\!\!\!
\sum_{T'\,<\,t\,<\,T}\!\!\! \nul\,\mathrm{Det}\,\foD([T',t])=\sum_{T'\,<\,t\,<\,T}\!\!\! \nul
\,\mathsf{B}(t\,,T')
\label{Morse-Morse-local:Dirichletindex}
\end{eqnarray}
while the nullity is
\begin{eqnarray}
\nul \foD([T',T])\,=\,\nul\,\mathsf{B}(T\,,T')
\label{Morse-Morse-local:Dirichletnullity}
\end{eqnarray}
The same construction can be repeated for other local boundary conditions.
When the spectral flow encounters a zero mode, the Bott representation
of the boundary conditions (\ref{Morse-geometry:Bott}) permits to write
\begin{eqnarray}
\frac{d\lift\entau}{d\tau\,}(T')&=&\Bott_1\,
\frac{d y_n}{d \tau}(\tau) \equiv(\de_\tau\lift\entau)(T')  
\label{Morse-Morse-local:generalone}\\
\frac{d\lift \entau}{d\tau\,}(\tau)&=&\Bott_2\,\frac{d y_n}{d \tau}(\tau)
=\frac{d \lift \entau}{dt}(\tau)+(\de_{\tau}\lift \entau)(\tau)
\label{Morse-Morse-local:generaltwo}
\end{eqnarray}
and therefore
\begin{eqnarray}
\de_{\tau}\ellntau|_{\ellntau=0}
=-\left.\left(\liftd \entau\,\sym\,\frac{d\lift \entau}{dt}\right)(\tau)\right|
_{\ellntau=0}
\label{Morse-Morse-local:generalspectralflow}
\end{eqnarray}
It might happen that the derivative (\ref{Morse-Morse-local:generalspectralflow}) is zero. 
In such a case the direction of the spectral flow should be inferred 
using higher order derivatives. 
The method has been thoroughly investigated by M\"ohring Levit and Smilansky 
in ref.'s~\cite{LevitMoehringSmilanskyDreyfus,MoehringLevitSmilansky}. 
However the information carried by the index is of topological nature and 
does not depend on the details of the Sturm-Liouville operator. 
Thus it is possible to rule out degeneration by hypothesis resorting 
if necessary to small positive definite perturbations as in the previous section. 
In this sense it is not restrictive to consider only operators begetting a spectral
flow with only {\em regular crossings} characterised by a {\em non-singular} 
{\em crossing form}
\begin{eqnarray}
\daleth(\fu(t,T'),\bc)&:=&\,-\,\left.\Bott_1^{\dagger}\,
\left(\fu^{\dagger}\,\sym\,\frac{d\fu}{d\tau}\right)\!(t,T')\Bott_1
\right|_{\gph\,\fu(t,T')\cap\,\bc}
\nonumber\\
&=&-\,\left.\Bott_2^{\dagger}\,\mathsf{H}(t)\,\Bott_2
\right|_{\gph\,\fu(t,T')\cap\,\bc}
\label{Morse-Morse-local:crossingform}
\end{eqnarray} 
In practice, one evaluates the restriction of the matrix on the right hand to the subspace 
where the crossing occurs. The resulting symmetric matrix has by hypothesis non-vanishing
eigenvalues.\\
The Morse index reads
\begin{eqnarray}
\ind \fobc([T',T])=\ind \qac_{[T',T'+dt]}-\!\!\!
\sum_{\substack{ T'<t<T\,\\\tau |\gph\,\fu(t,T')\cap\bc}}\!\!\!\!\!\!\!\!
\mathrm{sign}\daleth(\fu(t,T'),\bc)
\label{Morse-Morse-local:localindex}
\end{eqnarray}
The $\mathrm{sign}$ function must be understood as the difference between positive and
negative eigenvalues whenever multidimensional crossings occur.\\
An immediate consequence of (\ref{Morse-Morse-local:crossingform}) is that
each time the Bott pair comprises
\begin{eqnarray}
\Bott_2=\left[\begin{array}{cc} 0\quad & \quad 0 \\ 0\quad &\quad \id_{d}\end{array}\right]
\label{Morse-Morse-local:Bott2}
\end{eqnarray}
the Morse index is just the number of crossings counted with their degeneration
\begin{eqnarray}
\ind &&\!\!\fobc([T',T])=\sum_{\substack{T'<t<T\,\\ t |\gph\,\fu(t,T')\in\bc 
}}\!\!\!\!\!\!\!\!\!\!\nul\{0\oplus\id_{d}-\fu(t,T')\Bott_1\}
\label{Morse-Morse-local:kinetic}
\end{eqnarray}
In such cases the crossing form (\ref{Morse-Morse-local:crossingform}) is 
determined by the kinetic energy block of the Hamiltonian matrix 
(\ref{Morse-geometry:Hamiltonian}) at $\tau$ equal to zero.\\
In practice Dirichlet boundary conditions play a role of preeminence.
The reason is that Dirichlet fields are $\qac$-orthogonal to any Jacobi field $\jmath$ 
(\ref{Morse-geometry:Jacobi})
\begin{eqnarray}
\qac(\jmath,\xi)=0\,,\qquad\qquad\forall\, \xi(t)\,\in\,\sbcD
\label{Morse-Morse-local:orthogonality}
\end{eqnarray}
Exploiting such property, Morse indices for general boundary conditions can be 
evaluated proceeding in analogy to section~\ref{Forman:quadratic} where 
the semiclassical approximation was accomplished by separating
classical trajectories from quantum fluctuations.
Here the idea is to write any vector field in $\chi$ in $\sbc$ as a linear combination 
of a Dirichlet $\xi$ and a Jacobi field $j$
\begin{eqnarray}
\chi(t)=\xi(t)+\jmath(t)
\label{Morse-Morse-local:decomposition}
\end{eqnarray}
The Jacobi field is used to match the boundary conditions.
If the kinetic energy is positive definite the number of negative 
eigenvalues is finite. Hence, the evaluation of the Morse index is 
reduced to the evaluation of the index of the restriction of 
$\qac$ to a finite dimensional vector space.
A general theorem in linear algebra (see for example \cite{Artin} pag. 120)
establishes the properties of the index of a symmetric form $\mathcal{F}$ 
in a finite dimensional real vector space $\mathbb{V}$. 
In particular for any $\mathbb{W}\,\subset\,\mathbb{V}$ the index 
is given by
\begin{eqnarray}
\ind \mathcal{F}&=&\ind \mathcal{F}|_{\mathbb{W}}+
\ind \mathcal{F}|_{\mathbb{W}^{\perp}}+
\mathrm{dim}\{\mathbb{W}\cap\mathbb{W}^{\perp}\}-
\mathrm{dim}\{\mathbb{W}\cap\mathrm{Ker}\mathcal{F}\}
\label{Morse-Morse-local:finitedim}
\end{eqnarray}
where $\mathbb{W}^{\perp}:=\{v\,\in\,\mathbb{V}\,|\,\mathcal{F}(v\,,w)\,=\,0\,\,\,
\forall\,w\,\in\,\mathbb{W}\}$. The theorem formalises the obvious observation that
the index is the dimension of the subspace of $\mathbb{V}$ where the quadratic
form $\mathcal{F}$ is {\em semi}-negative definite minus the intersection of such 
subspace with the kernel of $\mathcal{F}$. 
Combined with the decomposition (\ref{Morse-Morse-local:decomposition}) the
theorem provides the way to evaluate the index of Sturm-Liouville operators
with both local and non-local boundary conditions. The procedure is illustrated
by the periodic case.

\subsubsection{Periodic boundary conditions}
\label{Morse-Morse:periodic}

Periodic boundary conditions 
\begin{eqnarray}
\chi(t+T-T')=\chi(t)
\label{Morse-Morse-periodic:periodicityone}
\end{eqnarray}
can be considered in $[T',T]$ only if the differential 
operation $\fo$ itself is periodic in interval $[T',T]$
\begin{eqnarray}
\fo(t+T-T')=\fo(t)\,,\qquad\qquad \forall\, t
\label{Morse-Morse-periodic:periodicity}
\end{eqnarray}
A periodic field $\chi$ admits the representation 
(\ref{Morse-Morse-local:decomposition}) for $\xi$ a Dirichlet field and 
$\jmath$ a recurrent Jacobi field:
\begin{eqnarray}
\left(\,\id_{2 d}-\fu(T,T')\right)
\left[\begin{array}{c} \jmath(T') \\ (\nabla \jmath)(T')
\end{array}\right]\,=\,
\left[\begin{array}{c} 0 \\ (\nabla \xi)(T)-(\nabla \xi)(T')
\end{array}\right]
\label{Morse-Morse-periodic:recurrence}
\end{eqnarray}
Periodicity can always be enforced in the sense of $\si$ by a discontinuity of the momentum 
of the Dirichlet fields at one boundary.
The sets of recurrent Jacobi fields and Dirichlet fields are non intersecting if
\begin{eqnarray}
\mathrm{det}\mathsf{B}(T,T')\,\neq\,0
\label{Morse-Morse-periodic:Bmatrix}
\end{eqnarray}
where the $d\times d$-dimensional real matrix $\mathsf{B}(T,T')$ is specified
by the block representation (\ref{Morse-Morse-local:flow})
of the linear flow. Whenever (\ref{Morse-Morse-periodic:Bmatrix}) holds true, 
one has
\begin{eqnarray}
\ind \foP([T',T])=\ind\left.\qac(\jmath,\jmath)\right|_{\mathrm{rec.}}+
\ind \foD([T',T])
\label{Morse-Morse-periodic:index}
\end{eqnarray}
The first term on the right hand is amenable to the more explicit form
\begin{eqnarray}
\ind\left.\qac(\jmath,\jmath)\right|_{\mathrm{rec.}}\!\!\!=\ind
(\mathsf{D}\,\mathsf{B}^{-1}+\mathsf{B}^{-1}\mathsf{A}-\mathsf{B}^{-1}
-\mathsf{B}^{-1\,\dagger})(T,T')
\label{Morse-Morse-periodic:orderofconcavity}
\end{eqnarray}
The result follows from the recurrence condition 
(\ref{Morse-Morse-periodic:recurrence}) and the use of the properties of a 
symplectic matrix given in appendix~\ref{Hamilton:linear}. 
The same properties insure that 
(\ref{Morse-Morse-periodic:orderofconcavity}) is the
index of a symmetric matrix and therefore it is well defined.
Morse referred to (\ref{Morse-Morse-periodic:orderofconcavity}) 
as {\em the order of concavity} of a periodic extremal, see 
ref.~\cite{Morse} pag.~$71$.\\
It remains to analyse the physical meaning of (\ref{Morse-Morse-periodic:Bmatrix}).
From the point of view of the second variation (\ref{Morse-Morse-periodic:Bmatrix})
is not a restrictive condition. The stability of a periodic orbit coincides with
the stability of any of the trajectories covering the orbit. 
A trajectory starting at time $T'$ in generic position will fulfill the 
condition.\\
Non generic situations when (\ref{Morse-Morse-periodic:Bmatrix}) does not hold true 
can treated using the general formula (\ref{Morse-Morse-local:finitedim}) with 
$\mathbb{W}$, $\mathbb{W}^{\perp}$ denoting respectively the sets of Dirichlet and 
recurrent Jacobi fields in $[T',T]$, \cite{BallmannThorbergssonZiller}.

Morse indices appear from the above examples to be determined by the oscillations
of the Jacobi fields. An intuitive explanation is provided by variational calculus.
Jacobi fields are extremals of the second variation. Because the kinetic energy is 
positive definite, the second variation as function of the upper time $T$ 
cannot become negative on any functional subspace  until a Jacobi field does not 
encounter a conjugate point.\\
The Morse index for periodic extremals can be also derived by purely symplectic geometrical
methods based on the analysis of the canonical form of the linearised flow 
\cite{Klingenberg,Klingenberg2}.

\subsection{Elementary applications}
\label{Morse:examples}

The simplest applications of Morse index theorems are offered 
by one dimensional Hamiltonian systems.

{\em i Particle in a potential well}

\begin{figure}
\begin{center}
\mbox{\includegraphics[height=6cm,width=10cm]{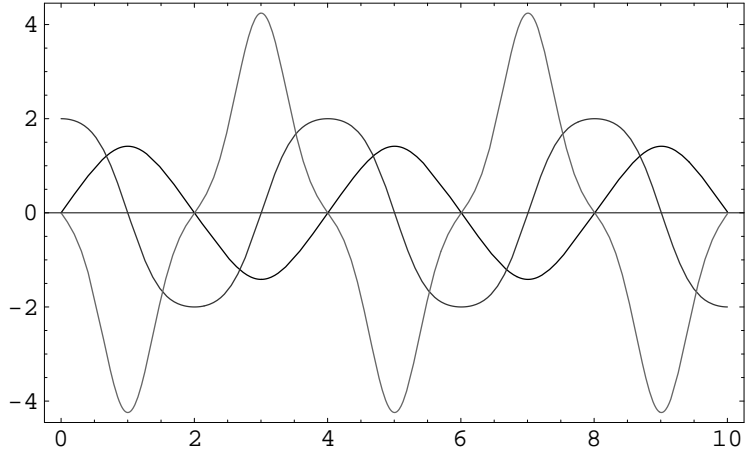}}
\end{center}
\caption{In decreasing gray level position, velocity and acceleration of
a particle of unit mass in the quartic potential $V(q)=q^2/2+q^4/4$.
The acceleration is minimal (maximal) and negative (positive) at  
odd (even) turning points of the velocity.}
\label{fig:anharmonic}
\end{figure}

\begin{figure}
\centering
\mbox{\includegraphics[height=6cm,width=10cm]{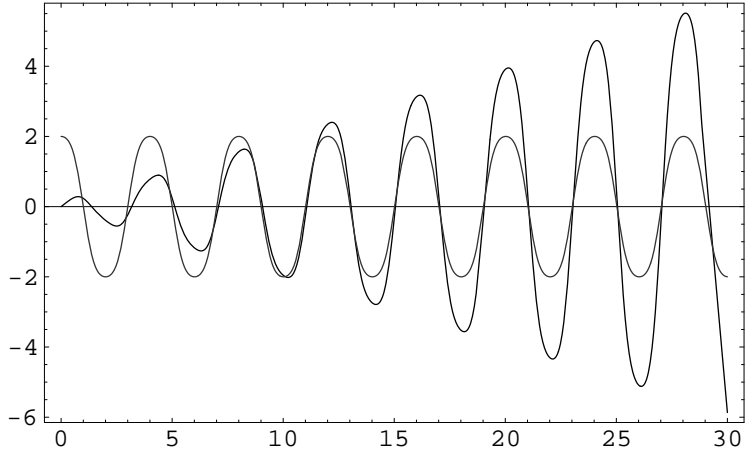}}
\caption{ The Jacobi fields $\dot{q}_{\cl}$ and $\de_{E} q_{\cl}$
relative to the motion of a particle of unit mass in the potential 
$V(q)=q^2/2+q^4/4$.  
The envelope of the amplitude peaks of the non periodic Jacobi field
$\de_{E} q_{\cl}$ is linear with slope $-dT_{\cl}/dE$. 
The conservation of the skew product imposes the 
non periodic Jacobi fields to be positive at odd turning 
points of the velocity.}
\label{fig:Jacobi}
\end{figure}

Consider the action functional
\begin{eqnarray}
\ac(\dot{q}\,,q)\,=\,\int_{0}^{T}dt\,\left[\frac{m\,\dot{q}^2}{2}-U(q)\right]
\label{Morse-examples:action}
\end{eqnarray}
The potential $U$ is chosen to be a positive definite analytic function of 
$q$. It is also supposed to give rise, at least within a certain energy range,
to bounded motion. The case of a quadratic potentials is ruled out by hypothesis 
and it will be treated apart.
A concrete example is provided by the 
anharmonic oscillator
\begin{eqnarray}
U(q)\,=\,\frac{m\,\omega^2\,q^2}{2}+\frac{\varpi\,q^4}{4}
\label{Morse-examples:anharmonic}
\end{eqnarray}
Classical trajectories are extremals of the action occurring 
on the curves of level of the Hamiltonian
\begin{eqnarray}
E\,=\,\ha(p\,,q)\,=\,\frac{p^2}{2\,m}+U(q)\,,
\quad\quad\quad p\,=\,\frac{\de \la}{\de \dot{q}}\,=\,m\,\dot{q} 
\label{Morse-examples:energy}
\end{eqnarray}
which are eventually closed in consequence of the postulated shape of the potential. 
In fig.~\ref{fig:anharmonic} a typical solution is plotted for the anharmonic quartic 
potential (\ref{Morse-examples:anharmonic}) in rescaled adimensional units.\\
Let $x_{\cl}(t;E)$ be the phase space lift of a trajectory on a periodic orbit of 
prime period $T_{cl}(E)$. 
A suitable basis for the linearised dynamics is represented by the Jacobi fields $\dot{x}(t;E)$,
$\de_E x(t;E)$. Linear independence is guaranteed by the invariance of the skew 
product
\begin{eqnarray}
(\de_{E}x_{\cl}^{\dagger}\,\sym
\dot{x}_{\cl})(t;E):=
\,(\de_E p_{\cl}\,\dot{q}_{\cl}-\,\de_E q_{\cl}\,\dot{p}_{\cl})(t;E)=
\left.\frac{d\ha}{dE}\right|_{x_{\cl}(t;E)}\!\!\!\!\!\!\!\!=1
\label{Morse-examples:Wronskian}
\end{eqnarray}
The skew product coincides with the Wronskian determinant of the linearised 
dynamics. 
In fig.~\ref{fig:Jacobi} the two Jacobi fields are plotted for 
(\ref{Morse-examples:anharmonic}). The Jacobi field of the second kind 
$\de_E x(t;E)$ describes a linear instability 
\cite{Poincare,DeWittMoretteMaheshwariNelson}
\begin{eqnarray}
\de_E x(t;E):=\kappa\,\frac{t-T'}{T_{\cl}}\,\dot{x}_{\cl}(t;E)+
\left.\left(\frac{\de x}{\de E}\right)(t;E)\right|_{\frac{t-T'}{T_{cl}}=\mathrm{const.}}
\label{Morse-examples:nonperiodic}
\end{eqnarray}
with
\begin{eqnarray}
\kappa:=-\,\frac{dT_{\cl}}{dE}
\end{eqnarray}
The linearised flow in phase space is (compare with formula 
(\ref{Hamilton-periodic:blocks}) in appendix~\ref{Hamilton}) 
\begin{eqnarray}
\fu(t,T')=[\dot{x}_{\cl}(t;E),\de_{E}x(t;E)]\,[\dot{x}_{\cl}(T';E),\de_{E}x(T';E)]^{-1}
\label{Morse-examples:fundamental}
\end{eqnarray}
The Sturm-Liouville operator associated to the second variation
along a trajectory of energy $E$ is
\begin{eqnarray}
\fo\,=\,-\,m\,\frac{d^2}{dt^2}-\left(\frac{d^{2}U}{d q^{2}}\right)(q_{\cl}(t;E))
\label{Morse-examples:SturmLiouville}
\end{eqnarray}
By (\ref{Morse-examples:fundamental}) the Morse index for Dirichlet boundary conditions 
is determined by the number of zeroes of
\begin{eqnarray}
\mathsf{B}(t,T')= \de_E q_{\cl}(t;E)\,\dot{q}_{\cl}(T';E)-\de_E q_{\cl}(T',E)\,\dot{q}_{\cl}(t;E)
\label{Morse-examples:jacobi}
\end{eqnarray}
The right hand side is a linear combination of solutions of the linearised dynamics.
Hence it is itself a Jacobi field vanishing at $t$ equal $T'$.
The zeroes encountered by (\ref{Morse-examples:jacobi}) in $]T,T'[$ define the number of conjugate
points and therefore the Morse index associated to Dirichlet boundary conditions in the interval
$[T',T]$. By (\ref{Morse-examples:jacobi}) also follows that if  $q_{\cl}$ starts 
from a turning point, ($\dot{q}_{\cl}$ equal zero), the index will be specified by the 
number of turning points encountered in $]T',T[$. 
 
Let now $T$ be equal to $r\,T_{cl}$ with $r$ a positive integer. 
Furthermore suppose the trajectory $q_{\cl}$ to start form a generic point (not a turning point)
on the periodic orbit. Observe that $\mathsf{B}(t,T')$ starts with positive derivative 
and has the same signature of $\kappa$ over multiples of the period. 
Moreover from the Wronskian condition (\ref{Morse-examples:Wronskian}) on a turning point
\begin{eqnarray}
1=-\,(\de_E q_{\cl}\,\dot{p}_{\cl})(t_{\mathrm{t.p,}};E)
\label{Morse-examples:times}
\end{eqnarray}
follows that $\mathsf{B}(t,T')$ must change sign at least once over one period, 
since velocity and acceleration change sign twice. Hence
\begin{eqnarray}
\ind \foD([T',T'+r T_{\cl}])\,=\,2\,r-\frac{1-\mathrm{sign}\kappa}{2}
\label{Morse-examples:periodDirichletindex}
\end{eqnarray}
If instead the trajectory starts from a turning point then 
\begin{eqnarray}
\ind \foD([T',T'+r T_{\cl}])=2\,r-1
\label{Morse-examples:Dirichletdegenerate}
\end{eqnarray}
Associating  periodic boundary conditions  to the differential operation 
(\ref{Morse-examples:SturmLiouville}) in the same interval $[T',T'+r\,T_{\cl}]$
defines the self-adjoint Sturm-Liouville operator $\foP$. 
The periodic Morse index is determined from the sum of the Dirichlet index and the
order of concavity. This latter is read from the monodromy matrix
\begin{eqnarray}
\mon=\left[\begin{array}{cc} 
1+\frac{d T_{\cl}}{d E}(E)\,(\dot{p}_{\cl}\dot{q}_{\cl})(T_{cl};E)  \,\, & \,\, 
-\frac{d T_{\cl}}{d E}(E)\,\dot{q}_{\cl}^{2}(T_{cl};E)
\\
\frac{d T_{\cl}}{d E}(E)\,\dot{p}_{\cl}^{2}(T_{\cl};E) \,\, & \,\, 
1-\frac{d T_{\cl}}{d E}(E)\,(\dot{p}_{\cl}\dot{q}_{\cl})(T_{\cl};E)
\end{array}
\right]
\label{Morse-examples:monodromy}
\end{eqnarray} 
If the starting point of the periodic trajectory is 
taken in generic position, the order of concavity is specified by
(\ref{Morse-Morse-periodic:orderofconcavity}) and vanishes identically:
\begin{eqnarray}
&&\ind \foP([T',T'+ r T_{\cl}])=2\,r-\frac{1-\mathrm{sign}\kappa}{2}
\nonumber\\
&&\nul \foP([T',T'+ r T_{\cl}])=1
\label{Morse-examples:indexperiodic}
\end{eqnarray}
\begin{figure}
\centering
\mbox{\includegraphics[height=6cm,width=10cm]{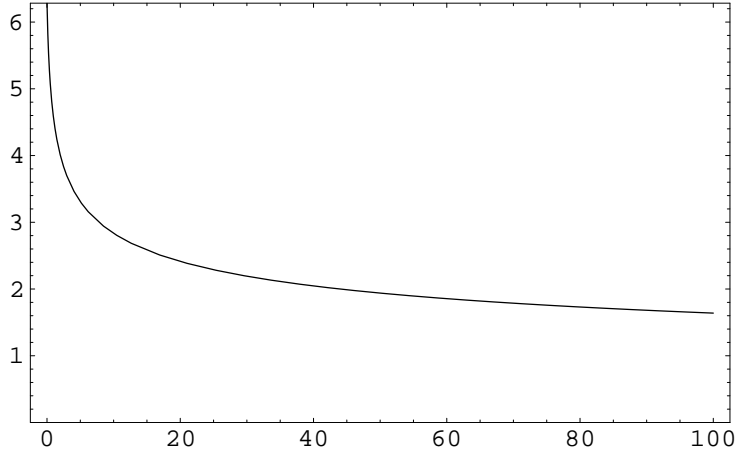}}
\caption{The period $T_{\cl}$ of a particle of unit mass in the potential 
$V(q)=q^2/2+q^4/4$ plotted versus the energy. The period decreases from 
the value $T_{\cl}(0)=2\,\pi$, the period of the harmonic oscillator, with
asymptotic behaviour $T_{\cl}(E)\sim E^{-\,1/4}$ for $E$ tending to infinity.} 
\label{fig:times}
\end{figure}
If instead the trajectory starts from a turning point, the general index formula 
(\ref{Morse-Morse-local:finitedim}) must be applied.
To the index (\ref{Morse-examples:Dirichletdegenerate}) must be added
a non zero contribution from the order of concavity. The
non zero contribution comes from the recurrent Jacobi field 
\begin{eqnarray}
\jmath_{\mathrm{rec.}}(t)=\dot{q}_{\cl}(t;E)\de_{E}p_{\cl}(T';E)-\de_{E}q_{\cl}(t;E)\dot{p}_{\cl}(T';E)
\end{eqnarray}
with now
\begin{eqnarray}
\dot{q}_{\cl}(T'+r T_{\cl};E)=0
\label{Morse-example:recurrent}
\end{eqnarray}
and it is equal to
\begin{eqnarray}
\qac(\jmath_{\mathrm{rec.}},\jmath_{\mathrm{rec.}})=
\mon_{2\,1}=\frac{d T_{\cl}}{d E}(E)\dot{p}_{\cl}^{2}(T_{\cl};E)
\label{Morse-examples:concavity}
\end{eqnarray}

The sum of the Dirichlet index and of the order of concavity recovers
(\ref{Morse-examples:indexperiodic}). In the case of the Hamiltonian 
(\ref{Morse-examples:anharmonic}) the index is seen to be even since the
period is a monotonically decreasing function of the energy (fig.~\ref{fig:times}).

{\em ii Harmonic oscillator}

The counting of the conjugate points in $[0,T]$ yields
(\ref{Morse-Morse-local:Dirichletindex})
\begin{eqnarray}
\ind\foD([0,T])=\#(\mbox{zeroes of}\sin(\omega t)\,\,\mbox{in}\,]0,T[)=
\mathrm{int}\left[\frac{\omega T}{\pi}\right] 
\label{Morse-examples-oscar:Dirichlet}
\end{eqnarray}
in agreement with the computations of section~\ref{Morse-geometry:oscar} and 
of appendix~\ref{formulae:oscar}. In consequence, the periodic Morse index is 
\begin{eqnarray}
\ind\foP([0,T])&=&\ind\foD([0,T])+\ind\frac{\cos(\omega\,T)-1}{\sin(\omega\,T)}
\nonumber\\
&=&1+2\,\mathrm{int}\left[\frac{\omega T}{2\,\pi}\right] 
\label{Morse-examples-oscar:periodicgeneric}
\end{eqnarray}
for $T$ generic.

\subsection{Crossing forms and Maslov indices}
\label{Morse:Maslov}
 
The constructions of Morse indices of the previous two sections
are reconciled if a more general topological framework is developed to 
investigate the spectral flow. 
This was done by Duistermaat in \cite{Duistermaat} who refined 
ideas introduced by Bott in \cite{Bott}.

First, the following heuristic argument helps to understand the topological 
invariance of the Morse index.
The Morse index of a self-adjoint $\fobc$ is the number of negative eigenvalues
\begin{eqnarray}
\ind \fobc=\sum_{\ell_{j}\,\in\,\spe \fobc} \theta(-\ell_{j})
\label{Morse-Maslov:Heaviside}
\end{eqnarray}
If $\fobc$ is embedded into a smooth $n$-parameters family of 
self-adjoint Sturm-Liouville operators including a positive 
definite element, the index can be thought as the 
line integral in the space of the parameters connecting $\fobc$ to the
positive definite element. Parametrising with $\tau$ such curve, the index 
takes the form  
\begin{eqnarray}
\ind \fobc&=&\,-\,\sum_{\ellntau\,\in\,\spe \fobctau}\int_{0}^{\tau^\prime}d\tau 
\,\de_\tau\,\theta(-\ellntau)
\nonumber\\
&=&\sum_{\ellntau\,\in\,\spe \fobctau}\int_{0}^{\tau^\prime}d\tau 
\,\frac{\de\ellntau}{\de \tau}\,\delta(\ellntau)
\label{Morse-Maslov:delta}
\end{eqnarray}
where it is assumed that zero and $\tau^\prime$ correspond respectively to $\fobc$
and the positive definite operator. In order the equality to hold true, only 
regular crossing of zero modes are supposed to occur. The topological invariance of 
the index stems from the identical vanishing of loop integrals over exact 
differentials which ensures the independence of the index of the curve used to
compute (\ref{Morse-Maslov:delta}).

The last equality in (\ref{Morse-Maslov:delta}) states that along any curve in the 
space of self-adjoint Sturm-Liouville operators, the Morse index is specified by the sign of 
the eigenvalue flow on zero modes.
Proceedings as in section~\ref{Morse-Morse:local} yields
\begin{eqnarray}
\de_{\tau}\ellntau=\langle \entau\,,(\de_{\tau}\fobctau) \entau\rangle
\label{Morse-Maslov:eigenvalue}
\end{eqnarray}
The integral localises on zero modes. Differentiating the Jacobi equation 
\begin{eqnarray}
(\fo_{\tau}\,\jmath_{\tau})(t)=0
\label{Morse-Maslov:Jacobi}
\end{eqnarray}
versus the curve parameter $\tau$ allows to prove the chain of equalities
\begin{eqnarray} 
\langle \jmath_{\tau}\,,(\de_{\tau}\fo_{\tau}) \jmath_{\tau}\rangle\,=\,
-\,\langle \jmath_{\tau}\,,\fo_{\tau} \de_{\tau}\jmath_{\tau}\rangle
\,=\,\beth(\jmath\,,\de_{\tau}\jmath_{\tau})
\label{Morse-Maslov:Jacobiform}
\end{eqnarray}
The restriction to a crossing point defines the {\em crossing form}
\cite{Duistermaat,RobbinSalamon,RobbinSalamon2,RobbinSalamon3}:
\begin{eqnarray}
\daleth(\fu_{\tau}(T,T'),\bc):=
\beth(\jmath_{\tau}\,,\de_{\tau}\jmath_{\tau})|_{\bc\,\cap\,\gph\fu_{\tau}(T,T')}
\label{Morse-Maslov:crossingform}
\end{eqnarray} 
The requirement of smoothness of the deformation can be weakened. Provided the
spectral flow is continuous, discontinuous {\em non vanishing} derivatives of the zero 
eigenvalues also allow to infer the direction of the spectral flow.\\
The Morse index of $\fobc$ is
\begin{eqnarray}
\ind\fobc\,=\,\!\!
\sum_{\{\,\tau\,>\,0|\bc\cap\gph\fu_{\tau}(T,T')\,\}}\!\!
\mathrm{sign}\daleth(\fu_{\tau}(T,T'),\bc)
\label{Morse-Maslov:index}
\end{eqnarray}
for any homotopy transformation with only regular crossings 
connecting $\fobc$ to a positive definite Sturm-Liouville operator.
The sign of the crossing form over a discontinuous derivative is the arithmetic 
average of the derivatives from the left and the right of the critical value 
of $\tau$.\\
The crossing forms (\ref{Morse-Morse-local:crossingform}) produced by 
deformations of the time interval are particular examples. 
The fact is recognised by looking at the explicit 
expression of crossing forms. A crossing occurs each time for some
$x,y \,\in\,\rn^{2 d}$ the equality holds
\begin{eqnarray} 
\left[\begin{array}{c} \Bott_1\,y \\ \Bott_2\,y\end{array}\right]
=\gph \fu_{\tau} x
\label{Morse-Maslov:crossing}
\end{eqnarray}
$(\Bott_1\,,\Bott_2)$ being the Bott pair describing $\bc$.
Hence (\ref{Morse-Maslov:crossingform}) is equivalent to
\begin{eqnarray}
\daleth(\fu_{\tau}(T,T'),\bc)\,=\,-\,\left.\Bott_1^{\dagger}\,
\left(\fu_{\tau}^{\dagger}\,\sym\,\frac{\de \fu_{\tau}}{\de \tau}\right)(T,T')\,
\Bott_1\,\,\right|_{\gph \fu_{\tau}(T,T')\,\cap\,\bc}
\label{Morse-Maslov:crossingmatrix}
\end{eqnarray}
The matrix on the right hand side is symmetric at glance since smooth 
deformation of symplectic matrices are governed by linear Hamiltonian 
equations (appendix~\ref{Hamilton:perturbative}). 

The formulae (\ref{Morse-Maslov:crossing}), (\ref{Morse-Maslov:crossingmatrix}) 
disclose the possibility to define the Morse index of a self-adjoint
Sturm-Liouville operator as the index of homotopy transformations of finite 
dimensional quadratic forms. 
Namely an arbitrary deviation from the crossing condition 
(\ref{Morse-Maslov:crossing}) can be written in the form
\begin{eqnarray} 
\left[\begin{array}{c} \Bott_1\,y \\ \Bott_2\,y\end{array}\right]
+\left[\begin{array}{c} \mathsf{V}_1\,z \\ \mathsf{V}_2\,z \end{array}\right]
=\gph \fu_{\tau}(T,T') x\,,\quad\quad x,y,z \,\in\,\rn^{2 d}
\label{Morse-Maslov:qequation}
\end{eqnarray}
provided the Bott pair $(\mathsf{V}_1\,,\mathsf{V}_2)$ spans in the sense of 
section~\ref{Morse:geometry}, some linear complement $\mathfrak{V}$ 
in $\mathbb{R}^{4\,d}$ of $\gph \fu_{\tau}$ and $\bc$. In such a case 
(\ref{Morse-Maslov:qequation}) admits always solution for $z$ 
\begin{eqnarray}
[\Bott_2-\fu_{\tau}\Bott_1\,]y=
[\mathsf{V}_2-\fu_{\tau}\mathsf{V}_1\,]z
\label{Morse-Maslov:qreduction}
\end{eqnarray}
On solutions $\bar{z}_{\tau}$ the symplectic form on $\bc$
\begin{eqnarray}
\beth|_{\bc}:=\left[\begin{array}{c} \mathsf{V}_1\,\bar{z}_{\tau} \\ \mathsf{V}_2\,\bar{z}_{\tau} 
\end{array}\right]^{\dagger}
\left[\begin{array}{cc} \sym & 0 \\ 0 & -\sym  \end{array}\right]
\left[\begin{array}{c} \Bott_1\,y \\ \Bott_2\,y\end{array}\right]
\label{Morse-Maslov:beth}
\end{eqnarray}
defines the family of symmetric matrices in $\rn^{2 d}\times\rn^{2 d}$:
\begin{eqnarray}
\mathcal{Q}&&\!(\bc,\mathfrak{V};\gph\,\fu_{\tau}):=
\nonumber\\
&&\frac{1}{2}\left\{[\Bott_2^{\dagger}\sym\mathsf{V}_2-\Bott_1^{\dagger}\sym\mathsf{V}_1]
[\mathsf{V}_2-\fu_{\tau}\mathsf{V}_1\,]^{-1}[\Bott_2-\fu_{\tau}\Bott_1\,]
+c.c.\right\}
\label{Morse-Maslov:quadratic}
\end{eqnarray}
The spectrum of $\mathcal{Q}(\bc,\mathfrak{V};\gph\,\mathsf{F}_{\tau})$ contains
a zero if and only if the parameter $\tau$ attains a value $\tau^{\star}$ such that 
$L_{\bc\,,\tau^{\star}}$ has a zero mode. 
Furthermore a direct computation shows that
\begin{eqnarray}
\mathrm{sign}\mathcal{Q}(\bc,\mathfrak{V};\gph\,\fu_{\tau^{\star}+d\tau})=
\mathrm{sign}\mathcal{Q}(\bc,\mathfrak{V};\gph\,\fu_{\tau^{\star}})+
\mathrm{sign}\daleth(\fu_{\tau^{\star}},\bc)
\end{eqnarray}
Proceeding in this way, the Morse index of a self-adjoint Sturm-Liouville operator 
is identified with the {\em Maslov index}, the index of the graph of a continuous 
curves in $Sp(2\,d)$ having {\em only regular crossings} with a given Lagrangian subspace 
$\bc$ of $\rn^{4 d}$ \cite{Arnold2,Hormander,Duistermaat,RobbinSalamon}. The 
identification permits to evaluate the Morse index without making direct reference to 
the corresponding operator. In contrast to the eigenvalues
of the Sturm-Liouville operator, the eigenvalues of the finite dimensional quadratic 
form (\ref{Morse-Maslov:quadratic}) as functions of $\tau$ can also change sign by 
diverging. The circumstance insures that arbitrary large values of the Morse index 
are eventually attained.

According to Robbin and Salamon \cite{RobbinSalamon}, the Maslov index of the graph 
$\gph\,\mathsf{S}_{\tau}$ of any one parameter family of matrices 
$\mathsf{S}(\tau)$ in $Sp(2 d)$ is defined as
\begin{eqnarray}
\aleph(\mathsf{S}(\tau),&\bc&,[\tau_1,\tau_2])=
\frac{1}{2}\,\mathrm{sign}\daleth(\mathsf{S}(\tau_1),\bc)
\nonumber\\
&+&\!\!\sum_{\substack{ \tau_1<\tau<\tau_2\,\\ 
\tau |\gph\mathsf{S}(\tau)\in\bc}}\!\!\!\!\!
\mathrm{sign}\daleth(\mathsf{S}(\tau),\bc)+
\frac{1}{2}\,\mathrm{sign}\daleth(\gph\mathsf{S}(\tau_2),\bc)
\label{Morse-Maslov:Maslov}
\end{eqnarray}
End-points contribute of course only if an intersection occurs there.
Maslov indices (\ref{Morse-Maslov:Maslov}) enjoy the following properties.
\begin{enumerate}
\item[i] {\em Naturality}: the index is invariant under a simultaneous symplectic 
transformation $\mathbf{\Psi}$ of $\bc$ and $\gph\mathsf{S}(\tau)$:
\begin{eqnarray}
\aleph(\Psi\,\mathsf{S}(\tau)\,,\Psi\,\bc\,,[\tau_1,\tau_2])=\aleph(\mathsf{S}(\tau)\,,
\bc\,,[\tau_1,\tau_2])
\label{Morse-Maslov:natuarality}
\end{eqnarray}
for all $\mathbf{\Psi}=(\Psi_1\,,\Psi_2)$ leaving invariant the symplectic form $\beth$ in 
$\rn^{4 d}$
\begin{eqnarray}
\Psi_1^{\dagger}\,\sym\,\Psi_1=\Psi_2^{\dagger}\,\sym\,\Psi_2
\label{Morse-Maslov:symplectic}
\end{eqnarray} 
\item[ii] {\em Catenation}: for $\tau_1\,<\,\tau_2<\,\tau_3$
\begin{eqnarray}
\aleph(\mathsf{S}(\tau)\,,\bc\,,[\tau_1,\tau_3])=
\aleph(\mathsf{S}(\tau)\,,\bc\,,[\tau_1,\tau_2])+
\aleph(\mathsf{S}(\tau)\,,\bc\,,[\tau_2,\tau_3])
\label{Morse-Maslov:catenation}
\end{eqnarray} 
\item[iii] {\em Product}: if the curve 
$\mathsf{S}(\tau)=\mathsf{S}^{(1)}(\tau)\oplus\mathsf{S}^{(2)}(\tau)$ 
and the Lagrangian subspace $\bc=\bc^{(1)}\oplus\bc^{(2)}$
\begin{eqnarray}
\aleph(\mathsf{S}^{(1)}(\tau)&&\!\!\oplus\mathsf{S}^{(2)}(\tau),\bc^{(1)}\oplus\bc^{(2)},
[\tau_1,\tau_2])=
\nonumber\\
&&\aleph(\mathsf{S}^{(1)}(\tau),\bc^{(1)},[\tau_1,\tau_2])+
\aleph(\mathsf{S}^{(2)}(\tau),\bc^{(2)},[\tau_2,\tau_3])
\label{Morse-Maslov:product}
\end{eqnarray} 
\item[iv] {\em Homotopy}: the Maslov index is invariant under fixed end-points
homotopies.
\end{enumerate}
Of the above properties the first three are immediate consequence of the definition.
Homotopy was first proven by Arnol'd in \cite{Arnold2}. The proof proceeds by showing
that the signature of the crossing form defines a {\em Brouwer degree} 
\cite{Milnor2,Frankel} for the crossing. The homotopy invariance
of the Brouwer degree is then a standard result of differential topology 
\cite{Milnor2}.

One can use Maslov indices to compute the Morse index of a self-adjoint  
Sturm-Liouville operator $\fo_{\bc^{(a)}}$ with respect to the Morse index of a second one 
$\fo_{\bc^{(b)}}$. Under the hypothesis that both $\bc^{(a)}$ and 
$\bc^{(b)}$ have no intersections with $\gph\,\mathsf{S}(\tau)$ at path end-points 
$\tau_1$, $\tau_2$, H\"ormander proved in ref.~\cite{Hormander} section~$3.3 $ the
equality
\begin{eqnarray}
&&\aleph(\mathsf{S}(\tau),\bc^{(a)}\,,[\tau_1,\tau_2])=
\aleph(\mathsf{S}(\tau),\bc^{(b)}\,,[\tau_1,\tau_2])+
\nonumber\\
&&\frac{1}{2}\left[\mathrm{sign}\mathcal{Q}(\bc^{(b)},\gph\mathsf{S}(\tau_2);\bc^{(a)})
-\mathrm{sign}\mathcal{Q}(\bc^{(b)},\gph\mathsf{S}(\tau_1);\bc^{(a)})\right]
\label{Morse-Maslov:Hormander}
\end{eqnarray}
Replacing $\mathsf{S}(\tau)$ with $\fu(\tau,T')$ shows that H\"ormander's relation 
is a generalisation of the relation (\ref{Morse-Morse-periodic:index}) between the indices 
of Dirichlet and periodic boundary conditions.

Homotopy and H\"ormander's identity (\ref{Morse-Maslov:Hormander}) provide all the needed 
tools to understand the origin of the different prescriptions given in the literature 
for the evaluation of Morse indices. The following sections are devoted to illustrate this 
point.\\ 
The reader interested in a more detailed presentation of Maslov index theory will 
enjoy the very accessible and comprehensive presentation by Robbin and Salamon 
in ref.~\cite{RobbinSalamon}.
The needed basic background in topology can be found in \cite{Milnor2}.

\subsection{Topological construction of Morse indices}
\label{Morse:topology}

Using the results of the previous section, the Morse index of $\fobc$, 
$\bc\sim(\Bott_1,\Bott_2)$  can be computed from any closed loop 
formed by varying $\fu_{\tau}(t,T')$ in the $(\tau,t)$-plane. It 
comprises the following oriented paths 
\begin{figure}
\begin{center}
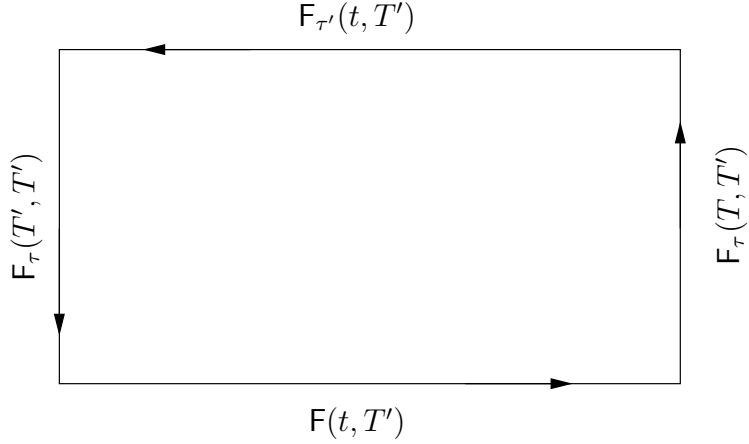
\end{center}
\caption{The homotopy loop} 
\label{fig:loop}
\end{figure}
\begin{enumerate}
\item[1] $\qquad\fu_{\tau}(T,T')\,,\qquad\qquad d\tau\,\leq\,\tau\,\leq\,\tau^{\prime}  
$\\
The flow $\fu_{\tau}(T,T')$ is solution of (\ref{Morse-geometry:Hamilton}) with 
the family of Hamiltonian matrices generated by adding a positive 
definite potential term as in section~\ref{Morse:geometry}.
The initial value $d\tau$ provides for an arbitrary small positive 
deformation which removes eventual zero modes. The upper value $\tau'$ is 
chosen large enough such that the Sturm-Liouville operator becomes positive 
definite. Thus, the Maslov index of the path coincides with Morse index of $\fobc$:
\begin{eqnarray}
\aleph_1(\fu,\bc)=\ind\fobc([T',T])
=\sum_{\substack{\tau>0\\
\bc\cap\gph\fu_{\tau}(T,T')}}\!\!\!\!\!\!
\nul\{\Bott_2-\fu_{\tau}(T,T')\Bott_1\}
\label{Morse-topology:branchone}
\end{eqnarray}
\item[2] $\qquad\fu_{\tau'}(t,T')\,, \qquad\qquad T\,\geq\,t\,\geq\,T'$\\
For $\tau$ large enough the operator is dominated by the positive definite
potential. No zero mode can occur until $t$ reaches $T'$ since the
quadratic form (\ref{Morse-geometry:perturbed}) is positive definite for all
$t$ strictly larger than $T'$. At initial time the condition
\begin{eqnarray}
\fu_{\tau}(T',T')=\id_{2 d}\,,\qquad\qquad \forall\,\tau
\label{Morse-topology:nochange}
\end{eqnarray}
may generate an intersection between $\bc$ and $\gph\,\id_{2 d}$.
The Maslov index (\ref{Morse-Maslov:Maslov}) is in such a case 
\begin{eqnarray}
\aleph_2(\fu,\bc)
=\,\frac{1}{2}\,\mathrm{sign}\left\{\left.\Bott_2^{\dagger}\,\mathsf{H}_{\tau^{\prime}}(T')
\Bott_2\,\,\right|_{\gph\, \id_{2 d}\,\cap\,\bc}\right\}
\label{Morse-topology:branchtwo}
\end{eqnarray}
since the crossing form can be directly expressed in terms of the Hamiltonian
matrix (\ref{Morse-geometry:Hamiltonian}). The sign is the opposite than in
(\ref{Morse-Maslov:crossingform}) due to the negative orientation 
of the path. 
\item[3] $\qquad\fu(t,T')\,, \qquad\qquad T'\,\leq\,t\,\leq\,T $\\
The corresponding Maslov index is
\begin{eqnarray}
\aleph_3&&\!\!(\fu,\bc)=\,-\,\frac{1}{2}\,\mathrm{sign}\left\{\left.\Bott_2^{\dagger}\,\mathsf{H}(T')
\Bott_2\,\right|_{\gph\,\id_{2 d}\,\cap\,\bc}\,\right\}
\nonumber\\
&&-\!\!\!\!\!\!\sum_{\substack{T'<t<T\,\\ t |\gph\,\fu(t,T')\cap\bc}}
\!\!\!\!\!\!\!\!
\mathrm{sign}\{\Bott_2^{\dagger}\,\mathsf{H}(t)\Bott_2\,\}-
\frac{1}{2}\,\mathrm{sign}\left\{\left.\Bott_2^{\dagger}\,
\mathsf{H}(T)\Bott_2\,\right|_{\gph\fu(T,T')\cap\bc}\right\}
\label{Morse-topology:branchthree}
\end{eqnarray}
\item[4] $\qquad\fu_{\tau}(T,T')\,,\qquad\qquad 0\,\leq\,\tau\,\leq\,d\tau$\\
By construction the last branch may encounter a Lagrangian intersection only
in the initial point. The Maslov index is therefore
\begin{eqnarray}
\aleph_4(\fu,\bc)&=&-\,\frac{1}{2}\,\mathrm{sign}\left\{
\left.\Bott_1^{\dagger}\,
\left(\fu_{\tau}^{\dagger}\,\sym\,\frac{\de \fu_{\tau}}{\de \tau}\right)(T,T')\,
\Bott_1\,\,\right|_{\tau=0\,,\gph \fu(T,T')\,\cap\,\bc}\right\}
\nonumber\\
&&=\frac{1}{2}\,\nul\{\Bott_2-\fu(T,T')\Bott_1\,\}
\label{Morse-topology:branchfour}
\end{eqnarray}
The last equality holds true because increasing $\tau$ begets 
by construction only positive definite crossing forms along 
this branch of the homotopy path.
\end{enumerate}
The Maslov index of the overall loop is zero
\begin{eqnarray}
\sum_{i=1}^{4}\aleph_i(\fu,\bc)=0
\label{Morse-topology:overall}
\end{eqnarray}
Therefore it follows
\begin{eqnarray}
\ind \fobc([T',T])= \aleph_1(\fu,\bc)\,=\,-[\aleph_2(\fu,\bc)+\aleph_3(\fu,\bc)+\aleph_4(\fu,\bc)]
\label{Morse-topology:Morsebranches}
\end{eqnarray}
The Maslov index of the second branch of the loop can be evaluated a priori.
For any $\tau$ the Hamiltonian matrix (\ref{Morse-geometry:Hamiltonian}) 
is similar to
\begin{eqnarray}
(\mathsf{O}\,\mathsf{H}\,\mathsf{O}^{-1})_{\tau}(t)=
\left[\begin{array}{cc} 
-\pot-\tau\,\mathsf{U}\quad&\quad  0 \\
0 \quad & \quad \mass^{-1}
\end{array}\right]\!(t)
\label{Morse-topology:Hamiltondiag}
\end{eqnarray}
for $\mathsf{O}$ a suitable non singular matrix.
For $\tau$ large enough, (\ref{Morse-topology:Hamiltondiag}) is the 
orthogonal sum of two $d\,\times\,d$ dimensional blocks with opposite sign 
definition. 
The observation recovers the result (\ref{Morse-Morse-local:kinetic}) 
obtained for local boundary conditions. If $\Bott_2=0\oplus\id_d$ the Morse 
index coincides with the number of crossings counted with their degeneration 
in the open time interval $]T',T[$. Any overlap of the Bott
matrix $\Bott_2$ with the position space produces a non trivial 
generalisation of the order of concavity defined in 
subsection~\ref{Morse-Morse:periodic} for the particular case of periodic 
boundary conditions. 
Homotopy invariance of Maslov indices and H\"ormander's identity
(\ref{Morse-Maslov:Hormander}) endow with great freedom in the choice of 
the representation of the Morse index which most suits the analysis of
the physical origin of the contributions.
Periodic boundary conditions well illustrate the general situation.

\subsection{Morse index and structural stability of periodic orbits}
\label{Morse:stability}

The difference $T-T'$ is in this section supposed to be equal or 
a multiple of the period of the Sturm-Liouville operator.\\
The Bott pair describing periodic boundary conditions is 
$(\Bott_1\,,\Bott_2)=(\id_{2 d},\id_{2 d})$. Therefore the Maslov index of 
the second branch of the homotopy loop is zero.\\
H\"ormander's identity can be applied directly to the catenation of the second
and fourth branch. The identification of $\bc^{(a)}$ with $(\id_{2 d},\id_{2 d})$ 
and of $\bc^{(b)}$ with any $\mathfrak{A}\sim (\Bott_1,\Bott_2)$ such that
\begin{eqnarray}
\left\{\begin{array}{c}
\mathfrak{A}\cap(\id_{2 d},\id_{2 d})=\emptyset
\\
\mathfrak{A}\cap\gph\fu(T,T')=\emptyset
\end{array}\right. \quad \Leftrightarrow \quad
\left\{\begin{array}{c}
\det(\Bott_2-\Bott_1)\,\neq\,0
\\
\det[\Bott_2-\fu(T,T')\Bott_1]\,\neq\,0
\end{array}\right.
\label{Morse-stability:condition}
\end{eqnarray}
yields
\begin{eqnarray}
\sum_{i=2}^{4}\aleph_{i}\{\fu,(\id_{2 d},\id_{2 d})\}\!=
\aleph_{3}(\fu,\mathfrak{A})+
\frac{1}{2}\,\mathrm{sign}\mathcal{Q}(\mathfrak{A},\gph\fu(T,T');(\id_{2 d},\id_{2 d})\,)
\label{Morse-stability:34}
\end{eqnarray}
A more detailed proof can be found in theorem $4.3$ of ref.~\cite{Duistermaat}.
The signature and the index of any symmetric form $\mathcal{F}$ in $2 d$ dimensions 
are connected by the relation
\begin{eqnarray}
\ind \mathcal{F}=\frac{1}{2}\,[2 d-\ke \mathcal{F}-\mathrm{sign}\mathcal{F}]
\label{Morse-stability:signature}
\end{eqnarray}
Thus one can recast the periodic index in the form
\begin{eqnarray}
\ind\foP([T,T'])=-\aleph_{3}(\fu,\mathfrak{A})
+\ind\mathcal{Q}(\gph\fu(T,T'),\mathfrak{A};(\id_{2 d},\id_{2 d})\,)-d
\label{Morse-stability:Duistermaat}
\end{eqnarray}
which is the final expression given by Duistermaat in ref.~\cite{Duistermaat}.

The topological expression of the periodic Morse index can be used to
analyse how the stability properties of the linear flow $\fu(t,T')$ are
reflected in the index. The use of homotopy simplifies the analysis.
The idea is to embed the linear flow $\fu(t,T')$ into a family
of curves in $Sp(2 d)$ with fixed end points. By property $iv$ of 
section~\ref{Morse:Maslov} the Maslov index is the same for any element 
of family. The Floquet representation of the linear flow 
\begin{eqnarray}
\fu(t,0)&=&\mathsf{Pe}(t)\,\exp\left\{\frac{t}{T}\,
\sym^{\,\dagger}\!\!\int_{0}^{T}dt'\,\mathsf{H}(t')\right\}
=:\mathsf{Pe}(t)\,e^{t\,\sym^{\dagger}\,\avh}
\label{Morse-stability:Floquet}
\end{eqnarray}
is suited for the construction. In order to streamline the notation $T'$ has been 
set to zero. 
The matrix $\mathsf{Pe}(t)$ is supposed to have prime period $T$, 
while $\avh$ is the average of the Hamiltonian over a period. 
The Floquet form is the element of the continuous family
\begin{eqnarray}
\mathsf{G}(t,s)=
\left\{\begin{array}{cc}\mathsf{Pe}[(1+s)\,t]\,
e^{(1-s)\,t\,\sym^{\dagger}\avh} \quad &\quad 0\leq t \leq \frac{T}{2}
\\
\mathsf{Pe}(t)\mathsf{Pe}^{-1}[s\,(t-T)]\,
e^{[s\,(t-T)+t]\sym^{\dagger}\avh} \quad&\quad \frac{T}{2} \leq t\leq T
\end{array}\right.
\label{Morse-stability:homotopy}
\end{eqnarray}
attained by setting $s$ equal to zero. The end points of the homotopy family are
independent of $s$. In order the Floquet representation to be fully specified,
it is necessary to fix a convention on the phase of the eigenvalues of the 
monodromy matrix: 
\begin{eqnarray}
\mon:=e^{T\sym^{\dagger}\avh}
\label{Morse-stability:monodromy}
\end{eqnarray}
A convenient choice is to restrict the phase to the interval $[0,2\,\pi[$.
The choice attributes to the exponential matrix in the Floquet representation 
a {\em winding number} equal to zero. Gel'fand and Lidskii introduced 
in \cite{GelfandLidskii} the winding number of a linear Hamiltonian flow 
as a topological characterisation of structural stability. 
In particular they proved that linear flows with the same monodromy and winding
can be deformed into each other. 

The tools forged above will now be used to compute the periodic Morse index
in two different ways. The first is meant to illustrate the relation of the index
with the stability properties of the flow, the second to recover the result
of section~\ref{Morse-Morse:periodic}.

\subsubsection{The Conley and Zehnder index}
\label{ConleyZehnder}

The most direct way to evaluate the periodic index is to 
consider intersections of the homotopy loop with the periodic pair 
$(\id_{2 d}\,,\id_{2 d})$.
The restriction to regular crossings imposes some conditions on the 
linear flows. In particular flows with parabolic normal forms
must be regularised by a positive definite perturbation in order to 
render non singular the crossing form.\\
With this proviso, the index is equal to the Maslov index of the third branch of the homotopy 
loop (\ref{Morse-topology:Morsebranches}). 
This latter is most conveniently computed at $s$ equal one for the family
of symplectic curves (\ref{Morse-stability:homotopy}):
\begin{eqnarray}
\aleph_3\{\fu,(\id_{2 d},\id_{2 d})\}&=&\aleph\{\mathsf{Pe}(2\,t),(\id_{2 d},\id_{2 d}),[0,T/2]\}
\nonumber\\
&+&\aleph\{e^{(2\,t-T)\,\sym^{\dagger}\,\avh},(\id_{2 d},\id_{2 d}),[T/2,T]\}
\label{Morse-stability-ConleyZehnder:contributions}
\end{eqnarray}
The first Maslov index is independent of the Lagrangian manifold associated to
the periodic Bott pair $(\id_{2 d}\,,\id_{2 d})$. The statement is proven by the 
following reasoning. Since the matrix $\mathsf{Pe}(t)$ is periodic its index does 
not change if the interval $[0,T/2]$ is replaced by $[t',T/2+t']$ for any $t'$ 
such that no intersection occurs. H\"ormander's identity
(\ref{Morse-Maslov:Hormander}) and periodicity then show
\begin{eqnarray}
\aleph\{\mathsf{Pe}(2\,t),(\id_{2 d},\id_{2 d}),[0,T/2]\}=
\aleph\{\mathsf{Pe}(2\,t),\mathfrak{A},[t',T/2+t']\}
\label{Morse-stability-ConleyZehnder:Hormander}
\end{eqnarray}
for any $\mathfrak{A}$ having no intersections with the graph of the flow at path 
end-points. In particular one can chose $\mathfrak{A}\sim(\id_{2 d}\,,-\,\id_{2 d})$
and observe that any symplectic matrix admits a unique {\em polar representation} 
\begin{eqnarray}
\mathsf{Pe}(2\,t)=(\mathsf{Sy}\,\mathsf{Or})(2\,t)\,,\qquad
\left\{\begin{array}{l}\mathsf{Sy}(2\,t):=(\mathsf{Pe}^{\dagger}\mathsf{Pe})^{\frac{1}{2}}(2\,t)\\
\mathsf{Or}(2\,t):=[(\mathsf{Pe}^{\dagger}\mathsf{Pe})^{-\,\frac{1}{2}}\mathsf{Pe}](2\,t)
\end{array}\right.
\label{Morse-stability-ConleyZehnder:polar}
\end{eqnarray}
with both $\mathsf{Or}$ and $\mathsf{Sy}$ in $Sp(2 d)$.
The polar representation can be embedded into a family of curves analogous to
(\ref{Morse-stability:homotopy}). Since  $\mathsf{Sy}$ is positive definite,
crossing occurs only when
\begin{eqnarray}
\det[\id_{2\,d}+\mathsf{Or}(2\,t)]=0
\end{eqnarray}
as $t$ ranges in $[0\,,T/2]$.
The Maslov index can be thus computed from the intersection of the orthogonal 
matrix $\mathsf{Or}(2\,t)$ with the anti-periodic Bott pair.\\ 
An orthogonal and symplectic matrix admits in general the block representation
\begin{eqnarray}
\mathsf{Or}=\left[\begin{array}{cc} \mathsf{X} \quad&\quad \mathsf{Y} \\ 
-\mathsf{Y} \quad&\quad \mathsf{X}\end{array}\right]\,,\qquad\qquad  
(\mathsf{X}+i \mathsf{Y})^{\dagger}(\mathsf{X}+i \mathsf{Y})=\id_{d}
\label{Morse-stability-ConleyZehnder:orthogonal}
\end{eqnarray}
and therefore it is similar to
\begin{eqnarray}
\frac{1}{\sqrt{2}}\left[\begin{array}{lr} \id_{d} \,\,&\,\, -\,i \id_{d} \\ 
\id_{d} \,\,&\,\, i\,\id_{d}\end{array}\right]
\left[\begin{array}{cr} \mathsf{X} \,\,&\,\, \mathsf{Y} \\ 
-\mathsf{Y} \,\,&\,\, \mathsf{X}\end{array}\right]
\frac{1}{\sqrt{2}}\left[\begin{array}{lr} \id_{d} \,\,&\,\, i \id_{d} \\ 
\id_{d} \,\,&\,\, -i\,\id_{d}\end{array}\right]=
\left[\begin{array}{lr} \mathsf{X}+i\,\mathsf{Y} \,\,&\,\, 0 \\ 
0 \,\,&\,\, \mathsf{X}-i\,\mathsf{Y}\end{array}\right]
\label{Morse-stability-ConleyZehnder:Hermitian}
\end{eqnarray}
The analysis of the crossing forms yields modulo four the equality 
\begin{eqnarray}
\aleph\{\mathsf{Pe}(2\,t)&,&(\id_{2 d},\id_{2 d}),[0,T/2]\}=
\nonumber\\
&&\frac{\mathrm{arg}\det[\mathsf{X}(T)+i\,\mathsf{Y}(T)]-
\mathrm{arg}\det[\mathsf{X}(0)+i\,\mathsf{Y}(0)]}{\pi}\,,\quad \mathrm{mod}\,4
\label{Morse-stability-ConleyZehnder:winding}
\end{eqnarray}
The right hand side is an integer by periodicity. It coincides with twice the 
Gel'fand and Lidskii \cite{GelfandLidskii} winding number of the linear periodic 
Hamiltonian flow $\mathsf{Pe}(2\,t)$. 

The only contribution to the second Maslov index in 
(\ref{Morse-stability-ConleyZehnder:contributions}) comes when $t$ is equal 
$T/2$. 
The crossing form can be straightforwardly evaluated from the normal forms of the 
stability blocks of the exponential flow.
Following the same conventions of appendix~\ref{Morse:examples} 
the Hamiltonian matrices associated to the blocks are
\begin{eqnarray}
&&\avh|_{\mathrm{ell.}}=\left[\begin{array}{lr} \omega \quad & \quad 0 \\ 0 & \omega
\end{array}\right]
\,,\qquad
\avh|_{\mathrm{hyp}}=\left[\begin{array}{lr} 0 \quad & \quad \omega \\ \omega & 0
\end{array}\right]
\nonumber\\
&&\avh|_{\mathrm{lox.}}=\left[\begin{array}{lr} 0 \quad & \quad \omega\,(\id_2+\sym_2) \\ 
\omega\,(\id_2+\sym_2)^{\dagger} & 0\end{array}\right]
\label{Morse-stability-ConleyZehnder:normalforms}
\end{eqnarray}
A consequence of the analysis at the end of section~\ref{Hamilton:periodic} is that 
inverse hyperbolic blocks behave at the origin as elliptic ones \cite{Ekeland,LeboeufMouchet}.
The simplest parabolic blocks have the normal form
\begin{eqnarray}
\left.\fu(t,T')\right|_{\mathrm{par.}}\,=\,\left[\begin{array}{cc}
1 \quad & \quad \kappa\, \frac{t}{T}
\\
0 \quad & \quad 1
\end{array}\right]
\label{Morse-stability-ConleyZehnder:parabolic}
\end{eqnarray}
In appendix (\ref{Hamilton:perturbative}) it is shown that 
(\ref{Morse-stability-ConleyZehnder:parabolic}) bifurcates under a generic 
positive definite perturbation to an elliptic or to an hyperbolic block
if, within the conventions adopted in this work, 
the sign of $\kappa$ is respectively negative or positive. 
In order to infer the sign of the infinitesimal characteristic frequencies 
of the elliptic blocks generated by the regularisation, one can reason as follows. 
A sufficiently small positive definite perturbation of the $\foP([0,T])$ does not affect 
the periodic Morse index because it removes the zero modes while preventing any 
eigenvalue from turning negative.
Therefore the value of the Maslov index associated to parabolic blocks by such
a perturbation of the $\foP([0,T])$ must coincide with the minimum over the set 
of the Maslov indices generated by arbitrary perturbations of the same system 
\cite{Long2,AnLong,AnLong2,Long}. 
This means that, once the Gel'fand-Lidskii winding number has been determined 
from the unperturbed system, the sign of the characteristic frequencies of elliptic blocks 
created by a positive definite perturbation must give rise to crossing forms with 
negative signature. A positive signature would produce a larger value of the 
Maslov index. A rigorous proof of the above perturbative argument can be found in 
\cite{Long2}.\\
Gathering the contributions of all blocks the resulting Maslov index is
\begin{eqnarray}
\aleph\{e^{(2\,t-T)\,\sym^{\dagger}\,\avh},(\id_{2 d},\id_{2 d}) ,[T/2,T]\}=
n_{\mathrm{ell}}+n_{\mathrm{i.h.}}-\sum_{i=1}^{n_{\mathrm{p}}}
\frac{1-\mathrm{sign}\,\kappa_i}{2}
\label{Morse-stability-ConleyZehnder:Maslov2}
\end{eqnarray}
where $n_{\mathrm{ell}}$, $n_{\mathrm{i.h.}}$ and $n_{\mathrm{p}}$ are respectively
the number of elliptic and parabolic blocks in the monodromy matrix
$\mon=\fu(T,0)$. 

The overall periodic Morse index is
\begin{eqnarray}
\ind \foP([0,T])=n_{\mathrm{ell}}+n_{\mathrm{i.h.}}-\sum_{i=1}^{n_{\mathrm{p}}}
\frac{1-\mathrm{sign}\,\kappa_i}{2}+2\,\mathfrak{w}
\label{Morse-stability-ConleyZehnder:ConleyZehnder}
\end{eqnarray}
with $\mathfrak{w}$ the Gel'fand-Lidskii winding number. Elliptic
and inverse hyperbolic blocks can therefore be associated with
the occurrence of an odd number of negative eigenvalues in the spectrum 
of a periodic Sturm-Liouville operator $\foP$.

In the mathematical literature 
\cite{RobbinSalamon,RobbinSalamon2,Salamon,McDuffSalamon} the periodic Morse index
expressed in the form (\ref{Morse-stability-ConleyZehnder:ConleyZehnder}) 
is often referred to as the {\em Conley and Zehnder index}. 
Conley and Zehnder \cite{ConleyZehnder} showed that the periodic Morse index 
(\ref{Morse-stability-ConleyZehnder:ConleyZehnder}) describes the foliation 
of the symplectic group $Sp(2 d)$ in leaves of different winding. 
Each leaf further decomposes in two connected components characterised by opposite 
sign of the determinant  
\begin{eqnarray}
\mathrm{sign}\,\mathrm{det}[\id_{2 d}-\mon]
=(-1)^{d-n_{\mathrm{ell}}-n_{\mathrm{i.h.}}}
\label{Morse-stability-ConleyZehnder:signature}
\end{eqnarray}
In their paper \cite{ConleyZehnder}, Conley and Zehnder analysed the stability properties of
periodic orbits of non-autonomous Hamiltonian systems without parabolic blocks and 
in dimension larger than two. General proofs of the equivalence of the periodic 
Morse index with the Conley and Zehnder index were later given in 
\cite{Viterbo,AnLong,AnLong2} where the case of conservative systems with parabolic blocks 
is also encompassed. 

It is straightforward to infer the behaviour of the index under iteration of a prime
periodic orbit.\\
The winding number increases linearly with the number of repetitions $r$ 
of the orbit. The sign (\ref{Morse-stability-ConleyZehnder:signature}) may instead change 
according to a nonlinear law. Namely under iteration elliptic blocks of frequency $\omega$ 
experience further intersections if $r\,\omega$ becomes a multiple of $2\,\pi$. 
Inverse hyperbolic blocks generate a crossing form at each iteration since 
they are turned by even powers of the monodromy matrix into direct hyperbolic 
ones. Parabolic blocks do not change signature. 
Hence one concludes
\begin{eqnarray}
\ind &&\foP([0,r\,T])=
\nonumber\\
&&\sum_{i=1}^{n_{\mathrm{ell}}}
\left[1+2\,\mathrm{int}\left(\frac{r\,\omega_i}{2\,\pi}\right)\right]
+r\,n_{\mathrm{i.h.}}-\sum_{i=1}^{n_{\mathrm{p}}}
\frac{1-\mathrm{sign}\,\kappa_i}{2}+2\,r\,\mathfrak{w}
\label{Morse-stability-ConleyZehnder:iteration}
\end{eqnarray}
The result holds for Hamiltonian systems without marginal degenerations. 
In \cite{Sugita}, Sugita derived formulae
(\ref{Morse-stability-ConleyZehnder:ConleyZehnder}), 
(\ref{Morse-stability-ConleyZehnder:iteration}) 
by computing quadratic phase space trace path integrals on a time lattice. 
Sugita also pointed out that the negative sign in front of the parabolic block 
contribution renders in general the phase of the Gutzwiller trace formula 
independent from the signature of the energy block.
A careful analysis of the behavior of the phase of the Gutzwiller trace formula
predicted by (\ref{Morse-stability-ConleyZehnder:iteration}) has been carried out in a 
recent paper by Pletyukhov and Brack \cite{PletyukhovBrack}
where several analytical and numerical examples are also investigated.\\
More general explicit expressions of the periodic Morse index in the Conley and Zehnder form,
encompassing marginal degenerations can be found in the paper \cite{Long}.

The behaviour of Morse indices of Sturm-Liouville operators of variational
origin was investigated in the topological setting in ref. 
\cite{CushmanDuistermaat}. The same results can also be obtained by means of pure 
algebraic methods \cite{Klingenberg2,Ekeland}.

Finally it is worth noting that the periodic Morse index can be identified with
the winding number of the linear flow by applying the polar 
decomposition directly to $\fu(t,0)$ \cite{Littlejohn2,Robbins}. 

\subsubsection{Focal point description}
\label{focal}

The periodic Morse index is needed in the Gutzwiller trace formula. 
Due to energy conservation at least one parabolic block is to be expected. 
Therefore in applications \cite{Robbins}
it is useful to exploit H\"ormander's identity (\ref{Morse-Maslov:Hormander})
to derive the index without need of perturbative arguments. Inserting
$\mathfrak{A}\sim(\id_{d}\oplus 0\,,0\oplus\id_{d})$ in Duistermaat's formula 
(\ref{Morse-stability:Duistermaat}) yields:
\begin{eqnarray}
\ind\foP([0,T])&=&\sum_{0\leq t\,<T} \nul \mathsf{D}(t,0)
\nonumber\\
&+&\ind\mathcal{Q}(\gph\fu(T,0),\id_{d}\oplus 0\,,0\oplus\id_{d};(\id_{2 d},\id_{2 d})\,)-d
\label{Morse-stability-focal:Duistermaat}
\end{eqnarray}
The matrix $\mathsf{D}$ is specified by the block representation of the flow 
(\ref{Morse-Morse-local:flow}). Some tedious algebra gives the expression of
the concavity form:
\begin{eqnarray}
\mathcal{Q}(\gph \fu(T,0)&,&\!(\id_{d}\oplus 0\,,0\oplus\id_{d});(\id_{2 d},\id_{2 d})\,)
\nonumber\\
&=&\left[\begin{array}{cc} 
(\mathsf{C} \mathsf{A}^{-1})(T,0) \,\,&\,\,\mathsf{A}^{\dagger\, -1}(T,0)-\id_{d}\\
\mathsf{A}^{-1}(T,0)-\id_{d} \,\,&\,\,-\,(\mathsf{A}^{-1}\mathsf{B})(T,0)
\end{array}\right]
\label{Morse-stability-focal:boundary}
\end{eqnarray}
It is straightforward to see that in any vector basis where 
$\mathsf{B}(T,0)$ and  $\mathsf{C}(T,0)$ vanish the index of 
(\ref{Morse-stability-focal:boundary}) is equal to $d$. 
Thus the periodic index coincides with the zeroes encountered in the same 
vector basis by $\mathsf{D}(t,0)$ through all the time interval $[0,T]$.\\
More generally if $\mathsf{B}(T,0)$ is nonsingular then one finds
\begin{eqnarray}
&&\ind\mathcal{Q}(\gph\,\fu(T,0),(\id_{d}\oplus 0\,,0\oplus\id_{d});(\id_{2 d},\id_{2 d})\,)=
\nonumber\\
&&\,\,\,\,\ind(\mathsf{D}\,\mathsf{B}^{-1}+\mathsf{B}^{-1}\mathsf{A}-\mathsf{B}^{-1}
-\mathsf{B}^{-1\,\dagger})(T,0)+\ind[-(\mathsf{A}^{-1}\mathsf{B})(T,0)]
\label{Morse-stability-focal:boundaryindex}
\end{eqnarray}
Under the same hypothesis the use of H\"ormander's identity gives
\begin{eqnarray}
\sum_{0\leq t\,<T}&&\!\!\nul \mathsf{D}(t,0)=
\nonumber\\
&&\aleph_3\{\fu,(0\oplus \id_{d}\,,0\oplus\id_{d})\}
+d-\ind[-(\mathsf{A}^{-1}\mathsf{B})(T,0)]
\end{eqnarray}
Since
\begin{eqnarray}
&\aleph_3\{\fu,(0\oplus \id_{d}\,,0\oplus\id_{d})\}=\sum_{0\,< t\,<T}\!\!\! \nul \mathsf{B}(t,0)
\end{eqnarray}
the expression of the periodic index given in
section~\ref{Morse-Morse:periodic} is finally recovered.

The focal point representation can also be applied in a coordinate independent way.
In particular it is possible to derive equation 
(\ref{Morse-stability-ConleyZehnder:ConleyZehnder}) (the Conley and Zehnder index), 
without need of the perturbative argument. \\
Consider for example a periodic Sturm-Liouville operator $\foP([0,T])$ such that
the associated linear flow consists of a parabolic and an elliptic block. 
The Floquet representation is then
\begin{eqnarray}
\fu(t,0)=\mathsf{Pe}(t)\,e^{t\,\sym^{\dagger}\,\avh}=\mathsf{Pe}_{\mathrm{par.}}(t)\,
e^{t\,\sym^{\dagger}\,\avh_\mathrm{par.}}\oplus
\mathsf{Pe}_\mathrm{ell.}(t)\,e^{t\,\sym^{\dagger}\,\avh_\mathrm{ell.}}
\label{addednote:Floquet}
\end{eqnarray}
As assumed above, phase $\omega t$ of the exponential matrix of the elliptic block 
is supposed to range in $[0,2\,\pi[$.

The {\em product} and {\em homotopy} properties of Maslov indices 
illustrated in section~\ref{Morse:Maslov}
permit to compute the periodic Morse index of $\foP([0,T])$ as the sum of the Maslov indices 
of the stability blocks. Moreover, H\"ormander equality (\ref{Morse-Maslov:Hormander})
can be separately applied to the periodic and exponential matrices entering the Floquet 
representation.  
The Maslov index of the periodic matrix is already known to coincide with twice the
Gel'fand and Lidskii winding number $\mathfrak{w}$. \\
In order to compute the Maslov index of 
$e^{t\,\sym^{\dagger}\,\avh_\mathrm{par.}}$, $e^{t\,\sym^{\dagger}\,\avh_\mathrm{ell.}}$
in $[0,T]$ one can apply Duistermaat formula (\ref{Morse-stability:Duistermaat})
with $\mathfrak{A}\sim(1\oplus 0,0\oplus 1)$. Consider first the elliptic block:
\begin{eqnarray}
e^{t\,\sym^{\dagger}\,\avh_\mathrm{ell.}}=\left[\begin{array}{cc}
\cos(\omega t) \quad & \quad \sin(\omega t)
\\
-\,\sin(\omega t) \quad & \quad  \cos(\omega t)
\end{array}\right]
\label{addednotes:elliptic}
\end{eqnarray}
Crossings occur when
\begin{eqnarray}
\det[(0\oplus 1)-e^{t\,\sym^{\dagger}\,\avh_\mathrm{ell.}}(1\oplus 0)]=-\cos(\omega t)=0
\label{addednote:ellipticcrossing}
\end{eqnarray}
The condition is satisfied if $\omega t$ is equal to $\pi/2$ or to $3\,\pi/2$. The corresponding 
crossing forms are equal to one. By equation (\ref{Morse-stability-focal:boundary})
the order of concavity is 
\begin{eqnarray}
&&\ind\mathcal{Q}(\gph e^{T\,\sym^{\dagger}\,\avh_\mathrm{ell.}},(1\oplus 0\,,0\oplus 1);(\id_{2},\id_{2})\,)-1=
\nonumber\\
&&\quad \quad \ind \left[\begin{array}{cc}
-\frac{\sin (\omega T)}{\cos(\omega T)} \quad & \quad \frac{1}{\cos(\omega T)}-1
\\
\frac{1}{\cos(\omega T)}-1 \quad & \quad  -\frac{\sin (\omega T)}{\cos(\omega T)}
\end{array}\right]-1
\label{addednote:ellipticconcavity}
\end{eqnarray}
It is equal to one for $\omega T$ taking values in $[0\,,\pi/2]$, is zero for 
$\omega T$ in $]\pi/2\,,3\pi/2]$ and finally minus one for $\omega T$ in $]3\,\pi/2\,,2\,\pi]$.
Adding up the contributions of the crossing and of the concavity forms one has:
\begin{eqnarray}
\aleph\{ e^{t\,\sym^{\dagger}\,\avh_\mathrm{ell.}},(1\oplus 0\,,0\oplus 1),[0,T]\}=1
\end{eqnarray}
independently of the value $\omega T$ in $[0,2\,\pi[$.\\
Turning to the parabolic block eq. (\ref{Morse-stability-ConleyZehnder:parabolic})
, it is seen that the only non zero contribution comes from the order of concavity:
\begin{eqnarray}
&&\ind\mathcal{Q}(\gph e^{T\,\sym^{\dagger}\,\avh_\mathrm{par.}},(1\oplus 0\,,0\oplus 1);(\id_{2},\id_{2})\,)-1=
\nonumber\\
&&\quad \quad \ind \left[\begin{array}{cc}
0 \quad & \quad 0
\\
0 \quad & \quad  -\kappa \end{array}\right]-1=-\frac{1-\mathrm{sign}\kappa}{2}
\label{addednote:parabolicconcavity}
\end{eqnarray}
Gathering all terms one gets into
\begin{eqnarray}
\ind\foP([0,T[)=1-\frac{1-\mathrm{sign}\kappa}{2}+2\mathfrak{w}
\end{eqnarray}
which is formula (\ref{Morse-stability-ConleyZehnder:ConleyZehnder})
applied to the example considered here. The block by block analysis of the Maslov index,
has been used by An and Long \cite{AnLong2} to prove the equivalence of the
Duistermaat formula (\ref{Morse-stability:Duistermaat}) with the Conley and Zehnder 
index.

\subsection{Phase space path integrals and infinite dimensional Morse theory}
\label{Morse:Hamilton}

Classical mechanics has its natural formulation in phase space. The lift 
of configuration space quantities to phase space was reiteratively used 
in the above presentation of Morse index theory. 
It is therefore natural to wonder whether the treatment of the 
semiclassical approximation can be simplified by starting from phase
space path integrals.

Formally, phase space path integrals can be written directly in the continuum  
by replacing the quadratic kinetic energy term in the configuration space 
Lagrangian with its Fourier transform 
\cite{DeWittMoretteMaheshwariNelson,LevitSmilansky,Roepstorff,Zinn}. 
The propagator is then represented in the form
\begin{eqnarray}
\feo(Q,T|Q',T')=\int_{q(T')=Q'}^{q(T)=Q}\mes[q(t)p(t)]e^{i\int_{T'}^{T}dt
[p_{\alpha}\dot{q}^{\alpha}-\ha(p,q)]}
\label{Morse-Hamilton:pathintegral}
\end{eqnarray}
The domain of integration is asymmetric between position and momentum variables as
the boundary conditions impose a constraint only on position variables. 
More general phase space path integral expression are obtained if 
(\ref{Morse-Hamilton:pathintegral}) is used to compute the time evolution 
of quantum observables other than the propagator.\\
Formally the derivation of the semiclassical approximation of section~\ref{Forman:quadratic} 
goes through also for phase space path integrals. In the presence of boundary conditions $\bc$
of the form (\ref{Forman-quadratic:bc}) it leads to the quadratic action
\begin{eqnarray}
\qac=\int_{T'}^{T}dt\,\delta x^{\alpha}\mathfrak{J}_{\alpha\,\beta}\delta x^{\beta}\,,
\qquad\qquad \alpha\,,\beta=1,...,2\,d
\label{Morse-Hamilton:fluctuations}
\end{eqnarray}
where $\delta x=(\flu,\delta p)$ denotes the fluctuations around the phase space 
classical trajectory $x_{\cl}$ while 
\begin{eqnarray}
\mathfrak{J}_{\alpha\,\beta}:=\,\sym_{\alpha\,\beta}\,\frac{d}{dt}-\mathsf{H}_{\alpha\,\beta}
\label{Morse-Hamilton:Jacobi}
\end{eqnarray}
with $\mathsf{H}_{\alpha\,\beta}$ the Hamiltonian matrix (\ref{Morse-geometry:Hamiltonian})
at $\tau$ equal zero. 
The operator $\mathfrak{J}_{\bc}$ is then self-adjoint with respect to the phase space 
scalar product
\begin{eqnarray}
\langle x,y\rangle=\frac{1}{T-T'}\int_{T'}^{T}dt x^{\alpha}\delta_{\alpha,\beta}
y^{\beta}\equiv\frac{1}{T-T'}\int_{T'}^{T}dt x^{\dagger} y
\label{Morse-Hamilton:scalarproduct}
\end{eqnarray} 
The scalar product presupposes a suitable rescaling of positions and momenta. 
The fact is not at variance with classical mechanics where position and momenta 
lose their physical interpretation under canonical transformations \cite{Arnold,LandauLifshitz}. 
\\
Forman theorem can be applied to recover the configuration space result
\begin{eqnarray}
|\mathrm{Det}  \mathfrak{J}_{\bc}([T',T])|=|\varkappa_{\bc}\,
\det[\mathsf{Y}_{1}+\mathsf{Y}_{2}\fu(T,T')]|
\label{Morse-Hamilton:Forman}
\end{eqnarray}
The operator $\mathfrak{J}_{\bc}$ is a first order Dirac operator 
\cite{LevitSmilansky,SalamonZehnder,RobbinSalamon2}. At variance with the Sturm-Liouville 
operators of the Lagrangian formulation the eigenvalue spectrum is unbounded from below 
and from above. Hence the Morse index cannot be defined as in configuration space.
However, the lattice approximation identifies the index of any {\em non singular}
discrete version of $\mathfrak{J}_{\bc}$ as the difference between
the positive and negative eigenvalues. It is possible to
prove \cite{RobbinSalamon3,RobbinSalamon4} that in the {\em infinite time lattice} limit 
such quantity converges to the Morse index found above working in configuration space 
(see also \cite{McDuffSalamon}). Difficulties may arise if one works in directly in the
continuum.
The discussion can be made more concrete for periodic boundary conditions.
The eigenvalue spectrum of limit fluctuation operator $\mathfrak{J}_{\mathrm{Per.}}$ is 
invariant under periodic transformations in $Sp(2 d)$ continuously connected to the identity:
\begin{eqnarray}
\mathsf{S}^{\dagger}\mathfrak{J}\mathsf{S}=\sym\,\frac{d}{dt}-
(\mathsf{S}^{\dagger}\mathsf{H}\mathsf{S}-\mathsf{S}^{\dagger}\sym\dot{\mathsf{S}})
\end{eqnarray}
The choice
\begin{eqnarray}
\mathsf{S}(t,T')=\fu(t,T')\,e^{-\sym^{\dagger} \avh\,(t-T')}
\end{eqnarray}
with $\avh$ defined as in (\ref{Morse-stability:Floquet}) replaces the Hamiltonian matrix
in $\mathfrak{J}_{\mathrm{Per.}}$ with its average over one period. The eigenvalue problem 
is then explicitly solvable in general 
\cite{DashenHasslacherNeveu2,DashenHasslacherNeveu3,DittrichReuter,Sugita}. The result is
evidently independent of the winding number of the monodromy of the linearised flow.
This means that the explicit diagonalisation of $\mathfrak{J}_{\mathrm{Per.}}$ 
will give only the correct sign of the functional determinant but not the Morse
index. The semiclassical approximation needs the square root of the 
determinant. Thus, in the phase space continuum limit the phase factor of 
the path integral {\em cannot} be defined by simply diagonalising the fluctuation 
operator. Instead, it is defined by the the index of 
the eigenvalue (Fredholm) flow of the homotopy transformation between 
$\mathfrak{J}_{\bc}$ and another Dirac operator whereof the index is a 
priori known. This is the content of the {\em infinite dimensional Morse theory}
developed in ref.'s \cite{SalamonZehnder,RobbinSalamon2} where the topological methods 
of sections~\ref{Morse:Maslov}, \ref{Morse:topology} are systematically applied.
Thus, the evaluation of the periodic Morse index turns out to be an example in
the simplest physical context of a general Fredholm flow theory which 
has found in recent years wide applications in field theory starting 
with the treatment of the $SU(2)$ anomaly given by Witten in \cite{Witten2}. 
Recently Witten's approach has been applied by Sugita in \cite{Sugita} to 
the derivation of the phase of the Gutzwiller trace formula.

Once the more subtle continuum limit rules of calculus are understood, 
phase space path integrals offer the advantage of an elegant formalism able 
to deal with a wider class of problems than their configuration space 
homologues. A typical case is the quantisation of systems classically governed
by singular Lagrangians, see for example \cite{Faddeev}.


%% file: homotopy.pstex_t
\begin{picture}(0,0)%
\includegraphics{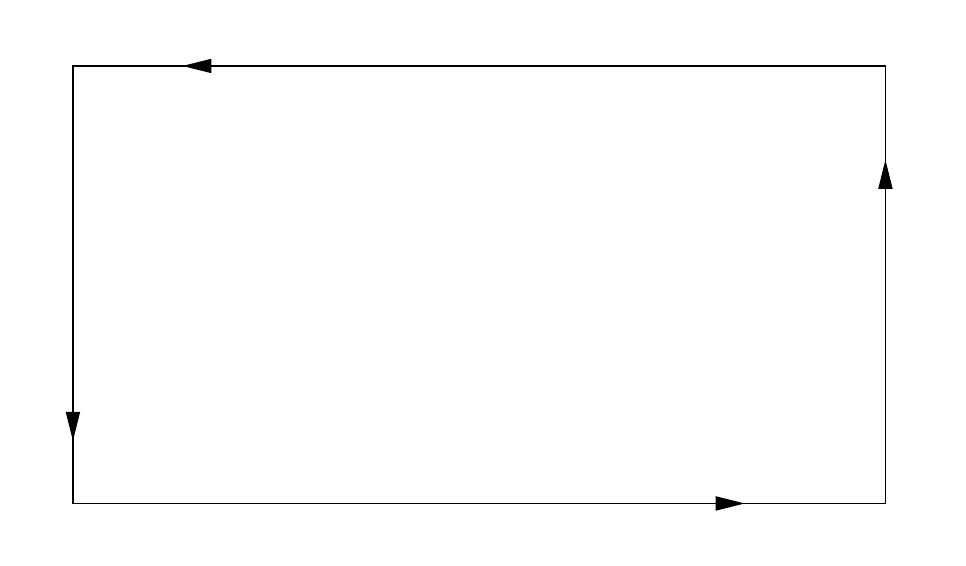}%
\end{picture}%
\setlength{\unitlength}{3947sp}%
\begingroup\makeatletter\ifx\SetFigFont\undefined%
\gdef\SetFigFont#1#2#3#4#5{%
  \reset@font\fontsize{#1}{#2pt}%
  \fontfamily{#3}\fontseries{#4}\fontshape{#5}%
  \selectfont}%
\fi\endgroup%
\begin{picture}(4620,2706)(106,-2311)
\put(4726,-886){\rotatebox{90.0}{\makebox(0,0)[b]{\smash{\SetFigFont{12}{14.4}{\rmdefault}{\mddefault}{\updefault}\special{ps: gsave 0 0 0 setrgbcolor}$\mathsf{F}_{\tau}(T,T')$\special{ps: grestore}}}}}
\put(2326,-2311){\makebox(0,0)[b]{\smash{\SetFigFont{12}{14.4}{\rmdefault}{\mddefault}{\updefault}\special{ps: gsave 0 0 0 setrgbcolor}$\mathsf{F}(t,T')$\special{ps: grestore}}}}
\put(301,-961){\rotatebox{90.0}{\makebox(0,0)[b]{\smash{\SetFigFont{12}{14.4}{\rmdefault}{\mddefault}{\updefault}\special{ps: gsave 0 0 0 setrgbcolor}$\mathsf{F}_{\tau}(T',T')$\special{ps: grestore}}}}}
\put(2326,239){\makebox(0,0)[b]{\smash{\SetFigFont{12}{14.4}{\rmdefault}{\mddefault}{\updefault}\special{ps: gsave 0 0 0 setrgbcolor}$\mathsf{F}_{\tau^{\prime}}(t,T')$\special{ps: grestore}}}}
\end{picture}

%% file: moduli.tex
\section{Stationary phase approximation of trace path integrals}
\label{moduli}

If the Lagrangian admits periodic extremals and is invariant under the action of a Lie group, 
the periodic Sturm-Liouville operator associated to the second variation is singular. 
The quadratic approximation of chapter~\ref{Forman} diverges.
As it is well known in soliton and instanton calculus \cite{BelavinPolyakovSchwartzTyupkin,Polyakov,Polyakov3,Polyakov2,tHooft,tHooft2,Coleman,VainshteinZakharovNovikovShifman,DoreyHollowoodKhozeMattis},
the divergence can be cured using the Faddeev and Popov method \cite{FaddeevPopov,Faddeev}.
Gutzwiller's trace formula follows from a direct application of the method
both for Abelian or non Abelian Lie symmetries.
The expressions of the degenerate trace formulae derived 
by Creagh and Littlejohn \cite{Creagh,CreaghLittlejohn} 
with traditional WKB techniques are recovered in the more general path integral formalism.

\subsection{Trace path integrals and zero modes}
\label{moduli:finite}

In chapter~\ref{Forman} the semiclassical approximation of path integrals
was derived under the hypothesis of a non singular fluctuation operator.
If the action is invariant under time translations, the trace path integral
\begin{eqnarray}
\Tr K=\int_{\lmf}\mes[\sqrt{g}q(t)]e^{\frac{i}{\hbar}\ac}
\label{moduli-finite:trace}
\end{eqnarray}
violates the assumption. A time translation maps a solution of 
the classical equations of the motion into another solution. Thus, the set of 
trajectories on a periodic orbit defines a degenerate extremal of the action. 
The infinitesimal generator of time translations is the time 
derivative $d/dt$. The induced vector field along a $T$-periodic trajectory $q_{\cl}(t)$ 
is the velocity $\dot{q}_{\cl}(t)$ which is therefore also a periodic Jacobi field 
for the second variation. 
To see this, it is enough to differentiate the Euler-Lagrange equations 
specifying the periodic trajectory $q_{\cl}(t)$:
\begin{eqnarray}
\frac{d}{d t}\left[\frac{\de \la}{\de q^\alpha(t)}-\frac{d}{dt}\frac{\de \la}
{\de \dot{q}^{\alpha}(t)}\right]_{q_{\cl}(t)}=0
\label{moduli-finite:actionextremum}
\end{eqnarray} 
and observe that the result can be rewritten as
\begin{eqnarray}
(\foP \dot{q}_{\cl})(t)=0
\label{moduli-finite:timezeromode}
\end{eqnarray}
Hence the velocity field is a zero mode of the second variation operator $\foP$ 
around the periodic extremal $q_{\cl}(t)$.\\
N\"other theorem provides a general mechanism to associate zero modes
to continuous symmetries of the action. Let for example 
$G$ be an unbroken non-trivial, compact, connected finite-dimensional group $G$ 
of symmetries of the action constituted by isometries which also leave invariant 
vector and scalar potentials. Furthermore, assume $G$ to be globally parametrised 
by coordinates $(\mo^1,...,\mo^N)$. In such coordinates the origin is supposed to
coincide with the identity transformation. 
Any dynamic extremal $q_{\cl}(t)$ of the path integral action which breaks completely $G$ 
is embedded in an $N$-parameter family of solutions of the 
Euler-Lagrange equations. 
The generic element of the family is obtained from $q_{\cl}(t)$ by the symmetry 
transformation
\begin{eqnarray}
q_{\cl\,[\mo^1,...,\mo^N]}^{\alpha}(t):=\psi^{\alpha}(q_{cl}(t),\mo^1,...,\mo^N)
\label{moduli-moduli:Hamilton}
\end{eqnarray} 
The Jacobi fields are spawned by the vector fields induced along the trajectory 
by the generators $\{\de / \de \mo^i\}_{i=1}^{N}$ of infinitesimal transformations
around the identity:
\begin{eqnarray}
\jmath^{\alpha}_{a}(t)=\lim_{\mathfrak{t}^{a} \downarrow 0}
\frac{ \psi^{\alpha}(q_{\cl}(t),0,...,0,\mathfrak{t}^{a},0,...,0)-q_{\cl}^{\alpha}(t)}
{\mathfrak{t}^{a}} 
\label{moduli-finite:Jacobi}
\end{eqnarray}
The Jacobi fields are $T$-periodic if $q_{\cl}(t)$ is $T$-periodic. 
Thus they are zero modes of $\foP$, the fluctuation operator of 
$q_{cl}(t)$. The velocity field can be as well encompassed by the definition 
(\ref{moduli-finite:Jacobi}). This is done by shifting $t$ to $t+\mo^0$ and by
re-defining the transformation law $\varphi$ as the composition of the group action 
on the trajectory coordinates with the flow solving the Euler-Lagrange equations. 
The set of parameters $(\mo^0,...,\mo^N)$, collectively denoted by $\mo$, are 
the moduli or collective coordinates of the degenerate stationary point. 

The occurrence of zero modes among the quadratic fluctuations is  not a 
path integral peculiarity. The same problem is present already for ordinary integrals. 
A classical example \cite{Zinn} is the ``zero-dimensional'' field theory 
\begin{eqnarray}
\iota(\hbar)=\int_{R^d}d^dQ\,e^{-\frac{U(|Q|)}{\hbar}}\,
\label{moduli-finite:example}
\end{eqnarray}
with action
\begin{eqnarray}
U(|Q|)=U(|\mathsf{R}Q|)
\label{moduli-finite:action}
\end{eqnarray}
invariant under the group of $d$-dimensional rotations $SO(d)$ and growing
to infinity for large value of the radial coordinate $|Q|$.\\ 
In the small $\hbar$ limit the integral can be performed by means of the 
steepest descent method. The first two derivatives of the exponent 
in Cartesian coordinates 
\begin{eqnarray}
&&\frac{\de U}{\de Q^{\alpha}}(|Q|)=\frac{Q_{\alpha}}{|Q|}\frac{d U}{d |Q|}(|Q|)
\nonumber\\
&&\frac{\de^{2} U}{\de Q^{\alpha}\de Q^{\beta}}(|Q|)=
\frac{Q_{\alpha} Q_{\beta}}{|Q|^2}\frac{d^{2} U}{d^{2} |Q|}(|Q|)+
\frac{1}{|Q|}\left[\delta_{\alpha\,\beta}-\frac{Q_{\alpha} Q_{\beta}}{|Q|^2}\right]
\frac{d U}{d |Q|}(|Q|)
\label{moduli-finite:extremum}
\end{eqnarray}
show that quadratic fluctuations around the any extremum $Q^{\star}$ different from zero 
are governed by the projector 
\begin{eqnarray}
\frac{\de^{2} U}{\de Q^{\alpha}\de Q^{\beta}}(|Q^{\star}|)=
\frac{Q_{\alpha}^{\star}\,Q_{\beta}^{\star}}{|Q^{\star}|^2}\frac{d^{2} U}{d^{2} |Q|}(|Q^{\star}|)
\label{moduli-finite:projector}
\end{eqnarray}
The projector has $d-1$ zero eigenvalues. The number of zero modes is equal to 
the dimension of the quotient $SO(d)/SO(d-1)$ between the original group of 
symmetry and the subgroup $SO(d-1)$ of transformations leaving an extremum 
invariant. 
The phenomenon is general and well known in field theory as it provides the mechanism
for spontaneous symmetry breaking. The subgroup $G_s$ of a symmetry group $G$ leaving 
invariant a ``vacuum state'' is then called the {\em stationary group} or 
{\em stabilizer} of the vacuum. The dimension of the quotient $\mathcal{V}=G/G_s$ 
yields the number of Goldstone modes, emerging from the symmetry breaking 
\cite{Goldstone,GoldstoneSalamWeinberg,Zinn}.\\
The breakdown of the quadratic approximation is due to the violation of the original 
symmetry of the action by its extremal point. 
The problem is obviated by the use of the symmetry to eliminate the ignorable
degrees of freedom from the approximation through a suitable change of variables. 
In the example (\ref{moduli-finite:example}) this is simply done by turning to 
spherical coordinates.
The steepest descent - stationary phase approximation is then performed on the reduced 
action while the ignorable degrees of freedom are integrated non-perturbatively. 

The same method can be applied to path integrals. For example 
in ref.~\cite{RajaramanWeinberg} the Edwards and Gulyaev 
construction (appendix~\ref{changeofvar}) was used to write quantum mechanical 
and field theoretic trace path integrals in spherical coordinates. 
The energy density of systems governed by Lagrangians invariant under rotations
was then evaluated by applying the semiclassical approximation to the radial 
coordinate. The original symmetry is thus always restored in the final expression of the 
steepest descent approximation for all systems with a finite number of degrees of freedom.
In other words, no spontaneous symmetry breaking can occur in quantum mechanics 
\cite{Parisi}.

\subsection{Faddeev and Popov method and moduli space}
\label{moduli:moduli}

Performing as in ref.~\cite{RajaramanWeinberg} the removal of ignorable coordinates 
by change of variables in the path integral may turn to be 
unwieldy. A viable alternative is provided by the Faddeev and Popov method
\cite{FaddeevPopov} originally devised for the quantisation of gauge field theories. 
In order to streamline the discussion, the hypothesis is made that the dynamic 
extremals of the path-integral break completely the $G\times[0,T]$ symmetry 
of the trace path integral.\\
Symmetry transformations acting on a path in the loop space $\lmf$
define an equivalence class of configurations or group orbit which leaves 
the action functional invariant. A periodic degenerate stationary point of 
the action represents an equivalence class of periodic classical trajectories. 
Hence the path integral is expected to be proportional to the volume of the
group orbit. For a compact group $G\times[0,T]$ such volume is finite and it is
natural to extract its contribution before proceeding to the semiclassical approximation.
In order to do so, one recipe is to restrict the path integral to an ``hyper-surface''
in $\lmf$ which intersects any group orbit only once. This means that if 
\begin{eqnarray}
\mathcal{F}_{a}[q(t)]=0\,,\qquad\qquad \forall\,a=0,...,N
\label{moduli-moduli:condition}
\end{eqnarray}
is the equation specifying the hyper-surface, then the equation
\begin{eqnarray}
\mathcal{F}_{a}[\psi(q(t),\mo)]=0\,,\qquad\qquad \forall\,a=0,...,N
\label{moduli-moduli:unique}
\end{eqnarray}
has only one solution in $G\times[0,T]$. The path integral constrained to the
hyper-surface ranges then over $\lmf/(G\times[0,T])$ and does not contain 
any longer the zero modes produced by the symmetry.\\
The decomposition of any closed path in $[0,T]$ into a classical $T$-periodic 
trajectory and quantum fluctuation is accomplished by setting
\begin{eqnarray}
q^{\alpha}(t)=q^{\alpha}_{\cl\,[\mo]}(t)+\sqrt{\hbar}\,\flu^{\alpha}_{[\mo]}(t)
\label{moduli-moduli:decomposition}
\end{eqnarray}
The quantum path on the left hand side of (\ref{moduli-moduli:decomposition}) is
by definition independent of the moduli $\mo$.
The simplest choice of the constraint $\mathcal{F}$ is then
\begin{eqnarray}
\mathcal{F}_{a}[q(t)]:=\int_{0}^{T}dt\,[q^{\alpha}(t)-q^{\alpha}_{\cl\,[\mo]}(t)]
\frac{\de q_{\cl\,[\mo] \, \alpha}}{\de \mo^a}(t)
\,,\qquad\qquad \forall\,a=0,...,N
\label{moduli-moduli:constraint}
\end{eqnarray}
Applying Lie's first fundamental theorem (appendix~\ref{Lie}) the 
derivatives of the classical trajectory with respect to the moduli are expressed
in terms of the Jacobi fields (\ref{moduli-finite:Jacobi}) induced by the infinitesimal 
generators of the group transformations:
\begin{eqnarray}
\frac{\de q^{\alpha}_{\cl\,[\mo]}}{\de \mo^a}(t)=
\mathfrak{R}^{b}_{a}(\mo)\,\frac{\de q^{\alpha}_{\cl\,[\mo]}}{\de q_{\cl}^{\beta}}(t)
\,\jmath^{\beta}_{b}(t)
\label{moduli-moduli:Lie}
\end{eqnarray}
The matrix $\mathfrak{R}^{a}_{b}(\mo)$ depends only on the moduli 
and coincides with the identity when $\mo$ is equal to zero. 
The constraint (\ref{moduli-moduli:constraint}) is inserted in 
the path integral in the guise of the Faddeev-Popov expression of the unity
\begin{eqnarray}
1\equiv\int_{G\times[0,T]}\prod_{i=0}^{N}d\mo^{i}\left|\mathrm{det}
\left\{\frac{\de \mathcal{F}_{a}}{\de \mo^{b}}\right\}\right|\,\prod_{j=0}^{N}
\delta(\mathcal{F}_{j}[q(t)])
\label{moduli-moduli:unity}
\end{eqnarray}
The matrix elements of the Jacobian,
\begin{eqnarray}
\frac{\de \mathcal{F}_{a}}{\de \mo^{b}}=\int_{0}^{T}dt\,
\left\{[q^{\alpha}(t)-q^{\alpha}_{\cl\,[\mo]}(t)]
\frac{\de^2 q_{\cl\,[\mo] \, \alpha}}{\de^2 \mo^a}(t)-
\frac{\de q^{\alpha}_{\cl\,[\mo]}}{\de \mo^{b}}(t)
\frac{\de q_{\cl\,[\mo] \, \alpha}}{\de \mo^a}(t)\right\}
\label{moduli-moduli:determinant}
\end{eqnarray}
are computed on the intersection of the group orbit with the manifold 
(\ref{moduli-moduli:condition}). 
Once (\ref{moduli-moduli:unity}) has been inserted in the path integral the order of integration
between the moduli and the path integral measure is exchanged. Quantum fluctuations are then 
written as the $\si$-series 
\begin{eqnarray}
\flu^{\alpha}_{[\mo]}(t)=
\sum_{n} c_{n} \,\lambda_{n}^{[\psi]\,\alpha}(t)
\label{moduli-moduli:fluctuations}
\end{eqnarray}
extended over the eigenstates of the fluctuation operator $\foP^{[\psi]}$ around the 
classical trajectory $q^{\alpha}_{\cl\,[\mo]}(t)$. The geometrical meaning of 
the constrain (\ref{moduli-moduli:constraint}) becomes manifest: it restricts the span 
of quantum fluctuations to ``massive'' eigenstates, orthogonal with respect to the 
$\si$-scalar product to the nullspace of $\foP^{[\psi]}$.  
Actually, an orthonormal basis for this latter is composed by vector fields of the
form
\begin{eqnarray}
\lambda_{\mathrm{z.m.}\,,a}^{[\psi]\,\alpha}(t)=
(\mathsf{Z}^{-1})^{b}_{a}\,\frac{\de q^{\alpha}_{\cl\,[\mo]}}{\de q_{\cl}^{\beta}}(t)
\,\jmath^{\beta}_{b}(t)\,,\qquad\qquad a=0,...,N
\label{moduli-moduli:zeromodes}
\end{eqnarray}
as it follows from the use of the Gram-Schimdt recursion method.
Since the symmetry is an isometry for the metric,
\begin{eqnarray}
g_{\alpha\,\beta}(q)=\frac{\de q^{\gamma}_{[\mo]}}{\de q^{\alpha}}
\frac{\de q^{\delta}_{[\mo]}}{\de q^{\beta}}g_{\gamma\,\delta}(\psi(q,\mo))
\label{moduli-moduli:isometry}
\end{eqnarray}
it is enough to compute the triangular matrix $\mathsf{Z}^{b}_{a}$ 
around the identity transformation using the Gram-Schmidt recursion 
relations:
\begin{eqnarray}
&&\lambda_{\mathrm{z.m.}\,,0}^{\alpha}(t)=\frac{\jmath_{0}^{\alpha}(t)}
{\sqrt{\langle\jmath_{0}\,,\jmath_{0} \rangle}}
\nonumber\\
&&\lambda_{\mathrm{z.m.}\,,1}^{\alpha}(t)=\frac{\jmath_{1}^{\alpha}(t)
-\frac{\langle\jmath_{0} \,,\jmath_{1} \rangle}
{\langle \jmath_{0}\,,\jmath_{0} \rangle}
\jmath_{0}^{\alpha}(t)}{\sqrt{\langle \jmath_{1}\,,\jmath_{1} \rangle-
\frac{\langle \jmath_{0}\,,\jmath_{1} \rangle^2}{
\langle \jmath_{0}\,,\jmath_{0} \rangle}}}
\nonumber\\
&&\mathrm{etc.}
\label{moduli-moduli:GramSchmidt}
\end{eqnarray} 
The path integral measure was argued in chapter~\ref{Forman} to be proportional to the
infinite product of integrals over the amplitudes $c_{n}$ of the $\si$-expansion 
of the quantum paths. Therefore, carried under path integral sign 
(\ref{moduli-moduli:unity}) becomes
\begin{eqnarray}
&&\int_{G\times[0,T]}\prod_{i=0}^{N}d\mo^{i}\left|\mathrm{det}
\left\{\frac{\de \mathcal{F}_{a}}{\de \mo^{b}}\right\}\right|\,\prod_{j=0}^{N}
\delta(\mathcal{F}_{j}[q(t)])=
\nonumber\\
&&\int_{G\times[0,T]}\prod_{i=0}^{N}d\mo^{i}
\left|\det\mathsf{Z}\det\mathfrak{R}\right|^{2}\,\prod_{j=0}^{N}
\delta(\mathfrak{R}_{j}^{a}\,\mathsf{Z}_{a}^{b}\,c_{\mathrm{z.m.}\,,b})+o(\hbar)
\end{eqnarray}
with $a,b=0,...,N$.
The $\delta$-functions remove the divergences from the semiclassical approximation:
\begin{eqnarray}
&&\Tr K=
\nonumber\\
&&\sum_{o\in \mathrm{p.o.}}e^{i \frac{\ac_{\cl}^{o}}{\hbar}}\!
\int_{G\times[0,T]}\!\!\!\!\!\!\!\!\!\!\!\!d\mo^0dG\!
\int_{\mathcal{T}\lmf}\!\!\!\!\mes(\prod_n dc_n)
\,\det\mathsf{Z}^{(o)}\!\!\!\!\!\!\prod_{a\in\nul\foP}\!\!\!\!\!\!\delta(c_a)\,
e^{i \frac{\qac^{(o)}}{2}}\!+o(\sqrt{\hbar})
\label{moduli-moduli:insertion}
\end{eqnarray}
The sum over $o$ ranges over all the distinct degenerate stationary points 
of the action. The measure $dG$ 
\begin{eqnarray}
dG=\prod_{a=1}^{N}d\mo_a \det\mathfrak{R}(\mo^1,...,\mo^N) 
\label{moduli-moduli:invariant}
\end{eqnarray}
is invariant over the compact group $G$ (\cite{Bernard} and appendix~\ref{Lie}). 

The path integral (\ref{moduli-moduli:insertion}) yields
\begin{eqnarray}
\Tr K\cong\!
\sum_{o \in \mathrm{p.o.}}\!
e^{i\,\frac{\ac_{\cl}^{(o)}}{\hbar}}\!\!
\int_{G\times[0,T]}\!\!\!\!\!\!\!\!\!\!d\mo^0\,dG\,
\frac{e^{-i\,\frac{\pi}{2}\,\left[\ind \foP^{(o)}([0,T])+\frac{N+1}{2}\right]}}
{(2\,\pi\,\hbar)^{\frac{N+1}{2}}
\left|\frac{\mathrm{Det}_{\perp}\foP^{(o)}([0,T])}{\det\mathsf{Z}^{(o)\,2}}\right|^{\frac{1}{2}}}
\label{moduli-moduli:trace}
\end{eqnarray}
$\mathrm{Det}_{\perp}$ is the functional determinant produced by the restriction 
to massive modes.
Symmetry transformations are by (\ref{moduli-moduli:isometry}) isometries also for the 
$\si$-scalar product. The integrand in (\ref{moduli-moduli:trace}) is therefore 
function only of the degenerate stationary point of the action. One can rewrite
\begin{eqnarray}
\Tr K\cong
\sum_{o \in \mathrm{p.o.}}\frac{T\,|G|\,e^{i\,\frac{\ac_{\cl}^{(o)}}{\hbar}-
i\,\frac{\pi}{2}\,\left[\ind \foP^{(o)}([0,T])+\frac{N+1}{2}\right]}}
{(2\,\pi\,\hbar)^{\frac{N+1}{2}}
\left|\frac{\mathrm{Det}_{\perp}\foP^{(o)}([0,T])}{\det\mathsf{Z}^{(o)\,2}}\right|^{\frac{1}{2}}}
\label{moduli-moduli:trace2}
\end{eqnarray}
with
\begin{eqnarray}
|G|:=\int_{G}dG
\label{moduli-moduli:volume}
\end{eqnarray}
The factor $(2\,\pi\,\hbar\,e^{i\,\frac{\pi}{2}})^{\frac{N+1}{2}}$ in the denominator
of (\ref{moduli-moduli:trace2}) comes from the definition of the path integral measure
(compare with  appendix~\ref{formulae}).
Namely the integrals over the amplitudes of the $\si$ expansion of the quantum paths
are normalised to $(2\,\pi\,\hbar\,e^{i\,\frac{\pi}{2}})^{\frac{1}{2}}$.\\
In the following section it will be shown how the use of Forman's theorem recasts the
last formula in terms of invariant quantities of the classical periodic orbits.

An historical remark before closing the section.
The Faddeev-Popov method was applied to the treatment of constrained
quantum mechanical system by Faddeev himself in ref.~\cite{Faddeev} using phase-space 
path integrals. However, in ref.~\cite{Polyakov3}, Polyakov credits Bogoliubov and Tyablikov 
\cite{BogoliubovTyablikov} to have been the first ones to treat non Gaussian fluctuations 
emerging in the steepest descent - stationary phase approximation with the method expounded above.\\
The Faddeev-Popov method has found wide application for the quantisation of quantum, 
statistical mechanical and field theories in a soliton or instanton background. 
Classical references are 
\cite{GervaisJevickiSakita,GervaisSakita,ChristLee,BelavinPolyakovSchwartzTyupkin,Polyakov,Polyakov3,tHooft,tHooft2,Coleman,VainshteinZakharovNovikovShifman}. A recent review is \cite{DoreyHollowoodKhozeMattis}.
The treatment of collective coordinates in problems with dynamic extremals only 
partially breaking the symmetry group has been analysed in \cite{GervaisNeveuVirasoro}. 
The application to scattering problems is discussed 
in \cite{VasilevKuzmenko}. \\ 
The consistency of the Faddeev-Popov method with a systematic asymptotic perturbation theory 
beyond leading order has been checked for example in references 
\cite{AleinikovShuryak,WoehlerShuryak,AntonovKorepin}. The investigation
of higher order corrections is sometimes simplified if the Faddeev-Popov method is 
implemented by means of ghost fields and the imposition of BRST conditions \cite{BecchiRouetStoraTyutin}.
The procedure is reviewed in \cite{BesKurchan,GarrahanKruczenskiBes,Zinn}.

\subsection{Zero mode subtraction}
\label{moduli:transversal}

The explicit expression of the functional determinant of the massive modes 
can be computed adopting the same strategy pursued in the evaluation of the Morse 
index of Sturm-Liouville operators with zero modes. 
A positive definite $T$-periodic generic infinitesimal perturbation $\tau \dU$ 
is introduced to subtract the zero modes. Forman's theorem can be applied directly to the 
non-singular operator $\foPtau([0,T])$. Finally, the functional determinant of massive modes
is recovered by taking the limit
\begin{eqnarray}
|\mathrm{Det}^{\perp}\foP([0,T])|=\lim_{\tau\downarrow0}\left|
\frac{\det[\id_{2\,d}-\fu_{\tau}(T,0)]}{\prod_{n=0}^{N}\ellntau}\right|
\label{moduli-transversal:subtraction}
\end{eqnarray}
The product in the denominator runs over the eigenvalues of $\foPtau([0,T])$ flowing
into zero modes at $\tau$ equal zero. In order to evaluate (\ref{moduli-transversal:subtraction})
it is enough to compute the eigenvalue product up to first order in $\tau$: 
\begin{eqnarray}
\prod_{n=0}^{N}\ellntau&=&
\tau^{N+1}\,\frac{
\det \{\langle \jmath_a\,,\mathsf{U}\,\jmath_b \rangle\}}
{\det\{\langle \jmath_a\,,\jmath_b \rangle\}}
+o(\tau^{N+1})
\label{moduli-transversal:eigenvalues}
\end{eqnarray}
The denominator cancels out with the Gram-Schmidt Jacobian $\det\mathsf{Z}$. 
The cancellation is not accidental since the functional determinant restricted 
to massive modes by definition cannot depend on zero modes. 
For the same reason the averages of the perturbation over the periodic 
Jacobi fields should cancel out. The mechanism of the cancellation is governed by 
the parabolic blocks of the monodromy matrix of the periodic orbit. The monodromy
matrix is obtained from the linearisation of the classical flow around any trajectory 
$q_{\cl}^{\alpha}(t)$ on the periodic orbit by setting
\begin{eqnarray}
\mon=\fu(T,0)
\label{moduli-transversal:monodromy}
\end{eqnarray}
The monodromy matrix admits an orthogonal decomposition 
\begin{eqnarray}
\mon\,=\,\mon^{\parallel}\,\oplus\,\mon^{\perp}
\label{moduli-transversal:decomposition}
\end{eqnarray}
with $\mon^{\parallel}$, $\mon^{\perp}$ governing respectively the
stability of degrees of freedom longitudinal or transversal to the orbit 
in phase space. The decomposition corresponds to a partial reduction to normal form
of the monodromy. The continuity of the eigenvalue flow of a symplectic matrix under
continuous parametric perturbation permits to write at $\tau$ different from zero
\begin{eqnarray}
\mon_{\tau}=\mon^{\parallel}_{\tau}\,\oplus\,\mon^{\perp}_{\tau}
\label{moduli-transversal:taudecomposition}
\end{eqnarray}
The perturbation shifts the stability of the blocks away from the marginal case. 
Hence, the block $\mon^{\parallel}_{\tau}$ pairs up to degrees of freedom longitudinal 
to the orbit only at $\tau$ equal zero. 
The functional determinant factorises as
\begin{eqnarray}
\det[\id_{2\,d}-\mon_{\tau}]= \det[\id_{l}-\mon^{\parallel}_{\tau}]
\det[\id_{2\,d-l}-\mon^{\perp}_{\tau}]
\label{moduli-transversal:determinant}
\end{eqnarray}
with $l\,\leq\,2\,d$ the dimension of the parabolic subspace of phase space.
The limit
\begin{eqnarray}
\lim_{\tau\rightarrow0}\det[\id_{2\,d-l}-\mon^{\perp}_{\tau}]=
\det[\id_{2\,d-l}-\mon^{\perp}]:=\mathrm{det}_{\perp}[\id_{2 d}-\mon]
\label{moduli-transversal:finitedeterminant}
\end{eqnarray}
is non zero and independent from the moduli. Linearised flows along different 
trajectories belonging to the same degenerate extremum are related by  
similarity transformations which leave $\det_{\perp}[\id_{2 d}-\mon]$ 
invariant.

\subsubsection{Abelian symmetries}
\label{moduli-transversal:abelian}

The symmetry group is Abelian if the generating functions of the symmetry transformations
are in involution (appendix~\ref{Hamilton:symmetries}):
\begin{eqnarray}
\{\ha_{a},\ha_{b}\}_{\mathrm{P.b.}}=0\,,\qquad\qquad 0\,\leq\, a,b\,\leq\, N\, <\,d
\label{moduli-transversal-abelian:involution}
\end{eqnarray}
The Poisson brackets (\ref{moduli-transversal-abelian:involution}) are equivalent to 
skew-orthogonality relations between the phase space lifts
\begin{eqnarray}
\mathcal{J}_a(t)=\left[\begin{array}{c}\jmath_a(t)\\ (\nabla\jmath_a)(t)\end{array}\right]
\label{moduli-transversal-abelian:lift}
\end{eqnarray}
of the $N+1$ periodic Jacobi fields. By assumption the Jacobi fields are linearly independent. 
Hence $N+1$ is at most equal to $d$. The construction of a symplectic basis 
requires the introduction of other $N+1$ linear independent generalised 
eigenvectors of the monodromy matrix
\begin{eqnarray}
\begin{array}{l}
\mon^{\parallel\,\alpha}_{\,\,\,\beta}\,
\mathcal{J}^{\beta}_{a}(t)=
\mathcal{J}^{\alpha}_{a}(t)
\\
\mon^{\parallel\,\alpha}_{\,\,\,\beta}\,\mathcal{X}^{\alpha}_{a}(t)=
\mathcal{X}^{\alpha}_{a}(t)+\mathsf{V}^{\alpha}_{\beta}\,
\mathcal{J}^{\beta}_{a}(t)
\end{array}\,\qquad\qquad a=0,..,N
\label{moduli-transversal-abelian:diagonalisation}
\end{eqnarray}
fulfilling the normalisation conditions
\begin{eqnarray}
\mathcal{X}^{\alpha}_{a}\,\sym_{\alpha\,\beta}\mathcal{J}^{\beta}_{b}=\delta_{a\,b}\,,\qquad\qquad a,b=0,..,N
\label{moduli-transversal-abelian:normalisation}
\end{eqnarray}
The generalised eigenvectors are mapped by the symplectic matrix $\sym$ 
into a basis dual to the $\mathcal{J}_a$'s.
By means of a linear symplectic transformation it is also possible to reduce 
$\mon^{\parallel}$ to the orthogonal product of $N+1$ two-dimensional parabolic blocks 
of the form given in appendix~\ref{Hamilton:periodic}. Otherwise phrased, 
involution renders the normalisation conditions (\ref{moduli-transversal-abelian:normalisation}) 
compatible with the existence of a reference frame where each of the configuration space 
projections of the periodic eigenvectors of $\mon$ coincides with one of the 
$\si$-{\em orthonormal} zero modes of $\foP$.
Introducing an infinitesimal periodic perturbation, the same calculation of 
appendix~\ref{Hamilton:perturbative} yields
\begin{eqnarray}
\mon^{\parallel}_{\tau}=
\left[\begin{array}{cc}\id_{N+1} \quad&\quad \mathsf{V} \\ \tau^{N+1} 
\{\langle \jmath_i\,,\mathsf{U}\,\jmath_j \rangle \} 
\quad&\quad \id_{N+1}
\end{array}\right]+O(\tau^{N+1})
\label{moduli-transversal-abelian:blocks}
\end{eqnarray}
The terms of order $O(\tau^{N+1})$ or higher neglected in (\ref{moduli-transversal-abelian:blocks})
do not contribute to the limit 
\begin{eqnarray}
\lim_{\tau\downarrow 0}\left|\frac{\det[\id_{2\,(N+1)}-\mon^{\parallel}_{\tau}]}{\tau^{N+1}\,
\det \{\langle \jmath_i\,,\mathsf{U}\,\jmath_j \rangle\}}
\right|
=|\det \mathsf{V}|
\label{moduli-transversal-abelian:limit}
\end{eqnarray}
The determinant on the right hand side is just the product of the non-diagonal elements
of the normal forms of the parabolic blocks of the monodromy matrix represented in the
canonical symplectic basis 
\begin{eqnarray}
|\det \mathsf{V}|=|\prod_{i=0}^{N}\kappa_i|
\label{moduli-transversal-abelian:nondiagonal}
\end{eqnarray}  
Physically $\kappa_i$'s are the characteristic growth rates of the linear independent marginal 
instabilities of the degenerate periodic orbit. The functional determinant is therefore independent on
the moduli. Provided all the orbits contributing to the stationary phase are prime, 
the semiclassical trace path integral with Abelian symmetries is within leading order
\begin{eqnarray}
\Tr K\cong
\sum_{o \in \mathrm{p.p.o.}}
\frac{T |G|\,e^{i\,\frac{\ac_{\cl}^{(o)}}{\hbar}-\frac{i\,\pi}{2}\,\left(
\ind \foP^{(o)}([0,T])+\frac{N+1}{2}\right)}}
{(2\,\pi\,\hbar)^{\frac{N+1}{2}}\,
\sqrt{|\,\det\mathsf{V_{o}}\,\det_{\perp}[\id_{2\,d}-\mon_{o}]\,|}}
\label{moduli-transversal-abelian:trace}
\end{eqnarray}
Were the stationary phase on a time interval $r\,T$ with $r$ integer 
dominated only by the iteration of prime orbits, the above formula would admit 
the extension:
\begin{eqnarray}
\Tr K\cong
\sum_{o\in \mathrm{p.p.o.}}\sum_{r=1}^{\infty}
\frac{T |G| e^{i \frac{r \ac_{\cl}^{(o)}}{\hbar}-\frac{i \pi}{2}\left(
\ind \foP^{(o)}([0,r\,T])+\frac{N+1}{2}\right)}}
{(2 \pi \hbar\, r)^{\frac{N+1}{2}}
\sqrt{|\det\mathsf{V_{o}}\,\det_{\perp}[\id_{2 d}-\mon^r_o]|}}
\label{moduli-transversal-abelian:iterate}
\end{eqnarray}
since
\begin{eqnarray}
\mon(r\,T)=\mon^r(T)
\label{moduli-transversal-abelian:monodromyiterate}
\end{eqnarray}

\subsubsection{Non Abelian symmetries}
\label{moduli-transversal:nonabelian}

A symmetry group is non-Abelian if the generating functions $\{\ha_a\}_{a=1}^{N}$ 
of the group transformations form a non trivial Poisson brackets algebra 
(appendix~\ref{Hamilton})
\begin{eqnarray}
\{\ha_a\,,\ha_b\}_{\mathrm{P.b.}}\,=\,-\,C_{a\,b}^{c}\,\ha_c+D_{a\,b}\,,\qquad\qquad 
a,b=1,...,N\,<\, 2\,d
\label{moduli-transversal-nonabelian:Poissonalgebra}\end{eqnarray}
In consequence, the skew products among the right periodic eigenvectors of the 
monodromy matrix
\begin{eqnarray}
\mathcal{J}^{\alpha}_{a}=
\sym^{\dagger\,\alpha\,\beta}\,\frac{\de \ha_a}{\de x^{\beta}}\,,\qquad\qquad a=0,...,N
\label{moduli-transversal-nonabelian:vectorfields}
\end{eqnarray}
with $\ha_0$ coinciding with the Hamiltonian $\ha$, are now fixed
by the Lie algebra of the group:
\begin{eqnarray}
\mathcal{J}^{\alpha}_{a}\,\sym_{\alpha\,\beta}\,\mathcal{J}^{\beta}_{b}=-
\{\ha_a\,,\ha_b\}_{\mathrm{P.b.}}
\label{moduli-transversal-nonabelian:skeworthogonal}
\end{eqnarray}
For any trajectory on a generic orbit of period $T$, the zero eigenvalues of the 
resulting $N+1\,\times\,N+1$ antisymmetric matrix correspond to the existence 
of Casimir operators in $G\times[0,T]$. The Hamiltonian $\ha$ furnishes an obvious 
example of a generating function commuting with all the elements of the Poisson 
algebra.\\
Let $k+1$ be the number of eigenvectors skew orthogonal mutually and to the remaining 
$N-k$. As in the Abelian case the diagonalisation of the block monodromy matrix 
spanned by such eigendirections requires the introduction of $k+1$ skew-orthonormal 
generalised eigenvectors 
\begin{eqnarray}
\begin{array}{l}
\mon^{\alpha}_{\beta}\,\mathcal{X}^{\beta}_{a}=\mathcal{X}^{\alpha}_{a}-
\kappa_{(a)} \mathcal{J}^{\alpha}_{a}
\\
\mathcal{X}^{\alpha}_{a}\sym_{\alpha\,\beta}\,\mathcal{J}^{\beta}_{b}=\delta_{a\,b}
\end{array}\,,\qquad\qquad a=0,...,k\,\,,\quad b=0,...,N
\label{moduli-transversal-nonabelian:generalised}
\end{eqnarray}
The remaining $N-k$ eigenvectors have non-degenerate skew products. From their
linear combinations it is possible to form a symplectic basis for the linear 
subspace they span. 
The limit
\begin{eqnarray}
\lim_{\tau\downarrow 0}\left|\frac{\det[\id_{N+k+2}-\mon^{\parallel}_{\tau}]}{\tau^{N+1}\,
\det \{\langle \jmath_i\,,\mathsf{U}\,\jmath_j \rangle\}}
\right|=|\det \mathsf{W}\,\det \mathsf{V}|
\label{moduli-transversal-nonabelian:limit}
\end{eqnarray}
yields two contributions. The determinant $\det \mathsf{V}$ retains the same 
definition as in the Abelian case. The new term $\det \mathsf{W}$ accounts
for the different normalisation, dictated by (\ref{moduli-transversal-nonabelian:skeworthogonal}), 
of the configuration space projections of the periodic eigenvectors of the monodromy matrix 
from the corresponding Jacobi fields describing the zero modes of $\foP$. 
In consequence, $\det \mathsf{W}$ coincides with absolute value 
of the inverse product of the non-zero eigenvalues of the antisymmetric matrix 
specified by (\ref{moduli-transversal-nonabelian:skeworthogonal}).
If only prime periodic orbits contribute, the leading order result is
\begin{eqnarray}
\Tr K\cong
\sum_{o \in \mathrm{p.p.o.}}
\frac{T\,|G|\,e^{i\,\frac{\ac_{\cl}^{(o)}}{\hbar}-\frac{i\,\pi}{2}\,
\left(\ind \foP^{(o)}([0,T])+\frac{N+1}{2}\right)}}
{(2 \pi \hbar)^{\frac{N+1}{2}}\sqrt{|\det\mathsf{W}_{o}\det\mathsf{V}_{o}\det_{\perp}[\id_{2 d}
-\mon_{o}]|}}
\label{moduli-transversal-nonabelian:trace}
\end{eqnarray}
The result is illustrated by an example common in applications.

Consider an action functional with time autonomous Lagrangian invariant 
under $SO(3)$. For the sake of simplicity in what follows the Darboux variable 
$x=(q,p)$ describes a six dimensional phase space. The hypothesis is non restrictive 
as it corresponds to the choice of a frame of coordinates such that $x$ describes
the parabolic degrees of freedom in the monodromy matrix.
The first integrals associated to the symmetry are the Hamiltonian and the three 
components $\mathcal{M}_{a}$, $a=1,...,3$ of the angular momentum with Poisson algebra  
\begin{eqnarray}
\{\mathcal{M}_{a},\mathcal{M}_{b}\}_{\mathrm{P.b.}}=\varepsilon_{a\,b}^{c}\,\mathcal{M}_{c}\,,
\qquad\qquad a,b,c=1,...,3 
\label{moduli-transversal-nonabelian:Poisson}
\end{eqnarray}
with $\varepsilon_{a\,b\,c}$ the completely antisymmetric tensor.
Let $\mathsf{R}_a$, ($a=1,...,3$) denote a rotation around the $a$-th Cartesian axis.
A canonical parametrisation \cite{SattingerWeaver} of $SO(3)$ is provided by the 
angular momentum versor 
\begin{eqnarray}
\frac{(\mathcal{M}_1,\mathcal{M}_2,\mathcal{M}_3)^{\dagger}}
{||\mathcal{M}||}&=&\mathsf{R}_3(\mo^1)\,\mathsf{R}_3(\mo^2)\,(0,0,1)^{\dagger}
\nonumber\\
&:=&(\cos \mo^1 \sin \mo^2,\sin \mo^1 \sin \mo^2,\cos \mo^2)^{\dagger}
\label{moduli-transversal-nonabelian:versor}
\end{eqnarray}
and by the angle $\mo^{3}$ of rotation around the direction of the angular momentum.
In terms of the moduli $(\mo^1,\mo^2,\mo^3)$, the generic element of $SO(3)$ admits 
the representation
\begin{eqnarray}
\mathsf{R}(\mo^1,\mo^2,\mo^3)=
\mathsf{R}_3(\mo^1)\mathsf{R}_2(\mo^2)\mathsf{R}_3(\mo^3)
[\mathsf{R}_3(\mo^1)\mathsf{R}_2(\mo^2)]^{\dagger}
\label{moduli-transversal-nonabelian:so3}
\end{eqnarray}
with
\begin{eqnarray}
0\,\leq\,\mo^1\,,\mo^3\,\leq\,2\,\pi\,,\qquad\qquad 0\,\leq\,\mo^2\,<\,\pi
\label{moduli-transversal-nonabelian:ranges}
\end{eqnarray}
By (\ref{moduli-moduli:Lie}) the Jacobian entering the definition
of the invariant measure can be extracted from the left invariant differential
over $SO(3)$
\begin{eqnarray}
(\mathsf{R}^{-1}d\mathsf{R})(\mo^1,\mo^2,\mo^3)=d\mo^{a}\mathfrak{R}_{a}^{b}(\mo^1,\mo^2,\mo^3)
\mathsf{r}_{b}
\label{moduli-transversal-nonabelian:leftinvariant}
\end{eqnarray}
The differential is written in matrix notation in the basis of the infinitesimal 
generators of rotations around the three coordinate axes
\begin{eqnarray}
\mathsf{r}_{a}:=
\left.\frac{d \mathsf{R}_a}{d \mathfrak{s}}(\mathfrak{s})\right|_{\mathfrak{s}=0}
\label{moduli-transversal-nonabelian:generators}
\end{eqnarray}
The invariant measure reads
\begin{eqnarray}
dSO(3)=\prod_{a=1}^{3}d\mo^{(a)} \det\mathfrak{R}(\mo^1,\mo^2,\mo^3)=
\prod_{a=1}^{3}d\mo^{(a)} \,4\,\sin\mo^{(2)}\,\sin^2\frac{\mo^{(1)}}{2}\,
\label{moduli-transversal-nonabelian:measure}
\end{eqnarray}
The result does not depend on the use of the left-invariant differential 
(\ref{moduli-transversal-nonabelian:leftinvariant}) to construct the measure since $SO(3)$
is compact and connected: the invariant measure is therefore unique 
\cite{Gilmore,SattingerWeaver}. \\
The squared modulo of the angular momentum defines the generating function associated
to the Casimir operator of $SO(3)$. It is therefore convenient to choose
the periodic eigenvectors of the monodromy matrix in the guise
\begin{eqnarray}
&&\mathcal{J}^{\alpha}_{0}=\sym^{\dagger\,\alpha\,\beta}\,\frac{\de \ha}{\de x^{\beta}}
\nonumber\\
&&\mathcal{J}^{\alpha}_{a}=
\sym^{\dagger\,\alpha\,\beta}\hat{n}^{b}_{a}\,
\frac{\de \mathcal{M}_b}{\de x^{\beta}}
\label{moduli-transversal-nonabelian:so3vectorfields}
\end{eqnarray}
with $a,b=1,...,3$. The $\{\hat{n}_{a}\}_{a=1}^{3}$ are time independent orthonormal 
versors in $\mathbb{R}^{3}$ such that
\begin{eqnarray}
&&\hat{n}^{b}_{a}\delta_{b,b'}\hat{n}^{b'}_{a'}=\delta_{a\,a'}
\nonumber\\
&&\hat{n}^{a}_{3}=\frac{\mathcal{M}^{a}}{||\mathcal{M}||}
\label{moduli-transversal-nonabelian:versors}
\end{eqnarray}
The explicit form of the versors, modulo a normalisation factor, can be derived directly 
from the expression (\ref{moduli-moduli:Lie}) of Lie's first fundamental theorem applied to
$SO(3)$. With the choice (\ref{moduli-transversal-nonabelian:so3vectorfields}) the last three 
periodic eigenvectors have projections in configuration space mutually orthogonal with
$\hat{n}_{3}$ always pointing in the direction of the angular momentum.\\ 
The Poisson algebra of the generating functions yields the skew orthogonality relations
\begin{eqnarray}
&&\mathcal{J}^{\alpha}_{0}\,\sym_{\alpha\,\beta}\,\mathcal{J}^{\alpha}_{a}=0\,,\qquad\qquad
\nonumber\\
&&\mathcal{J}^{\alpha}_{a}\,\sym_{\alpha\,\beta}\,\mathcal{J}^{\beta}_{b}=-\mathcal{M}_{c}
\varepsilon^{c}_{d\,e}\hat{n}^{d}_{a}\,\hat{n}^{e}_{b}=-||\mathcal{M}||\,\hat{n}^{c}_{3}
\varepsilon_{c\,d\,e}\hat{n}^{d}_{a}\,\hat{n}^{e}_{b}
\label{moduli-transversal-nonabelian:skewproducts}
\end{eqnarray}
The diagonalisation of the monodromy matrix requires the introduction of two
generalised eigenvectors
\begin{eqnarray}
&&\mon^{\alpha}_{\beta}\,\mathcal{J}^{\beta}_{E}=\mathcal{J}^{\alpha}_{E}-
\frac{d T}{d E}\mathcal{J}^{\alpha}_{0}
\nonumber\\
&&\mon^{\alpha}_{\beta} \,\mathcal{J}^{\beta}_{||\mathcal{M}||}
=\mathcal{J}^{\beta}_{||\mathcal{M}||}+\kappa_{||\mathcal{M}||}\,\mathcal{J}^{\alpha}_{3}
\label{moduli-transversal-nonabelian:so3generalised}
\end{eqnarray}
where $\kappa_{||\mathcal{M}||}$ physically gives the variation of the total angle of rotation
around the angular momentum versus a change of the absolute value of the angular momentum itself.
Finally for any generic periodic orbit described by the system one can apply 
(\ref{moduli-transversal-nonabelian:trace}) with
\begin{eqnarray}
&&|\det \mathsf{W}|=\frac{1}{||\mathcal{M}||^2} 
\nonumber\\
&&|\det \mathsf{V}|=\left|\frac{d T}{d E}\,\kappa_{||\mathcal{M}||}\right|
\label{moduli-transversal-nonabelian:identifications}
\end{eqnarray}

\subsection{Gutzwiller trace formula}
\label{moduli:Gutzwiller}

Gutzwiller's trace formula is a direct application of the general theory 
of path integration over loop spaces.

The relation (\ref{density-qm:density}) between the energy spectrum of a time autonomous 
quantum mechanical system and the propagator is amenable to the path integral 
expression
\begin{eqnarray}
\rho(E)=-\left.
\lim_{\ima\,Z\,\downarrow 0}\,
\ima\,\int_{0}^{\infty}\frac{dT}{i\,\pi\,\hbar} \,e^{i\,\frac{Z\,T}{\hbar}}\,
\int_{\lmf}\mes[\sqrt{g}q(t)]e^{\frac{i}{\hbar}\ac}\right|_{E=\rea\,Z}
\label{moduli-Gutzwiller:density}
\end{eqnarray}
The semiclassical approximation must be performed on the Fourier transform 
of the propagator trace. The integration over the time variable entails the evaluation 
of the propagator for infinitesimally small times $T$. At variance with the finite $T$ case,
in such a limit the semiclassical approximation cannot be identified with the stationary 
phase approximation. The propagator becomes proportional to a $\delta$-Dirac distribution 
for zero time increments:
\begin{eqnarray} 
\feo_z(Q,T|\,Q',0)\sim
\left(\frac{m}{2\,\pi\,i\,\hbar\,T}\right)^{\frac{d}{2}}
e^{-\frac{1}{i\,\hbar}\int_{0}^{T}\!\!\!dt\,\la_z(q_t,\dot{q}_t)}
\label{moduli-Gutzwiller:shorttime}
\end{eqnarray}
The divergence of the prefactor does not allow to neglect it in comparison to the 
fast variation of the phase for $\hbar$ going to zero. Thus, the stationary phase cannot 
be directly applied in this limit. 
In consequence, the semiclassical energy spectrum consists of the separate contribution
coming from short orbits with period tending to zero and from long orbits of finite period $T$.

\subsubsection{Short orbits}
\label{moduli-Gutzwiller:short}

The short orbit contribution to the trace formula was investigated in 
detail long ago in the review by Berry and Mount \cite{BerryMount}. 

The short orbit contribution is obtained starting from the 
Fourier transform of the short time representation of the 
propagator on closed paths 
\begin{eqnarray}
K(Q,T|Q)\sim 
\int_{\mathbb{R}^d}\frac{d^dP}{(2 \pi \hbar)^{d}}\,
e^{-i\,\frac{\ha(P,Q)\,T}{\hbar}}\,,\qquad T\,,\hbar\,\downarrow\,0
\end{eqnarray}
The insertion of the short time approximation into the energy density 
(\ref{moduli-Gutzwiller:density}) provides in an Euclidean space the 
asymptotic expression
\begin{eqnarray}
\rho_{\mathrm{s.o.}}(E)& \sim& \lim_{\mathrm{Im}Z \downarrow 0}\mathrm{Im}\,\frac{i}{\pi\,\hbar} 
\int_{\mathbb{R}^{2 d}}\frac{d^dPd^dQ}{(2 \pi \hbar)^{d}}\int_{0}^{\infty}
\,e^{i\,\frac{Z-\ha(P,Q)}{\hbar}\,T-\frac{\eta \,T}{\hbar}}
\nonumber\\
&&=\int_{\mathbb{R}^{2 d}}\frac{d^dPd^dQ}{(2 \pi \hbar)^{d}}\delta^{(d)}(E-\ha(P,Q))
\label{moduli-Gutzwiller-short:microcanonic}
\end{eqnarray}
The short orbit contribution brings about a microcanonic average over the energy
surface which dominates the asymptotic of the energy density at large values of $E$
\cite{Berry2,Parisi}.\\
In the typical case of a kinetic plus potential energy Hamiltonian 
\begin{eqnarray}
\ha=\sum_{i=1}^{d}\frac{P^2_i}{2 m}+U(Q)
\end{eqnarray}
the short orbit contribution reduces to
\begin{eqnarray}
\rho_{\mathrm{s.o.}}(E)\sim\frac{2\,\pi^{d/2}}{\Gamma(d/2)}\,
\int_{\mathbb{R}^{d}}\frac{d^dQ}{(2 \pi \hbar)^{d}} [E-U(Q)]^{\frac{d-2}{2}}
\label{moduli-Gutzwiller-short:integrated}
\end{eqnarray}
the prefactor being the measure of the surface of a unit sphere 
in $d$-dimensions.

\subsubsection{Long orbits}
\label{moduli-Gutzwiller:long}

The long orbit contribution is just the Fourier transform of the semiclassical asymptotics 
of trace path integrals. The time-energy Fourier transform makes sense because generic 
conservative Hamiltonian systems are expected to exhibit periodic orbit in smooth families 
versus the energy. 
The time-energy Fourier transform in the semiclassical limit reduces then to 
a further stationary phase approximation. 
Alternatively, it is also possible to apply the stationary phase approximation directly
to the energy density formula (\ref{moduli-Gutzwiller:density}). In such a case the
functional to extremise is 
\begin{eqnarray}
\rac(q,\dot{q})=E\,T+\int_{0}^{T}dt\la(q,\dot{q})
\label{moduli-Gutzwiller-long:functional}
\end{eqnarray}
with the loop space condition 
\begin{eqnarray}
q^{\alpha}(t+T)=q^{\alpha}(t)\,\qquad\qquad\qquad \forall\, t
\label{moduli-Gutzwiller-long:periodic}
\end{eqnarray}
For differentiable paths this latter entails
\begin{eqnarray}
\frac{\de q^{\alpha}}{\de T}(t+T;...,T,...)=-\dot{q}^{\alpha}(t;...,T,...)
+\frac{\de q^{\alpha}}{\de T}(t;...,T,...)
\label{moduli-Gutzwiller-long:periodicdiff}
\end{eqnarray}
the partial derivative with respect to $T$ affecting only the parametric dependence of 
the path on its period.\\
On a $T$ periodic trajectory the functional (\ref{moduli-Gutzwiller-long:functional}) 
is equal to the classical reduced action \cite{Arnold,LandauLifshitz}:
\begin{eqnarray}
\rac(q_{\cl},\dot{q}_{\cl})=\oint_{0}^{T}(dq^{\alpha}p_{\alpha})(t)
\label{moduli-Gutzwiller-long:reduced}
\end{eqnarray}
The reduced action is function only of the orbit and not of individual trajectories
wending their path on the orbit.
By (\ref{moduli-Gutzwiller-long:periodic}) and (\ref{moduli-Gutzwiller-long:periodicdiff}) 
the functional is stationary when
\begin{eqnarray}
&&\delta_T\rac:=\delta T\,\frac{\de \rac}{\de T}=E+
\left[\la-\dot{q}^{\alpha}\frac{\de\la}{\de \dot{q}^{\alpha}}\right]_{t=T}+
\int_{0}^{T}dt\,\frac{\de q^{\alpha}}{\de T}\,D_{\alpha}\la
\nonumber\\
&&\delta_{q}\rac=\int_{0}^{T}dt\,\delta q^{\alpha}D_{\alpha}\la
\nonumber\\
&&D_{\alpha}\la:=\frac{\de \la}{\de q^{\alpha}}-\frac{d\,}{dt}\frac{\de \la}{\de \dot{q}^{\alpha}}
\label{moduli-Gutzwiller-long:stationary}
\end{eqnarray}
vanish.
Extrema are periodic orbit of now fixed energy $E$. The first of the equations 
(\ref{moduli-Gutzwiller-long:stationary}) establishes therefore an energy-period 
relation $E=E(T)$.\\
Evaluated over any classical periodic trajectory  $q_{\cl}$, 
the second variation does not contain mixed terms since $\de_{T}q_{\cl}$ is a (non-periodic) 
Jacobi field:
\begin{eqnarray}
&&\delta_T^{2}\rac(q_{\cl}(t),\dot{q}_{\cl}(t))=-\,\frac{d E}{d T}\,\delta T^{2}
\nonumber\\
&&\delta_q^{2}\rac(q_{\cl}(t),\dot{q}_{\cl}(t))=\int_{0}^{T}dt\,
\delta q^{\alpha} (\foP)_{\alpha\,\beta}\delta q^{\beta}
\label{moduli-Gutzwiller-long:second}
\end{eqnarray}
The integral over the fluctuations finally yields:
\begin{eqnarray}
&&\rho_{\mathrm{l.o.}}(E)\sim 
\nonumber\\
&&\ima\,\sum_{o\in \mathrm{p.p.o.}}\sum_{r=1}^{\infty}
\frac{i\,T_o\,|G|}{\pi\,\hbar\,(2\,\pi\,\hbar)^{N/2}}
\frac{e^{i\,\frac{r\,\rac_{\cl}^{o}}{\hbar}-
\frac{i\,\pi}{2}\aleph_{o,r}}}{|\frac{1}{r}\frac{d E_o}{d T}\,
\det\mathsf{W}_{o}\det\mathsf{V}_{o}^{(r)}\det_{\perp}(\id_{2 d}
-\mon_{o}^r)|^{\frac{1}{2}}}
\label{moduli-Gutzwiller-long:trace}
\end{eqnarray}
where the sum ranges on all orbits ``$o$'' at energy $E$ and their $r$-th iterates. 
The phase factor associated to extrema of the reduced action is
\begin{eqnarray}
\aleph_{o,r}=\ind\foP^{(o)}[0,r\,T_o]+
\frac{1}{2}\left(1+\mathrm{sign}\frac{d E_o}{dT\,}\,\right)+
\frac{N}{2}
\label{moduli-Gutzwiller-long:Maslov}
\end{eqnarray}   
The use of the Conley and Zehnder representation of the periodic Morse index 
(\ref{Morse-stability-ConleyZehnder:iteration}) shows that the signature of the 
energy block cancels out from the phase, \cite{Sugita}. 
If the energy is the only conserved quantity, the formula further simplifies (see also 
appendix~\ref{Hamilton:perturbative})
\begin{eqnarray}
&&\det\mathsf{W}_{o}=1
\nonumber\\
&&
\left|\frac{1}{r}\frac{d E_o}{d T}\det\mathsf{V}_{o}^{(r)}\right|=1
\label{moduli-Gutzwiller-long:onlyenergy}
\end{eqnarray}
so to retrieve Gutzwiller's original result given in chapter~\ref{density}. The opposite
limit of a separable system is treated by subtracting the $d$ zero modes associated
to the prime integrals in involution. In this way the result of Berry and Tabor 
\cite{BerryTabor} is recovered. The reader is referred to the existing literature 
\cite{RajaramanWeinberg,DittrichReuter,Creagh,Creagh2} for a detailed discussion of the 
semiclassical quantisation of the extra integrals of the motion. 

\subsubsection{Example}
 
The simplest application of the trace formula is the derivation of the 
semiclassical energy spectrum of a one dimensional particle with kinetic 
plus potential classical Hamiltonian. Moreover the potential is supposed to be a
single well.

The short orbit contribution is proportional to the period of the accessible 
orbits at energy $E$
\begin{eqnarray}
\rho_{\mathrm{s.o.}}(E)=\sum_{o\in \mathrm{p.o.}}
\int_{Q_{min}^{(o)}}^{Q_{max}^{(o)}}\frac{dQ}{\pi \hbar}\frac{m}{\sqrt{2\,m(E-U(Q))}}
=\sum_{o\in \mathrm{p.o.}}\frac{T_o(E)}{2 \pi \hbar}
\label{moduli-Gutzwiller-example:microcanonic}
\end{eqnarray}
Turning to long orbits one observes that (\ref{moduli-Gutzwiller-long:Maslov}) 
is equal to
\begin{eqnarray}
\aleph_{o,r}=2\,r
\label{moduli-Gutzwiller-example:Maslov}
\end{eqnarray}
as it follows from the analysis carried out in 
sections~\ref{Morse:examples} and~\ref{Morse:stability}.
Thus the oscillating term in the trace formula is 
\begin{eqnarray}
\rho_{l.o.}(E)=\sum_{o\in \mathrm{p.p.o.}}\frac{T_o(E)}{\pi\,\hbar}\,\sum_{r=1}^{\infty}
\cos\left(\frac{r\,\rac^{o}(E)}{\hbar}-\pi\,r\right)
\label{moduli-Gutzwiller-example:oscillating}
\end{eqnarray}
Gathering the short and long orbit contributions gives
\begin{eqnarray}
\rho(E)&\cong&\sum_{o\in \mathrm{p.p.o.}}\frac{T_o(E)}{\pi\,\hbar}
\left[\,1+2\,\sum_{r=1}^{\infty}
\cos\left(\frac{r\,\rac^{o}(E)}{\hbar}-\pi\,r\right)\right]
\nonumber\\
&=&\sum_{o\in \mathrm{p.p.o.}}\frac{T_o(E)}{\pi\,\hbar}\sum_{r=-\infty}^{\infty}
e^{2\,\pi\,r\left(\frac{\rac^{o}(E)}{2 \pi \hbar}-\frac{1}{2}\right)}
\label{moduli-Gutzwiller-example:sum}
\end{eqnarray}
The series over $r$ yields a discrete representation of a train delta functions peaked at
\begin{eqnarray}
 \frac{\rac^{o}(E)}{2 \pi \hbar}-\frac{1}{2}=n
\label{moduli-Gutzwiller-example:quantisation}
\end{eqnarray}
for $n$ an arbitrary integer. Thus the Bohr-Sommerfeld \cite{Bohr,Sommerfeld} 
quantisation rule is recovered.

\subsubsection{Non generic degenerations}

Zero modes occur generically in correspondence to continuous symmetries of the action.
However, it is not infrequent to encounter in applications 
marginal cases where unit eigenvalues of the monodromy matrix do not stem from the
invariance of the action under a Lie group. The zero mode subtraction method expounded above 
proves nevertheless useful in order to compute the functional determinant of the massive
modes. The contribution of the degrees of freedom associated to marginal zero modes can 
be then computed using approximations higher than quadratic. This is done, for example,
by finding the normal form around the periodic orbit of the Lagrangian and then retaining only
the leading order projection along the zero mode eigendirections. The procedure is discussed 
in details by Schulman in his classical monograph on path integration \cite{Schulman}.


%% file: conclusions.tex
\section{Conclusions}
\label{sec:conclusions}

The derivation of the Gutzwiller trace formula is notoriously difficult.\\
In the author's opinion the difficulty is considerably mitigated by the use 
of path integral methods. The opinion is based on two strictly intertwined reasons.

The first reason is that trace path integrals offer a global, canonically invariant 
approach to the semiclassical approximation of the energy density of a quantum system.
This is in contrast to the WKB methods usually applied in the literature of the
Gutzwiller trace formula. Broadly speaking, WKB methods proceed by repeating explicitly
in the case of the trace of the propagator the construction made 
in section~\ref{Forman:measure}. The phase of single orbit contributions is determined 
by keeping track of the caustics encountered along a reference trajectory and finally 
computing the order of concavity when the trajectory has covered the entire orbit.\\ 
Path integrals permit to concentrate directly on the properties of the propagator 
trace.  
The change of point of view is of particular advantage in the interpretation
and then practical computation of the phase factors of the orbit contributions. Here
the difference between path integral and WKB methods can be summarised as a shift
of the focus from the local geometrical properties of classical phase space to the 
global topological properties of second variation self-adjoint operators.
Morse's theory of variational calculus in the large provides all the information 
to compute the index associated to periodic orbits. It is worth stressing 
that the topologically invariant Morse index for closed extremals was already 
computed by elementary methods in Morse classical monograph 
\cite{Morse} (see also \cite{Morse2}).\\ 
The Lagrangian manifolds techniques later developed by Bott, Arnol'd and Duistermaat  
establish an extremely powerful connection between Morse index theory and Fredholm 
flow theory. Local geometry of phase space then re-emerges but only as {\em a result} 
of topological invariance.\\ 
The index of the second variation defines as well the index of the functional determinant 
of the self-adjoint operator associated to the second variation. Whenever the classical 
kinetic energy is strictly positive definite, configuration and phase space path integrals  
provide equivalent representations of the propagator trace. In consequence
the Morse index for closed extremals defined in configuration space must coincide with the
Conley and Zehnder index \cite{ConleyZehnder,SalamonZehnder,RobbinSalamon2} 
associated to the Dirac operators governing the second variation in phase space. 
This latter is widely used in recent mathematical investigation
(see overview in \cite{Salamon} and the monograph \cite{McDuffSalamon}) aimed
at establishing the general conditions which guarantee the existence of periodic orbits on
a manifold equipped with a symplectic structure.   

The second reason dwells in the general use of the path integral methods wielded in 
the derivation. They provide a common language for the treatment of quantum and 
statistical finite dimensional and field theoretic models. In this framework, beside 
its intrinsic importance, the Gutzwiller trace formula acquires also a valuable 
pedagogical significance. It provides a paradigm for the application of general 
mathematical ideas which have proven of large use and relevance in different parts of physics. 
The author's hope is that the present may also serve as an illustration of such ideas
accessible to a broad physical and mathematical audience.

\section{Acknowledgements}
\label{sec:acknowledgements}

The original incentive to write the present work was given to me by the experience as 
teaching assistant in the Ph.D. course on ``Quantum Chaos'' held by P.~Cvitanovi\'c at 
Niels Bohr Institute in the fall 1999. I want to thank P.~Cvitanovi\'c for encouraging 
me to revisit the classical derivation of the trace formula in order to unravel the 
connections with concepts and tools of main stream modern theoretical physics.

My special gratitude goes to E.~Aurell and G.~Travaglini for innumerable and very 
helpful discussions at various stages of the preparation of this work. 
It is also a pleasure to express my gratitude to A.~Kupiainen and E.~Ravaioli for 
many clarifying discussions. I want to express my gratitude A.~Sugita and M.~Pletyukhov 
for a clarifying mail exchange which allowed my to correct a mistake in the discussion of the
Conley and Zehnder index. Moreover I want to thank M.~Pletyukhov and M.~Brack for sending me 
a preprint version of their paper. 
I also benefitted from discussions with K.H.~Andersen, 
M.~Cencini, P.~Codasco da Cortina, S.~Creagh, K.~Eloranta, T.~H\"anninen, T.~Korvola, 
M.~van Hecke, J.~Lukkarinen, R.~Manieri, A.~Mazzino, K.~Montonen, A.~Niemi, P.~Olla, 
M.~Stenlund, M.~Vergassola and A.~Vulpiani. 

Finally, I would like to thank N.V.~Antonov and H.~Arponen for their critical reading 
of the manuscript and L.~Bakker at Elsevier for his kind cooperation and supervision of the
type-setting process.


%% file: measure-appendix.tex
\section{Elementary geometric concepts}
\label{geometry}

A smooth, torsion-free manifold with strictly positive symmetric metric tensor 
$g_{\alpha\,\beta}$ is said to be a Riemann manifold. The metric with its inverse
are used to lower and raise indices
\begin{eqnarray}
&&\upsilon_{\alpha}=g_{\alpha\,\beta}\,\upsilon^{\beta}
\nonumber\\
&&\upsilon^{\alpha}=g^{\alpha\,\beta}\,\upsilon_{\beta}
\label{geometry:inverse}
\end{eqnarray}
The covariant derivative of a vector field $\upsilon^{\alpha}$
\begin{eqnarray}
&&\nabla_{\beta}\upsilon^{\alpha}:=
\de_\beta\upsilon^{\alpha}+\Gamma^{\alpha}_{\beta\,\gamma}\upsilon^{\gamma}
\nonumber\\
&&\de_\beta:=\frac{\de\,\,}{\de Q^{\beta}}
\label{geometry:derivative}
\end{eqnarray}
is compatible with the metric if for any pair of vector fields $\upsilon^{\alpha}$, 
$\chi^{\alpha}$ evaluated along an arbitrary curve $q^{\alpha}(t)\in\mf$ one has
\begin{eqnarray}
\frac{d\,}{dt}\left(\upsilon^{\mu}\,g_{\mu\,\nu}\chi^{\nu}\right)
=\left(\frac{\nabla\upsilon}{dt}\right)^{\mu}g_{\mu\,\nu}
\chi^{\nu}+\upsilon^{\mu}\,g_{\mu\,\nu}
\left(\frac{\nabla\chi}{dt}\right)^{\nu}
\label{Geometry:curve}
\end{eqnarray}
The identity is always satisfied if the connection
$\Gamma^{\alpha}_{\beta\,\gamma}$ satisfies
\begin{eqnarray}
&&\nabla_\alpha g_{\mu\,\nu}:=\de_\alpha  g_{\mu\,\nu}-\Gamma^{\beta}_{\alpha\,\mu}
g_{\beta\,\nu}-\Gamma^{\beta}_{\alpha\,\nu}g_{\mu\,\beta}=0
\nonumber\\
&&\Gamma^{\alpha}_{\mu\,\nu}=\Gamma^{\alpha}_{\nu\,\mu}
\label{geometry:compatibility}
\end{eqnarray}
On a Riemann manifold the above compatibility condition is uniquely solved
by the Christoffel symbols : 
\begin{eqnarray}
\Gamma^{\alpha}_{\mu\,\nu}\,=\,g^{\alpha\,\beta}(\de_{\mu}g_{\beta\,\nu}+
\de_{\nu}g_{\beta\,\mu}-\de_{\beta}g_{\mu\,\nu})
\label{geometry:Christoffel}
\end{eqnarray}
A non trivial metric tensor may arise from the parametrisation of an Euclidean
space in non Cartesian variables. The curvature tensor, the commutator of two
covariant derivatives, discriminates between this latter case and that of 
genuinely non-Euclidean space
\begin{eqnarray}
R^{\alpha}_{\beta\,\mu\,\nu}\,\upsilon^{\beta}:=
(\nabla_{\mu}\,\nabla_{\nu}-\nabla_{\nu}\,\nabla_{\mu})\upsilon^{\alpha}\,,
\qquad \forall\,\, \upsilon^{\alpha}
\label{geometry:curvature}
\end{eqnarray}  
since it vanishes identically in an Euclidean space. The Ricci tensor and the curvature 
scalar are defined as\footnote{Here the same definitions
of the curvature and Ricci tensor are adopted as in \cite{Frankel,Carroll,AnderssonDriver}.
In \cite{Schulman,Watanabe} the curvature tensor has opposite sign but the Ricci tensor 
as well as the curvature scalar are the same as here.}
\begin{eqnarray}
&&R^{\alpha}_{\,\,\beta}\,:=g^{\mu\,\nu}\,R^{\alpha}_{\,\,\mu\,\beta\,\nu}
\label{geometry:Ricci}
\\
&&\nabla_{\alpha}\,\upsilon^{\beta}=
[\de_{\alpha}\delta^{\beta}_{\mu}+\Gamma^{\beta}_{\alpha\,\mu}\,]
\upsilon^{\mu}\,,\quad\quad  \forall\,\, \upsilon^{\alpha}
\label{geometry:scalar}
\end{eqnarray}
The metric can also be described through the introduction on every point of $\mf$ 
of an orthonormal basis of $d$ vielbeins $\{\sigma^{\alpha}_{k}\}_{k=1}^{d}$.
The vielbeins satisfy the properties
\begin{eqnarray}
&&\sigma^{\alpha}_{k}\,g_{\alpha\,\beta}\,\sigma^{\beta}_{l}\,\equiv\,
\sigma_{\alpha\,k}\,\sigma^{\alpha}_{l}\,=\,\delta_{k\,l}
\nonumber\\
&& \sigma_{\alpha\,k}\,\sigma^{\beta}_{k}\,=\,\delta^{\alpha}_{\beta}
\label{geometry:orthonormality}
\end{eqnarray}
Latin indices are associated to an Euclidean metric. In terms of the vielbeins the metric reads
\begin{eqnarray}
g_{\alpha\,\beta}\,=\,\sigma_{\alpha\,k}\,\sigma_{\beta\,k}
\label{geometry:vielbeins}
\end{eqnarray}
Combined with the compatibility conditions this last equation bares the
dependence of the Christoffel symbols on the vielbeins. In particular in the  
Euclidean case when
\begin{eqnarray}
\de_{\beta}\,\sigma_{\alpha\,k}=\de_{\alpha}\,\sigma_{\beta\,k}
\label{geometry:euclidean}
\end{eqnarray}
the identity holds
\begin{eqnarray}
\Gamma^{\alpha}_{\mu\,\nu}\,=\,\sigma^{\alpha}_k\,\de_{\mu}\sigma_{\nu\,k}\,
=\,-\,\sigma_{\mu\,k}\de_{\nu}\sigma^{\alpha}_{k}
\label{geometry:euclideanconnection}
\end{eqnarray}
The Christoffel symbol is symmetric in the lower indices by (\ref{geometry:euclidean}).

\section{Covariant stochastic differential equations}
\label{SDE}

In general stochastic differential equations are not covariant under change of 
coordinates due to the $O(\sqrt{dt})$ increments of the Wiener process.
Nevertheless covariance can be achieved through a path-wise definition 
of the vielbeins.\\
In compact notation the system of stochastic differential equations 
(\ref{Forman-measure:SDE}) reads
\begin{eqnarray}
\begin{array}{lc}
dq^{\alpha}\,=\,\upsilon^{\alpha}\,dt+\sqrt{\frac{\hbar\,z}{m}}\,
\sigma^{\alpha}_{k}\diamond dw_{k}\,,
\quad 
&\quad q^{\alpha}(T')=Q^{\prime\,\alpha}
\\
d\sigma^{\alpha}_{k}\,=\,-\Gamma^{\alpha}_{\mu\,\nu}\,\sigma^{\mu}_{k}
\diamond dq^{\nu}\,,\quad 
&\quad g^{\alpha\,\beta}(Q')=(\sigma_k^\alpha\sigma^\beta_k)(T')
\\
d\varsigma\,=-\,\varsigma\,\frac{z\,\phi}{\hbar}\,dt\,,\quad & \quad\varsigma(T')=1
\end{array}
\label{SDE:SDE}
\end{eqnarray}
Geometrically the system describes the transport of an 
orthonormal frame of vectors $\sigma^{\alpha}_k$, $k=1,...,d$ parallel to the 
trajectories of the position process $q$ for any given realization of the 
Wiener process $w$. In fact, the second equation can be recast in the form
of a covariant derivative along a path $q^{\alpha}(t)$
\begin{eqnarray}
\frac{\nabla\sigma^{\alpha}_{k}}{dt\,}=0
\label{SDE:parallel}
\end{eqnarray}
Hence if the metric compatibility condition is imposed at initial time, it will
hold true all along any given stochastic trajectory.

In the main text it was stated that solutions of (\ref{SDE:SDE}) are the 
characteristics of the Fokker-Planck equation (\ref{Forman-measure:scalar}). 
The statement is verified by differentiating the average over the Wiener
process
\begin{eqnarray}
\langle\, \varsigma(T)\,\delta^{(d)}\left(q(T)-Q\right)\rangle:=
\int\mes\mu(w(t))\,e^{-\frac{z}{\hbar}\int_{T'}^{T}dt\, 
\phi\left(q(t),t\right)}\delta^{(d)}\left(q(T)-Q\right)
\label{SDE:solution}
\end{eqnarray}
along the trajectories of (\ref{SDE:SDE}). Averages of stochastic
differentials are most conveniently performed if the infinitesimal 
time increment of the Wiener process is independent on the current
state of the system. {\em Ito} stochastic differentials 
\begin{eqnarray}
\sigma^{\alpha}_{k}(t)\,dw^{k}(t):=\lim_{dt\,\downarrow\, 0}
\sigma^{\alpha}_{k}\left(t\right)\,\left(w^{k}(t+dt)-w^{k}(t)\right)
\label{SDE:Itodiscrretisation}
\end{eqnarray}
implement the condition 
\cite{DeWittMoretteElworthy,KaratzasShreve,IkedaWatanabe,Oksendal}.
The conversion of Stratonovich differential into Ito can always be
accomplished by expanding around the pre-point discretisation and retaining
terms up to order $O(dt)$ with the proviso
\begin{eqnarray}
dw_k\,dw_l\,=\delta_{k\,l}\,dt+o(dt)
\label{SDE:proviso}
\end{eqnarray}
A straightforward but tedious computation yields
\begin{eqnarray}
&&dq^{\alpha}=\left(\upsilon^{\alpha}-\frac{\hbar\,z}{2\,m}\,
\Gamma^{\alpha}_{\mu\,\nu}\,g^{\mu\,\nu}\right)\,dt+
\sqrt{\frac{\hbar\,z}{m}}\,\sigma^{\alpha}_{k}\,dw_{k}
\nonumber\\
&&d\sigma^{\alpha}_{k}=\frac{\hbar\,z}{2\,m}
\left[R^{\alpha}_{\,\beta}-\de_{\beta}\left(\Gamma^{\alpha}_{\mu\,\nu}\,g^{\mu\,\nu}
\right)\right]\,\sigma^{\beta}_{k}dt-\Gamma^{\alpha}_{\mu\,\nu}\,\sigma^{\mu}_{k}\,dq
\label{SDE:Ito}
\end{eqnarray}
The third equation is discretisation independent. Note that in the Ito equations 
there appear {\em non}-covariant quantities.
The time differentiation of (\ref{SDE:solution}) yields
\begin{eqnarray}
&&\frac{\de}{\de T}\langle\, \varsigma(T)\,
\delta^{(d)}\left(q(T)-Q\right)\rangle=
\langle \varsigma(T) \upsilon^{\alpha}_{\mathrm{Ito}}(q(T),T)
\frac{\de\quad}{q^\alpha(T)}\delta^{(d)}\left(q(T)-Q\right)\,\rangle
\nonumber\\
&& +\langle \varsigma(T)\left[
\frac{z\,\hbar}{2\,m}g^{\alpha\,\beta}(q(T))\frac{\de\quad}{q^\alpha(T)}
\frac{\de\quad}{q^\beta(T)}-\frac{z}{\hbar}\,\phi(q(T),T)\right]
\delta^{(d)}\left(q(T)-Q\right)\rangle
\nonumber\\
\label{SDE:timeder}
\end{eqnarray}
with the Ito drift
\begin{eqnarray}
\upsilon^{\alpha}_{\mathrm{Ito}}(q(T),T):=
\upsilon^{\alpha}(q(T),T)-\frac{\hbar\,z}{2\,m}\,\Gamma^{\alpha}_{\mu\,\nu}(q(T))\,
g^{\mu\,\nu}(q(T))
\label{SDE:Itodrift}
\end{eqnarray}
The term linear in $dw$ averages out due to statistical independence of 
the Wiener noise increments.
From (\ref{SDE:timeder}) straightforward algebra recovers the Fokker-Planck 
equation. The same result is also obtained by applying functional integrations by
parts on the original Stratonovich equations \cite{Bass,Norris}.

An analogous calculation evinces the equivalence in measure
of the system of stochastic differential equations (\ref{SDE:SDE}) with
a free Wiener motion on $\mf$ 
\begin{eqnarray}
dq^{\alpha}&=&\sqrt{\frac{\hbar\,z}{m}}\,\sigma^{\alpha}_{k}\diamond dw_{k}
\nonumber\\
d\sigma^{\alpha}_{k}&=&-\Gamma^{\alpha}_{\mu\,\nu}\,\sigma^{\mu}_{k}\diamond dq
\label{SDE:Wiener}
\end{eqnarray}
advecting the new potential term
\begin{eqnarray}
d\varrho\,&=&-\varrho\left[
\left(\frac{z\,\phi}{\hbar}+\frac{m\,||\upsilon||^{2}}{2\,\hbar\,z}+
\frac{1}{2}\nabla_{\alpha}\upsilon^{\alpha}\right)\,dt+
\sqrt{\frac{m}{\hbar\,z}}\,\upsilon^{\alpha}\sigma_{\alpha\,k}\diamond dw_{k}\right]
\nonumber\\
&=&-\varrho\,\left(\frac{z\,\phi}{\hbar}\,dt+
\sqrt{\frac{m}{\hbar\,z}}\,\upsilon^{\alpha}\sigma_{\alpha\,k}\,dw_{k}\right)
\label{SDE:transformationofthedrift}
\end{eqnarray}
Equivalence in measure means that the last two equations are associated to the
the same Fokker-Planck equation as (\ref{SDE:SDE})
\begin{eqnarray}
\sqrt{|g(Q)|}\, \feo_z(Q,T|\,Q',T')\,=\langle \varrho(T)\delta^{(d)}\left(q(T)-Q\right)\,\rangle
\label{SDE:Girsanov}
\end{eqnarray}
where the average is extended over the solutions of (\ref{SDE:Wiener}).
The representation (\ref{SDE:Girsanov}) of the transition probability density
is called the {\em Girsanov-Cameron-Martin formula}.

Finally it is worth noting that the Euclidean condition (\ref{geometry:euclidean})
is the integrability condition for the mapping
\begin{eqnarray}
d\tilde{q}_k=\sigma_{\alpha\,k}\,dq^{\alpha}
\label{SDE:exact}
\end{eqnarray}
which retrieves the natural Euclidean frame where the vielbeins are the versors
of the Cartesian axes. In the jargon of statistical mechanics, the 
integrability condition permits to map multiplicative into additive noise. 
In such a case, the Laplace-Beltrami operator on a scalar 
reduces to the Bochner's Laplacian
\begin{eqnarray}
g^{\mu\,\nu}\nabla_{\mu}\nabla_{\nu}\,\feo_z\,&=&\,\left[
\sigma^{\mu}_k\,\sigma^{\nu}_k\,\de_{\mu}\,\de_{\nu}-
\Gamma^{\alpha}_{\mu\,\nu}\,g^{\mu\,\nu}\,\de_\alpha\,\right]\feo_z
=\sigma^{\mu}_k\,\de_\mu\,(\,\sigma^{\nu}_k \de_\nu\,\feo_z\,)
\label{changeofvar:Bochner}
\end{eqnarray}
Hence if (\ref{SDE:exact}) holds, the Stratonovich equation
\begin{eqnarray}
dq^{\alpha}(t)\,=\,\upsilon^{\alpha}(q(t),t)\,dt+\sqrt{\frac{\hbar\,z}{m}}\,
\sigma^{\alpha}_{k}(q(t))\diamond dw_{k}(t)
\label{SDE:SDE2}
\end{eqnarray}
is covariant since it governs the characteristic curves 
of the Stratonovich-Bochner form of the Fokker-Planck 
equation \cite{IkedaWatanabe,Ito,Zinn} which is covariant whenever 
(\ref{geometry:euclidean}) holds.

\section{Path integrals from stochastic differential equations}
\label{shorttime}

Girsanov's formula (\ref{SDE:Girsanov}) reduces the construction of path 
integrals to the solution of the heat kernel equation on a Riemann 
manifold $\mf$ for short time intervals.  As in the main text, $\mf$ 
is restricted to be either compact and without boundaries or 
$\mathbb{R}^{d}$.
The problem can be solved by means of a generalisation of the L\'evy 
construction generally used to define the Wiener (Brownian) motion in 
Euclidean spaces \cite{Levy,KaratzasShreve}.
The derivation which follows is an explicit version of the
argument outlined in \cite{DeWittMoretteElworthyNelsonSammelman}.

On $\mathrm{R}^{d}$, the expectation of the Wiener process at any intermediate 
time $t\,\in\,[T',T]$ conditioned on its end points is
\begin{eqnarray}
\langle w_k(t)\,|w_k(T)=W_k, w_k(T')=W'_k\rangle\,
=\,\frac{(t-T')\,W_{k}+(T-t)\,W_{k}'}{T-T'}
\label{shorttime:expectation}
\end{eqnarray}
The variance is of the order $O(T-T')$ for $t$ of the order $(T-T')/2$. 
The L\'evy construction proceeds by dividing the time axis into small 
slices of duration $dT$. Within each time slice the Wiener process 
is interpolated with 
(\ref{shorttime:expectation}). The increments of the Wiener process 
over a time slice are Gaussian random variables
\begin{eqnarray}
b_{k}\,:=\,\frac{W_{k}-W_{k}^{\prime}}{dT}
\label{shorttime:whitenoise}
\end{eqnarray}
with zero average and unit variance.
The projection of the L\'evy construction over a Riemann manifold
$\mf$ gives 
\begin{eqnarray}
&&\dot{q}^{\alpha}(t)\,=\,\sqrt{\frac{z\,\hbar}{m}}\,
\sigma^{\alpha}_{k}(t)\,b_k
\nonumber\\
&&\dot{\sigma}^{\alpha}_{k}(t)\,=\,-\Gamma^{\alpha}_{\mu\,\nu}(q(t))\,\sigma^{\mu}_{k}(t)
\label{shorttime:interpolation}
\end{eqnarray}
The random equation is now differentiable and equivalent to the geodesic
problem
\begin{eqnarray}
\ddot{q}^{\alpha}_{\geo}(t)\,=\,-\,\Gamma^{\alpha}_{\mu\,\nu}(q_{\geo}(t))\,q^{\mu}_{\geo}(t)\,
q^{\nu}_{\geo}(t)
\label{shorttime:geodesic}
\end{eqnarray}
The probability distribution of the random system yields the short time expression
of the heat kernel 
\begin{eqnarray}
\sqrt{g(Q)}\feo_z(Q,T'+dT|Q',T')\sim\int_{\mathbb{R}^{d}}\!\frac{d^db}{(2\pi)^{\frac{d}{2}}}
e^{-\frac{\,b^2}{2}}\delta^{(d)}(q_{\geo}(T'+dT)-Q)
\label{shorttime:randomdensity}
\end{eqnarray}
asymptotically in 
\begin{eqnarray}
O(dT)\sim O(Q-Q')^2\,\downarrow\, 0
\label{shorttime:error}
\end{eqnarray}
The integral on the right hand side of (\ref{shorttime:randomdensity}) requires the solution
of the geodesic equation (\ref{shorttime:geodesic}) with the boundary conditions
\begin{eqnarray}
&&q^{\alpha}_{\geo}(T')=Q^{\prime\,\alpha}
\nonumber\\
&&q^{\alpha}_{\geo}(T'+dT)=Q^{\alpha}
\end{eqnarray}
On a Riemann manifold the solution is unique if $dT$ is short enough. Straightforward algebra 
gives
\begin{eqnarray}
\feo_z(Q,T'+dT|Q',T')\sim\frac{m^{\frac{d}{2}}
e^{-\frac{m\,||\dot{q}_{\geo}(T')||^2\,dT}{2\,z\,\hbar}}}
{(2\,z\,\hbar\,\pi)^{\frac{d}{2}}\det\vee(Q,Q',dT)}
\label{shorttime:asympotitcs}
\end{eqnarray}
where $\vee$ is
\begin{eqnarray}
\vee_{l\,k}(Q,Q',dT)&=&\sigma_{\alpha\,l}(Q)\left(\frac{\de q_{\geo}}{\de b_k}\right)(T'+dT)
\nonumber\\
&=&-\,\sigma_{\alpha\,l}(Q)\sigma_{\beta\,l}(Q')
\{q^{\alpha}_{\geo}(T'+dT),q^{\beta}_{\geo}(T')\}_{\mathrm{P.b.}}
\label{shorttime:determinant}
\end{eqnarray}
The Poisson brackets on the right hand side (see appendix~\ref{Hamilton} for further details) 
provide the $d$ linearly independent solutions of the linearised dynamics 
along the geodesic \cite{Carroll,Frankel}
\begin{eqnarray}
&&\frac{\nabla^2}{dt^2}\delta q_{\geo}^{\alpha}(t)+
R^{\alpha}_{\beta\,\mu\,\nu}(q_\geo(t))\,\dot{q}^{\beta}_{\geo}(t)
\delta q_{\geo}^{\mu}(t)\dot{q}^{\nu}_{\geo}(t)=0 
\nonumber\\
&&\delta q_{\geo}^{\mu}(T')=0\,,\qquad\qquad  \forall \mu
\label{shorttime:linearised}
\end{eqnarray}
Thus $\det \vee$ is the determinant of the scalarised linear dynamics. 
Since the vielbeins are parallel transported along the geodesic
\begin{eqnarray}
&&\sigma_{\alpha\,l}(t)\frac{\nabla^2}{dt^2}\delta q_{\geo}^{\alpha}(t)=\ddot{u}_{l}(t)
\nonumber\\
&&u_{l}(t):=\sigma_{\alpha\,l}(t)\delta q_{\geo}^{\alpha}(t)
\label{shorttime:fluctuations}
\end{eqnarray}
the scalar fluctuations fulfill the equations of 
the motion
\begin{eqnarray}
u_{l}(t)&=&\dot{u}_{l}(T')\,(t-T')
\nonumber\\
&+&\int_{T'}^{T}dt\,(T-t)\,
\sigma_{\alpha\,l}(t)R^{\alpha}_{\beta\,\mu\,\nu}(q_\geo(t))\,\dot{q}^{\beta}_{\geo}(t)
\delta q_{\geo}^{\mu}(t)\dot{q}^{\nu}_{\geo}(t)
\label{shorttime:iteration}
\end{eqnarray}
One gets into
\begin{eqnarray}
&&\vee_{l\,k}(Q,Q',dT)\,=\,\delta_{l,k}\,dT
\nonumber\\
&&\qquad -\frac{dT^2}{6}\,
\sigma_{\alpha\,l}(T')\sigma^{\mu}_{k}(T')R^{\alpha}_{\,\beta\,\mu\,\nu}
(q_\geo(T'))\,\dot{q}^{\beta}_{\geo}(T')\dot{q}^{\nu}_{\geo}(T')+o(dT^{3})
\label{shorttime:solution}
\end{eqnarray}
The determinant is computed by means of the identity
\begin{eqnarray}
\ln\,\det(\id-\epsilon\,\mathsf{X})\,=\,\tr\ln(\id-\epsilon\,\mathsf{X})\,=\,-
\epsilon\,\tr\mathsf{X}+o(\epsilon)
\label{shorttime:determinantidentity}
\end{eqnarray}
The approximations hold in probability, thus it is legitimate to replace 
in (\ref{shorttime:solution})
\begin{eqnarray}
\dot{q}^{\beta}_{\geo}(T')\dot{q}^{\nu}_{\geo}(T')\,\sim\,\frac{z\,\hbar}{m\,dT}\,
g^{\beta\,\nu}(q_\geo(T'))
\label{shorttime:replacement}
\end{eqnarray}
which is correct within the leading order in the asymptotics (\ref{shorttime:error})
\cite{DeWitt,LangoucheRoekaertsTirapegui}. 
The final result is
\begin{eqnarray}
\feo_z(Q,T'+dT|Q',T')\sim\left[\frac{m}{2\,z\,\hbar\,\pi\,dT}\right]^{\frac{d}{2}}\,
e^{-\frac{m\,||\dot{q}_{\geo}(T')||^2\,dT}{2\,z\,\hbar}+\frac{z\,\hbar\,R(Q')}{6\,m}\,dT}
\label{shorttime:shorttime}
\end{eqnarray}
In order to be elevated to the status of a proof the above derivation requires
a precise estimate of the errors done in the approximations. The interested reader
is referred to \cite{Molchanov,Bismut} for rigorous estimates of the heat kernel on Riemann
manifolds and to the recent paper \cite{AnderssonDriver} for a complete proof of the
path integral construction.

The short time approximation of the propagator is often given in the form
\cite{Schulman}
\begin{eqnarray}
\feo_z(Q,T'\!+\!dT|Q',T')=\feo_z(Q,T'\!+\!dT|Q,T')\,e^{-\int_{T'}^{T}dt\,\la_{O.M.}}
+o(dT)
\label{shorttime:Schulman}
\end{eqnarray}
where
\begin{eqnarray}
\la_{O.M.}=\frac{m}{2}\,||\dot{q}- \upsilon||^{2}+z^2\,\phi+\frac{z\,\hbar}{2}\,
\nabla_\alpha \upsilon^{\alpha}-\frac{(z\,\hbar)^2\,R}{12\,m}
\label{shorttime:OnsagerMachlup}
\end{eqnarray}
and \cite{Watanabe}
\begin{eqnarray}
\feo_z(Q,T'+dT|Q,T')\,=\,1+\frac{z\,\hbar\,R(Q)}{12\,m}dT+o(dT)
\label{shorttime:trace}
\end{eqnarray}
In the presence of curvature a rigorous probabilistic result supports the 
identification of (\ref{shorttime:OnsagerMachlup}) as ``classical'' Lagrangian. 
Namely for real values of the analytic continuation variable $z$ it has been shown 
\cite{Graham,TakahashiWatanabe,IkedaWatanabe,FujitaKotani} that the probability to 
find $q(t)$ in a tube of arbitrarily small radius $\epsilon$ around any smooth path  
$r(t)$ connecting $Q'$ to $Q$ in a time interval of arbitrary length $[T',T]$ 
reads
\begin{eqnarray}
p(|q(t)-r(t)|\,<\,\epsilon\,,\forall t\,\in\,[T',T])
\sim e^{-\frac{1}{z\,\hbar}\int_{T}^{T'}dt\, \la_{O.M.}(r,\dot{r})}
\label{shorttime:Watanabe}
\end{eqnarray}
asymptotically in $\epsilon$. Therefore (\ref{shorttime:OnsagerMachlup}) 
solves the Onsager-Machlup problem of determining the most probable smooth paths 
covered by the stochastic process. \\
The asymptotics of the diagonal component of the kernel (\ref{shorttime:trace})
describes the leading contribution to the square root to the Van-Vleck 
determinant associated to (\ref{shorttime:OnsagerMachlup})

\section{Non covariant path integrals}
\label{changeofvar}

Beside the covariant formalism discussed above, in the literature are often 
encountered non covariant constructions of path integrals. 
An example is the Edwards and Gulyaev treatment of a quantum mechanical 
two dimensional particle in a radial potential \cite{EdwardsGulyaev,RajaramanWeinberg}.
By (\ref{SDE:Ito}) the line element in radial coordinates
\begin{eqnarray}
ds^2\,=\,dr^2+r^2\,d\theta^2
\label{changeofvar:lineelement}
\end{eqnarray}
yields the Ito stochastic differential equations
\begin{eqnarray}
&&dr=\frac{\hbar\,z}{2\,r}\,dt+\sqrt{\frac{z\,\hbar}{m}}\,dw_1
\nonumber\\
&&d\theta=\,\sqrt{\frac{\hbar\,z}{m}}\,\frac{1}{r}\,dw_2
\nonumber\\
&&d\varsigma\,=-\,\frac{z\,U(r)}{\hbar}\,dt
\label{changeofvar:Ito}
\end{eqnarray}
The path integral action in the Ito representation is 
\begin{eqnarray}
\ac&=&\frac{1}{z\,\hbar}\int_{T'}^{T}dt
\frac{m}{2}\,\left[\,\left(\dot{r}-\frac{\hbar\,z}{2\,m\,r}\right)^2+
r^2\,\dot{\theta}^2\,\right]
\nonumber\\
&=&\frac{1}{z\,\hbar}\int_{T'}^{T}dt\left[
m\,\frac{\dot{r}^2+r^2\,\dot{\theta}^2}{2}+\frac{(\hbar\,z)^2}{8\,m\,r^2}-
\frac{\hbar\,z\dot{r}}{2\,r}+z^2\,U\,\right]
\label{changeofvar:Lagrangian}
\end{eqnarray}
In the Ito representation it is not possible to apply the rules of ordinary calculus.
In particular, the term in (\ref{changeofvar:Lagrangian}) linear in the radial velocity 
cannot be interpreted as an exact differential. Only 
the Stratonovich representation is compatible with ordinary calculus. 
For the term linear in the radial velocity, the relation between the two 
representations is 
\begin{eqnarray}
\left.\frac{\hbar\,z\dot{r}}{2\,r}\right|_{\mathrm{Ito}} \sim
\left.\frac{\hbar\,z\dot{r}}{2\,r}\right|_{\mathrm{Strat.}}-
\frac{\hbar\,z}{4\,m\,r^{2}}\langle \dot{r}^{2}\rangle dt\sim
\left.\frac{\hbar\,z\dot{r}}{2\,r}\right|_{\mathrm{Strat.}}-
\frac{(\hbar\,z)^2}{4\,m\,r^{2}}
\label{changeofvar:conversion}
\end{eqnarray}
since over infinitesimal increments
\begin{eqnarray}
\langle \dot{r}^{2}\rangle \sim \frac{z\,\hbar}{m\,dt} 
\label{changeofvar:average}
\end{eqnarray}
The first term on the right hand side of (\ref{changeofvar:conversion}) can be 
treated as an exact differential. The result is the path integral
\begin{eqnarray}
&&\feo_z(R,\Theta,T\,|R',\Theta',T')=\sqrt{\frac{R}{R'}}
\int\mes[r(t)\theta(t)]r(t)e^{-\frac{1}{z\,\hbar}\int_{T'}^{T}dt\,\la'}
\nonumber\\
&&\la'=m\,\frac{\dot{r}^2+r^2\,\dot{\theta}^2}{2}-\frac{(\hbar\,z)^2}{8\,m\,r^2}+z^2\,U
\label{changeofvar:EdwardsandGulyaev}
\end{eqnarray}
where the angular kinetic energy is still defined according to the Ito prescription.
The Edwards and Gulyaev result is finally recovered by setting $z=i$.

Change of variables in path integrals obtained in arbitrary discretisations 
are notoriously unwieldy.
The probabilistic interpretation supplies a useful guideline to understand 
the properties of the path integral. 
A thorough investigation of the relation between stochastic differential equations  
and path integrals can be found in ref.~\cite{YoungDeWittMorette}.


%% file: hamiltonian.tex
\section{Summary of some elementary facts in classical mechanics}
\label{Hamilton}

The appendix recalls some definitions and results of classical mechanics
used in the main text. 
The application of semiclassical methods in quantum mechanics requires
information not only about the solutions of the classical equations
of the motion but also about their local and structural stability.
While in the first case surveys of the material summarised in the present 
appendix can be found in monographes as
\cite{Arnold,ArrowsmithPlace,deAlmeida,McDuffSalamon} and \cite{Frankel} 
for a geometrical point of view, in the author's opinion the best introductions 
to stability problems in classical mechanics remain the classical research 
papers by Gel'fand and Lidskii \cite{GelfandLidskii} and Moser \cite{Moser}. 
Comprehensive presentations of linear Hamiltonian system 
are available in \cite{YakubovichStarzhinskii} and \cite{Ekeland}.

\subsection{Lagrangian versus Hamiltonian classical mechanics}

\subsubsection{Lagrangian formulation}
\label{Hamilton:config}

Classical trajectories are extremal curves on some $d$ dimensional 
manifold $\mf$ of the action functional
\begin{eqnarray}
\ac=\int_{T'}^{T}dt\,\la(q(t)\,,\dot{q}(t)\,,t)
\label{Hamilton-config:action}
\end{eqnarray}
The boundary conditions of the variational problem are determined by some given 
initial conditions. The Lagrangian $\la$ is a function defined on the tangent bundle 
$\tgs\,\times\,\mathbb{R}$ of $\mf$. A curve $\tilde{x}(t)$ on  
$\tgs\,\times\,\mathbb{R}$ is defined by the lift of a smooth curve $q(t)$ achieved 
by setting
\begin{eqnarray}
\tilde{x}(t)=\left[\begin{array}{c} q(t) \\ \dot{q}(t) \end{array}\right]\,:=\,
\left[\begin{array}{c} q(t) \\ dq(t)/dt \end{array}\right]
\label{Hamilton-config:lift}
\end{eqnarray}
Velocities $\dot{q}^{\alpha}$ transform as well as positions $q^{\alpha}$ as contra-variant 
vectors. Lagrangian of physical interest are quadratic in the velocities.

Any smooth extremal of the action must fulfill the Euler-Lagrange equations
\begin{eqnarray}
\frac{\de \la}{\de q^\alpha}-\frac{d}{d\,t}\frac{\de \la}{\de \dot{q}^\alpha}=0\,\quad 
\alpha=1,...,d 
\label{Hamilton-config:EulerLagange}
\end{eqnarray}
The ``ellipticity'' condition
\begin{eqnarray}
&&\spe\left\{\mass\right\}\,>\,0\,\quad\quad \forall\, q
\nonumber\\
&&(\mass)_{\alpha\beta}:=\frac{\de^{2}\la}{\de \dot{q}_\alpha \de\dot{q}_\beta}
\label{Hamilton-config:elliptic}
\end{eqnarray}
insures the equivalence of the Euler Lagrange equations to a first order 
system of differential equations in $\tilde{x}$ with fundamental solution 
specified by a diffeomorphism $\tilde{\Phi}$ 
\cite{ArrowsmithPlace}.
Given an initial condition $\tilde{x}'$ the time evolution of an extremal or classical 
trajectory is then obtained as
\begin{eqnarray}
\tilde{x}_{\cl}(t)\,=\,\tilde{\Phi}(t\,;\,\tilde{x}'\,,t')
\label{Hamilton-config:diffeomorphism}
\end{eqnarray}
The diffeomorphism $\tilde{\Phi}(t\,;\,\cdot\,,t')$ is a flow or one parameter group 
of transformations in $t$
\begin{eqnarray}
\begin{array}{ll}
\tilde{\Phi}(t';\tilde{x}',t')=x'\,,\quad & \forall\,x'
\\
\tilde{\Phi}(t;\tilde{\Phi}(t';x'',t''),t')=\tilde{\Phi}(t;x'',t'')\,,\quad & \forall\,x''
\end{array}
\label{Hamilton-config:flow}
\end{eqnarray}
The time evolution of infinitesimal perturbations of the initial conditions 
is governed by the linearised dynamics. 
\begin{eqnarray}
\delta \tilde{x}_{\cl}^{\alpha}(t)\,=
\left(\frac{\de \tilde{\Phi}^{\alpha}\,}{\de x^{\prime\,\beta}}\right)(t\,;\,x',\,t')\,
\delta \tilde{x}^{\prime\,\beta}:=\tfu^{\alpha}_{\beta}(t\,,t'\,;x')\,
\delta \tilde{x}^{\prime\,\beta}
\label{Hamilton-config:stability}
\end{eqnarray}
The columns of $\tfu$ are linearly independent extremals of the second variation
along a classical trajectory $q_{\cl}$:
\begin{eqnarray}
\left[\frac{\de \qla}{\de q^\alpha\,}-\frac{d}{d\,t}\frac{\de \qla}{\de \dot{q}^\alpha\,}
\right]_{\flu_{\cl}}\,\equiv\,(\fo\,\flu_{cl})_{\alpha}=0 
\label{Hamilton-config:fluctuations}
\end{eqnarray}
The ``classical fluctuations'' $\flu_{\cl}$ are referred to as Jacobi fields 
\cite{Poincare,Klingenberg,DeWittMoretteMaheshwariNelson}. The linearised dynamics 
is defined on the space $\qttgs$ tangent to the tangent space along the classical
trajectory.

\subsubsection{N\"other theorem}
\label{Hamilton:Noether}

Let $G$ be a Lie group with $N$ parameters $\{\mo^n\}_{n=1}^{N}$. 
Acting on configuration space, the Lie group generates smooth 
transformations of variables. 
In a local neighborhood of the identity, the transformations are 
spanned by the $N$ vector fields
\begin{eqnarray}
v_n^{\alpha}(q):=\left.\frac{\de \varphi^{\alpha}}{\de \mo^n}(\mo,q)\right|_{\mo=0}\,,\qquad\qquad
n=1,...,N
\label{Hamilton-Noether:infinitesimal}
\end{eqnarray}
induced by the Lie algebra of $G$ \cite{Gilmore,Frankel,Nakahara}.
The action functional is invariant under $G$ if
\begin{eqnarray}
&&\int_{0}^{T}dt\,\la(\varphi(t),\dot{\varphi}(t),t)=\int_{0}^{T}dt\,\la(q(t),\dot{q}(t),t)
\nonumber\\
&&\varphi(t)\equiv\varphi(\mo,q(t))
\label{Hamilton-Noether:action}
\end{eqnarray}
is satisfied. Rephrased in differential form, invariance means
\begin{eqnarray}
0=\left.\frac{\de \la}{\de \mo^n}(\varphi,\dot{\varphi})\right|_{\mo=0}\,,\qquad\qquad 
\forall\,n
\label{Hamilton-Noether:invariance}
\end{eqnarray}
Since Lagrangians of physical interest have the form
\begin{eqnarray}
\la=g_{\alpha\,\beta}\dot{q}^{\alpha}\dot{q}^{\beta}+\dot{q}^{\alpha}\,A_{\alpha}-U
\label{Hamilton-Noether:Lagrangian}
\end{eqnarray}
(\ref{Hamilton-Noether:invariance}) also entails that a symmetry is an isometry 
of the metric tensor which leaves invariant the vector and scalar potential
\begin{eqnarray}
&&v^{\gamma}_{n}\,\frac{\de g_{\alpha\,\beta}}{\de q^{\gamma}}+
g_{\alpha\,\gamma}\,\frac{\de v^{\gamma}_{n}}{\de q^{\beta}}+
g_{\beta\,\gamma}\,\frac{\de v^{\gamma}_{n}}{\de q^{\alpha}}=0
\nonumber\\
&&v^{\gamma}_{n}\,\frac{\de A_{\alpha}}{\de q^{\gamma}}+A_{\gamma}\,
\frac{\de v^{\gamma}_{n}}{\de q^{\alpha}}=0
\nonumber\\
&&v^{\gamma}_{n}\,\frac{\de U}{\de q^{\gamma}}=0
\label{Hamilton-Noether:isometry}
\end{eqnarray}
Combined with the Euler-Lagrange equations, (\ref{Hamilton-Noether:invariance}) yields
\begin{eqnarray}
0=\frac{d}{dt}\left[\frac{\de \la}{\de \dot{q}^{\alpha}}\,v_n^{\alpha}\right]
\label{Hamilton-Noether:Noether}
\end{eqnarray}
and consequently defines the conserved quantity
\begin{eqnarray}
\ha_n=\frac{\de \la}{\de \dot{q}^{\alpha}}\,v_n^{\alpha}
\label{Hamilton-Noether:conservationlaw}
\end{eqnarray}
The index contraction on the right hand side corresponds to a well defined scalar
product. To wit, under generic change of coordinates $q^{\prime\,\alpha}=q^{\prime\,\alpha}(q)$ 
the classical momentum transforms as a co-vector \cite{Arnold,Frankel}:
\begin{eqnarray}
p_{\alpha}=\frac{\de \la}{\de \dot{q}^\alpha}=\frac{\de q^{\prime\,\beta}}{\de q^{\alpha}}
\,\frac{\de \la^{\prime}}{\de q^{\prime\,\beta}}=
\frac{\de q^{\prime\,\beta}}{\de q^{\alpha}}\,p_{\beta}^{\prime}
\label{Hamilton-Noether:momentum}
\end{eqnarray}

\subsubsection{Elementary Hamiltonian formulation}
\label{Hamilton:elementary}

The ellipticity condition (\ref{Hamilton-config:elliptic}) permits to define the
Hamiltonian
\begin{eqnarray}
\ha(q,p,t)=\sup_{\dot{q}\,\in \,\tgs}\left\{p_\alpha\,\dot{q}^{\alpha}-\la(q,\dot{q},t)
\right\}
\label{Hamilton-elementary:legedretrensform}
\end{eqnarray}
A smooth curve satisfies the Euler-Lagrange equations if and only if it is 
solution of the Hamilton equations in canonical coordinates
\begin{eqnarray}
\dot{q}^\alpha=\frac{\de \ha}{\de p_\alpha}\,,\quad \quad
\dot{p}_\alpha=\,-\,\frac{\de \ha}{\de q^\alpha}\, \quad\quad \alpha=1,..,d
\label{Hamilton-elementary:Hamilton}
\end{eqnarray}
Let $\Phi$ denote the flow which solves the Hamilton equations. The flow is defined
in phase space, the geometrical cotangent bundle $\ctgs$.
The derivatives of $\Phi$  with respect to the initial positions and momenta 
specify the flow  $\fu$ of the linearised dynamics around a classical trajectory 
$q_{\cl}$. 
The space where the linearised dynamics is defined is tangent to the cotangent
bundle $\qtctgs$. A similarity transformation connects the linearised dynamics
around the same trajectory in phase and tangent space:
\begin{eqnarray}
\fu(t,t')\,=\,\mathfrak{T}(t)\,\tfu(t,t')\mathfrak{T}^{-1}(t')
\label{Hamilton-elementary:linear}
\end{eqnarray}
by $2\,d\,\times 2\,d$ dimensional matrix 
\begin{eqnarray}
\mathfrak{T}(t)\,=\,\left[\begin{array}{cc} \id_d \quad &\quad 0 \\ 
\vpt \quad & \quad \mass\end{array}\right]
\end{eqnarray}
the blocks being the second derivatives of the Lagrangian evaluated along 
the classical trajectory $q_{\cl}$. In particular the lower two blocks specify
the pull-back \cite{Frankel} of the momentum
\begin{eqnarray}
dp(\flu)\,=\,(\mass)_{\alpha\,\beta}\dflu^{\beta}+
(\vpt)_{\alpha\,\beta}\flu^{\beta}
\label{Hamilton-elementary:pullback}
\end{eqnarray}
In the main text the latter is denoted as
\begin{eqnarray}
\nflu\,:= \,dp(\flu)
\label{Hamilton-elementary:covariant}
\end{eqnarray}
to emphasise that the linear momentum transforms as a covariant vector.

\subsubsection{Symplectomorphisms}
\label{Hamilton:symplecto}

A general formulation of Hamiltonian dynamics is attained in arbitrary coordinates 
$x$:
\begin{eqnarray}
\sy_{\alpha\,\beta}\,\dot{x}^{\beta}=\,\frac{\de \ha}{\de x^{\alpha}}\,,
\quad\quad \alpha\,,\beta=1,...,2\,d
\label{Hamilton-symplecto:Hamilton}
\end{eqnarray}
where $\sy_{\alpha\,\beta}$ is a $2\,d\,\times\,2\,d$ tensor 
globally defined in $\ctgs$ and characterised by the properties 
\begin{eqnarray}
&&\begin{array}{ll}
\sy_{\alpha\,\beta}\,=\,-\, \sy_{\beta\,\alpha}\,,\quad & \quad\quad\quad  
\sy_{\alpha\,\gamma}\,\sy^{\gamma\,\beta}\,=\,\delta^{\beta}_{\alpha}
\\
\mathrm{det}\,\sy\,\neq\,0\,,\quad & \quad\quad\quad \forall\,x\,\in\,\ctgs
\end{array}
\nonumber\\
&&\frac{\de\,\sy_{\alpha\,\beta}}{\de x^{\gamma}\,}+ 
\frac{\de\,\sy_{\gamma\,\alpha}}{\de x^{\beta}\,}+
\frac{\de\,\sy_{\beta\,\gamma}}{\de x^{\alpha}\,}=0
\label{Hamilton-symplecto:symplectic}
\end{eqnarray}
In other words $\sy_{\alpha\,\beta}$ specifies a non degenerate closed two form 
\cite{Carroll,Frankel}. 
By Poincar\'e's lemma \cite{Frankel} any closed two form is locally the curl of
a vector potential  
\begin{eqnarray}
\sy_{\alpha\,\beta}\,=\,\de_{\alpha}\,\vartheta_{\beta}-\de_{\beta}\,\vartheta_{\alpha}
\label{Hamilton-symplecto:potential}
\end{eqnarray}
The elementary formulation of Hamiltonian dynamics is recovered from 
(\ref{Hamilton-symplecto:symplectic}), by means of Darboux's theorem. 
This latter states that on a differentiable manifold $\mf$
there exist local coordinates $x=(q\,,p)$ such that the symplectic matrix 
reduces to
\begin{eqnarray}
\sym\,:=\,\sy_{\mathrm{Darboux}}\,:=\,
\left[\begin{array}{cc} 0 \quad&\quad -\,\id_{d}\\ \id_d \quad&\quad 0\end{array}\right]\,,
\quad\quad\quad\quad \sym^{\dagger}\,\sym=\id_{2\,d}
\label{Hamilton-symplecto:Darboux}
\end{eqnarray}
A straightforward computation proves the Hamilton equations (\ref{Hamilton-symplecto:Hamilton}) 
are invariant in form under smooth transformations verifying the condition
\begin{eqnarray}
(\Psi^*\sy)_{\alpha\,\beta}\,:=\,
\frac{\de \Psi^{\mu}}{\de x^{\alpha}}\,\sy_{\mu\,\nu}(\Psi(x))\,
\frac{\de \Psi^{\nu}}{\de x^{\beta}}=\sy_{\alpha\,\beta}(x)
\label{Hamilton-symplecto:canonical}
\end{eqnarray}
Transformations which satisfy (\ref{Hamilton-symplecto:canonical}) are said canonical or 
symplectomorphisms.
One parameter symplectomorphisms continuous around the identity are characterised 
by the infinitesimal version of (\ref{Hamilton-symplecto:canonical}) around the identity:
\begin{eqnarray}
&&\Psi^{\alpha}(x,\mo)\,=\,x^{\alpha}+V^{\alpha}(x)\,\mo+O(\mo^2)
\nonumber\\
&&V^{\gamma}\,\de_{\gamma}\sy_{\alpha\,\beta}+\sy_{\gamma\,\beta}\,\de_{\alpha}\,V^{\gamma}
+\sy_{\alpha\,\gamma}\de_{\beta} V^{\gamma}\,=\,0
\label{Hamilton-symplecto:canonicalinfinitesimal}
\end{eqnarray}
The condition is satisfied by Hamiltonian vector fields  
\begin{eqnarray}
V^{\alpha}\,=\,\sy^{\alpha\,\beta}\,\frac{\de \ha_V}{\de x^{\beta}}
\label{Hamilton-symplecto:Hamiltonian}
\end{eqnarray}
the generating function $\ha_V$ being a phase space scalar.
Canonical transformations leave the Poisson brackets of two 
phase space scalar $\chi$ and $\phi$
\begin{eqnarray}
\{\phi\,,\chi\}_{\mathrm{P.b.}}\,:=\,\sy^{\alpha\,\beta}\,\frac{\de \phi}{\de x^{\alpha}}
\frac{\de \chi}{\de x^{\beta}}
\label{Hamilton-symplecto:Poisson}
\end{eqnarray}
invariant. Poisson brackets are antisymmetric, obey the Leibniz chain rule 
and satisfy the Jacobi identity \cite{Arnold,McDuffSalamon,LandauLifshitz} as follows 
from the properties (\ref{Hamilton-symplecto:symplectic}) of $\sy$. If $\chi$ and $\phi$ are 
generating functions the Poisson brackets are equivalent to the {\em skew product} 
of the corresponding Hamiltonian vector fields:
\begin{eqnarray}
\{\phi\,,\chi\}_{\mathrm{P.b.}}\,=\,\sy_{\alpha\,\beta}\,V^{\alpha}_{\chi}\,
V^{\beta}_{\phi}
\label{Hamilton-symplecto:skewproduct}
\end{eqnarray}
The attribute {\em skew} refers to the role of pseudo metric played 
by the antisymmetric tensor $\sy_{\alpha\,\beta}$. The skew product vanishes 
on pairs of linearly dependent vectors. As a result in a phase space of 
$2\,d$ dimensions the maximal number of linearly independent vectors which 
can be pairwise annihilated by the skew product is $d$ 
\cite{Arnold,McDuffSalamon}.

\subsubsection{Symmetries}
\label{Hamilton:symmetries}

Poisson brackets supply a convenient formalism to handle the
relation between continuous symmetries and conservation laws. 
Namely if $\ha_a$, $\ha_b$ are respectively the generating functions of the 
one-parameters symplectomorphisms $\Psi_a(\cdot,\mo^a)$ and $\Psi_b(\cdot,\mo^b)$, 
the derivatives along the flow obey the chain of identities:
\begin{eqnarray}
\left.\left(\frac{\de \ha_a}{\de \mo^b}\right)(\Psi_b(x,\mo^b))\right|_{\mo^b=0}\,=\,
\{\ha_b\,,\ha_a\}_{\mathrm{P.b.}}=
-\left.\left(\frac{\de \ha_b}{\de \mo^a}\right)(\Psi_a(x,\mo^a))\right|_{\mo^a=0}
\label{Hamilton-symmetries:differential}
\end{eqnarray}
An integral of the motion is therefore the generating function of a symplectomorphism 
leaving the Hamiltonian invariant.
In general, Frobenius theorem establishes integrability conditions 
in terms of the Lie algebras of  vector fields \cite{Frankel}. 
Poisson brackets are an integral version of the Lie brackets 
of the Hamiltonian vector fields with generating functions $\ha_a$, $\ha_b$ 
\cite{Arnold,McDuffSalamon}:
\begin{eqnarray}
&&\{V_a\,,V_b\}_{\mathrm{L.b.}}^{\alpha}\,:=V_a^{\mu}
\frac{\de V_b^\alpha}{\de x^{\mu}}-V_b^{\mu}
\frac{\de V_a^\alpha}{\de x^{\mu}}
\,=\,\,-\,\sy^{\alpha\,\beta}\frac{\de \,\,}{\de x^{\beta}}\{\ha_a\,,\ha_b\}_{\mathrm{P.b.}}
\label{Hamilton-symmetries:LiePoisson}
\end{eqnarray}
If the generating functions Poisson commute mutually and with the 
Hamiltonian $\ha$ they are simultaneously preserved by the dynamics. 
The corresponding Hamiltonian vector fields form an Abelian group 
of transformations. 
More generally, a symmetry group is realised by a Lie algebra of
Hamiltonian vector field $\{V_a\}_{a=1}^{N}$ with structure 
constants $C_{a\,b}^{c}$ (see appendix~\ref{Lie})
\begin{eqnarray}
\{V_a\,,V_b\}_{\mathrm{L.b.}}^{\alpha}\,=\,C_{a\,b}^{c}\,V_c^{\alpha}
\label{Hamilton-symmetries:Liealgebra}
\end{eqnarray}
yielding the Poisson bracket algebra
\begin{eqnarray}
\{\ha_a\,,\ha_b\}_{\mathrm{P.b.}}\,=\,-\,C_{a\,b}^{c}\,\ha_c+D_{a\,b}\,,
\qquad\qquad  D_{a\,b}\,=\,-\,D_{b\,a}
\label{Hamilton-symmetries:Poissonalgebra}
\end{eqnarray}
The $D_{a\,b}$'s are skew symmetric phase space constant depending only
on the structure of the Lie algebra \cite{Arnold}.
More light on the meaning of the $D_{a\,b}$'s is shed by the analysis of
the dependence of the $\{\ha_a\}_{a=1}^{N}$ on the Lie algebra vector fields. 
In consequence of (\ref{Hamilton-symplecto:symplectic}), (\ref{Hamilton-symplecto:potential}) 
the generating functions must admit the general representation
\begin{eqnarray}
\ha_a\,=\,\vartheta_{\alpha}\,V_a^{\alpha}+h_a
\label{Hamilton-symmetries:generating}
\end{eqnarray}
for $\{h_a\}_{a=1}^{N}$ some scalar functions and $\vartheta_{\alpha}$ the vector 
potential defined by (\ref{Hamilton-symplecto:potential}). 
The definition of Lie brackets (\ref{Hamilton-symmetries:Poissonalgebra}) entails 
the identity
\begin{eqnarray}
\vartheta_{\gamma}\{V_a\,,V_b\}_{\mathrm{L.b.}}^{\gamma}\,=\,
V^{\gamma}_b\,\frac{\de h_a}{\de x^{\gamma}}-V^{\gamma}_a\,\frac{\de h_b}{\de x^{\gamma}}
-\{\ha_a\,,\ha_b\}_{\mathrm{P.b.}}
\label{Hamilton-symmetries:oneformidentity}
\end{eqnarray}
which, combined with (\ref{Hamilton-symmetries:Liealgebra}) and 
(\ref{Hamilton-symmetries:generating}), yields
\begin{eqnarray}
D_{a\,b}\,=\,C_{a\,b}^{c}\,h_c\,-\,V_a^{\mu}
\frac{\de h_b^\alpha}{\de x^{\mu}}+V_b^{\mu}
\frac{\de h_a^\alpha}{\de x^{\mu}}
\label{Hamilton-symmetries:cocycle}
\end{eqnarray}
Thus it is only when the $D_{a\,b}$'s are zero for all $a$, $b$ that the scalar 
functions $\{h_a\}_{a=1}^{N}$ can be set to zero in the representation of the
generating functions of the Lie algebra. In such a case a symmetry is said {\em equivariant}. 
The denomination is justified by considering an example with globally defined canonical 
coordinates. In such a case the vector potential is
\begin{eqnarray}
\vartheta_{\alpha}\,=\,\frac{\sym_{\beta\,\alpha}\,x^{\beta}}{2}
\label{Hamilton-symmetries:vectorpotential}
\end{eqnarray}
while Hamiltonian vector fields have the form
\begin{eqnarray}
V^{\alpha}_a\,=\,\sym^{\alpha\,\beta}\frac{\de \ha_a}{\de x^{\beta}}
\label{Hamilton-symmetries:vectorfield}
\end{eqnarray}
Hence one has
\begin{eqnarray}
\vartheta_{\alpha} V^{\alpha}_a\,=\,\ha_a+\frac{1}{2}\,
\left(x^{\alpha}\frac{\de \,\ha_{a}}{\de x^{\alpha}}- 2\,\ha_a\right)
\label{Hamilton-symmetries:canonicalgenerating}
\end{eqnarray}
Equivariance then means that the constant of the motion is homogeneous of degree two 
in $x$. The rotation group provides an example of equivariant symmetry leading to the 
conservation of the angular momentum
\begin{eqnarray}
\ha_a\,=\,p_{m}\,\mathsf{r}_{m\,n}^{(a)}\,q_{n}
\label{Hamilton-symmtries:rotation}
\end{eqnarray}
with $\mathsf{r}_{m\,n}^{(a)}$ the infinitesimal rotation matrix 
around the $a$ axis.

\subsection{Linear Hamiltonian systems}

\subsubsection{General properties}
\label{Hamilton:linear}

Poisson brackets allow a covariant description of the classical linearised 
dynamics in the space $\qtctgs$ tangent to phase space  along a classical trajectory
\begin{eqnarray}
x_{\cl}(t)=\Phi(t\,;\,x',t')
\label{Hamilton-linear:trajectory}
\end{eqnarray}
The linear evolution matrix can be recast in the form
\begin{eqnarray}
\fu^{\alpha}_{\beta}(t,t')&=&\sy^{\gamma\,\delta}(x')\,\sy_{\beta\,\mu}(x')\,
\frac{\de x^{\prime\,\mu}}{\de x^{\prime\,\gamma}}\,
\left(\frac{\de \Phi^{\alpha}}{\de x^{\prime\,\delta}}\right)(t\,;\,x',t')
\nonumber\\
&=&\{x_{\cl}^{\mu}(t')\,,x_{\cl}^{\alpha}(t)\}_{\mathrm{P.b.}}\,\sy_{\beta\,\mu}(x_{\cl}(t'))
\label{Hamilton-linear:linearised}
\end{eqnarray}
In matrix notation, the linearised flow satisfy the linear Hamiltonian 
equations
\begin{eqnarray}
&&\sy(t) \frac{d \fu }{d t}(t,t')=\mathsf{H}(t) \fu(t,t')\,,
\nonumber\\
&& \fu(t',t')=\id_{2\,d}
\label{Hamilton-linear:equations}
\end{eqnarray}
having defined
\begin{eqnarray}
&&\sy(t):=\sy(x_{\cl}(t))
\nonumber\\
&&\mathsf{H}_{\alpha\,\beta}(t):=\left[\frac{\de^2\ha\,}{\de x^{\alpha}\de x^{\beta}}
-\sy^{\gamma\,\delta}\,\frac{\de \ha}{\de x^{\delta}}\,
\frac{\de \sy_{\alpha\,\gamma}}{\de x^{\beta}}\right]_{x_{\cl}(t)}
\label{Hamilton-linear:matrices}
\end{eqnarray}
 (\ref{Hamilton-linear:linearised}) provides for the 
invariance of the skew product of any particular solutions of the
linearised dynamics under canonical transformations. Namely any
canonical transformation of the flow
\begin{eqnarray}
&&\Psi^{\alpha}(x_{\cl}(t))\,=\,(\Psi\circ\Phi)^{\alpha}(t\,;x',t')
\nonumber\\
&&\frac{\de\Psi^{\alpha}}{\de x^{\prime\,\beta}}\,=\,
\frac{\de\Psi^{\alpha}}{\de \Phi^{\gamma}}(x_{\cl}(t))\,\fu^{\gamma}_{\beta}(t,t')
\label{Hamilton-linear:canonical}
\end{eqnarray}
satisfies the symplectic property
\begin{eqnarray}
\fu^{\gamma}_{\alpha}(t,t')\,\sy_{\gamma\,\delta}(t)\,\fu^{\delta}_{\beta}(t,t')
\,=\,\sy_{\alpha\,\beta}(t') 
\label{Hamilton-linear:symplectic}
\end{eqnarray}
Combined with (\ref{Hamilton-linear:linearised}), the equality specifies the
behaviour of the linearised dynamics under time reversal
\begin{eqnarray}
(\fu^{-1})^{\alpha}_{\beta}(t,t')\,=\,\fu^{\alpha}_{\beta}(t',t)
\label{Hamilton-linear:timereversal}
\end{eqnarray}
The property is inherited by the linear evolution matrix $\tfu$ in $\qttgs$.\\
By (\ref{Hamilton-symmetries:differential}) symmetry transformations $\Psi$ commute 
with the Hamiltonian flow $\Phi$ 
\begin{eqnarray}
(\Psi\circ\Phi)^{\alpha}(t\,;x',t')=\Phi^{\alpha}(t;\Psi(x'),t')
\label{Hamilton-linear:commutativity}
\end{eqnarray}
Hence the linearised dynamics must satisfy 
\begin{eqnarray}
\fu^{\alpha}_{\beta}(t,t';\Psi(x'))=\frac{\de\Psi^{\alpha}}{\de \Phi^{\gamma}}(x_{\cl}(t))
\fu^{\gamma}_{\delta}(t,t';x') \frac{\de\Phi^{\delta}}{\de \Psi^{\beta}}(x')
\label{Hamilton-linear:similarity}
\end{eqnarray}

\subsubsection{Linear periodic systems}
 \label{Hamilton:periodic}

Linearisation around a classical trajectory on a periodic orbit of period $T_{\cl}$
gives rise to a periodic Hamiltonian matrix
\begin{eqnarray}
\mathsf{H}(t)\,=\,\mathsf{H}(t+T_{\cl})\,,\quad\quad \forall\,t
\label{Hamilton-periodic:matrix}
\end{eqnarray}
The pseudo metric $\sy_{\alpha\,\beta}$ has the same periodicity.
The general form of the solution of a linear periodic system is dictated by
Floquet theorem \cite{ArrowsmithPlace,YakubovichStarzhinskii}:
\begin{eqnarray}
\fu(t,t')\,=\mathsf{Pe}(t\,,\,t')\,\exp\left\{\frac{t-t'}{T_{\cl}}\,
\int_{t'}^{t'+T_{\cl}}ds\,(\sy^{-1} \mathsf{H})(s)\right\}
\label{Hamilton-periodic:Floquet}
\end{eqnarray}
with $\mathsf{Pe}$ a periodic matrix such that
\begin{eqnarray}
\mathsf{Pe}(t'+n\,T_{\cl}\,;\,t')\,=\,\id_{2\,d}
\label{Hamilton-periodic:Floquetperiodic}
\end{eqnarray}
for all integer $n$. 

The stability of a classical periodic orbit is governed by the
monodromy matrix:
\begin{eqnarray}
\mon(T_{\cl})\,:=\,\fu(t'+T_{\cl},t')\,\equiv\,\exp\left\{
\int_{t'}^{t'+T_{\cl}}ds\,(\sy^{-1} \mathsf{H})(s)\right\}
\label{Hamilton-periodic:monodromy}
\end{eqnarray}
It is not restrictive to choose local coordinates such that the monodromy matrix 
satisfies  
\begin{eqnarray}
\mon^{\dagger}\,\sym\,\mon\,=\,\sym
\label{Hamilton-periodic:symplectic}
\end{eqnarray}
Together with 
\begin{eqnarray}
\mathrm{det}\,\mon\,=\,1
\label{Hamilton-periodic:normalization}
\end{eqnarray}
(\ref{Hamilton-periodic:symplectic}) defines the linear 
symplectic group $Sp(2\,d)$.
Any symplectic matrix has the square block form 
\begin{eqnarray}
\mon\,=\,\left[
\begin{array}{cc} \mathsf{A} \quad &\quad \mathsf{B} \\
\mathsf{C} \quad & \quad \mathsf{D}
\end{array}\right]\,,\quad\quad
\mon^{-1}\,=\,\left[
\begin{array}{cc} \mathsf{D}^{\dagger} \quad &\quad -\,\mathsf{B}^{\dagger} \\
-\,\mathsf{C}^{\dagger} \quad & \quad \mathsf{A}^{\dagger}
\end{array}\right]
\label{Hamilton-periodic:blocks}
\end{eqnarray}
Furthermore the symplectic condition (\ref{Hamilton-periodic:symplectic})
requires the $d\, \times\, d$ blocks to fulfill
\begin{eqnarray}
&&\mathsf{A}\mathsf{D}^{\dagger}-\mathsf{B}\mathsf{C}^{\dagger}=\id_{d}
\nonumber\\
&&\mathsf{A}\mathsf{B}^{\dagger}=\mathsf{B}\mathsf{A}^{\dagger}\,,\qquad
\mathsf{C}\mathsf{D}^{\dagger}=\mathsf{D}\mathsf{C}^{\dagger}
\label{Hamilton-periodic:algebra}
\end{eqnarray}
or equivalently
\begin{eqnarray}
&&\mathsf{D}^{\dagger}\mathsf{A}-\mathsf{C}^{\dagger}\mathsf{B}=\id_{d}
\nonumber\\
&&\mathsf{D}^{\dagger}\mathsf{B}=\mathsf{B}^{\dagger}\mathsf{D}\,,\qquad
\mathsf{C}^{\dagger}\mathsf{A}=\mathsf{A}^{\dagger}\mathsf{C}
\label{Hamilton-periodic:algebra2}
\end{eqnarray}
Every linear Hamiltonian flow in Darboux coordinates draws a curve in
the symplectic group.
The normal forms of elements of $Sp(2\,d)$ are also strongly constrained by 
(\ref{Hamilton-periodic:symplectic}). 
Left (generalised) eigenvectors of a symplectic matrix are specified by the 
right (generalised) eigenvectors through a complete set of skew orthogonality 
relations \cite{Moser,ConleyZehnder,Ekeland}. 
In this way it is possible to construct a {\em symplectic basis}
with elements $\{\mathfrak{e}_n,\mathfrak{f}_n\}_{n=1}^{d}$ satisfying
\begin{eqnarray}
&&\mathfrak{e}_m^{\dagger}\,\mathfrak{e}_n\,=\,\mathfrak{f}_m^{\dagger}\,
\mathfrak{f}_n\,=\,\delta_{m\,n}
\nonumber\\
&&\mathfrak{e}_m^{\dagger}\,\sym \,\mathfrak{f}_n\,=\,-\,\delta_{m,n}
\label{Hamilton-periodic:symplecticbasis}
\end{eqnarray}
which reduces simultaneously the symplectic matrix and the pseudo metric $\sym$ 
to normal form.\\
The eigenvalues of any element of $Sp(2\,d)$ are constrained in a rigid pattern 
of pairs or quartets in the complex plane. As a consequence the normal form
of a symplectic matrix typically (but not necessarily!) consists of the two or four 
dimensional blocks \cite{Moser,ConleyZehnder,YakubovichStarzhinskii} listed
below
\begin{enumerate}
\item[1] {\em Direct hyperbolic blocks} \\
The eigenvalues are $(e^{\tilde{\omega}}\,,e^{-\,\tilde{\omega}})$ 
with $\tilde{\omega}$ real. The simultaneous normal forms of an hyperbolic block
and of the associated pseudo-metric are:
\begin{eqnarray}
\mon_{\mathrm{d.h.}}\,=\,\left[\begin{array}{cc}
e^{\tilde{\omega}} \quad & \quad 0
\\
0 \quad & \quad e^{-\,\tilde{\omega}}
\end{array}\right]
\,,\quad\quad \sym_{\mathrm{d.h.}}\,=\,\left[\begin{array}{cc}
0 \quad & \quad -1
\\
1 \quad & \quad 0
\end{array}\right]
\label{Hamilton-periodic:hyperbolic}
\end{eqnarray}
\item[2] {\em Inverse hyperbolic blocks}
\\ The eigenvalues are 
$(-\,e^{\tilde{\omega}},,-\,e^{-\,\tilde{\omega}})$, $\tilde{\omega}$ real and
\begin{eqnarray}
\mon_{\mathrm{i.h.}}\,=\,\left[\begin{array}{cc}
-\,e^{\tilde{\omega}} \quad & \quad 0
\\
0 \quad & -\,e^{-\,\tilde{\omega}}
\end{array}\right]
\,,\quad\quad \sym_{\mathrm{i.h.}}\,=\,\left[\begin{array}{cc}
0 \quad & \quad -1
\\
1 \quad & \quad 0
\end{array}\right]
\label{Hamilton-periodic:inversehyperbolic}
\end{eqnarray}
Inverse hyperbolic blocks occur because the monodromy matrix may not have 
a real logarithm for all odd iterates \cite{LeboeufMouchet}.
\item[3] {\em Elliptic blocks}
\\The eigenvalues are 
$(e^{i\,\omega}\,,e^{-\,i\,\omega})$ with $\omega$ real.
The normal forms are 
\begin{eqnarray}
&&\mon_{\mathrm{ell.}}\,=\,\mathsf{R}_2(-\,\omega)
\,,\qquad\qquad \sym_{\mathrm{ell.}}\,=\,\frac{d\mathsf{R}_2}{d\omega}(0)\,=\,
\left[\begin{array}{cc}
0 \quad & \quad -1
\\
1 \quad & \quad 0
\end{array}\right]
\nonumber\\
&&\mathsf{R}_2(-\,\omega)\,:=\,\left[\begin{array}{cc}
\cos{\omega} \quad & \quad \sin{\omega}
\\
-\,\sin{\omega} \quad & \quad  \cos{\omega}
\end{array}\right]
\label{Hamilton-periodic:elliptic}
\end{eqnarray}
The eigenvectors 
\begin{eqnarray}
\mon\,\mathfrak{e}_{\omega}\,=\,e^{i\,\omega}\,\mathfrak{e}_{\omega}\,,
\quad \quad
\mon\,\mathfrak{f}_{-\,\omega}\,=\,e^{-\,i\,\omega}\,\mathfrak{f}_{-\,\omega}
\label{Hamilton-periodic:ellipticeigenvectors}
\end{eqnarray}
are at the same time orthogonal with respect to the skew and the standard scalar
products. Thus they satisfy
\begin{eqnarray}
\mathrm{sign}\left\{\frac{i}{2}\,
\mathfrak{e}_{\omega}^{\dagger}\,\sym\,\mathfrak{e}_{\omega}\,\right\}\,=\,1
\,,\quad\quad
\mathrm{sign}\left\{\frac{i}{2}\,
\mathfrak{f}_{-\,\omega}^{\dagger}\,\sym\,\mathfrak{f}_{-\,\omega}\,\right\}\,=\,-\,1
\label{Hamilton-periodic:ellipticnormalization}
\end{eqnarray}
The signature of the skew products is a constant of the motion known as 
{\em Krein invariant}. It provides an intrinsic characterisation of ``positive'' 
and ``negative''  frequencies \cite{Krein,GelfandLidskii,Moser,Ekeland}.
\item[4] {\em Loxodromic blocks}
\\ The eigenvalues are 
$(e^{\tilde{\omega}+i\,\omega}\,,e^{-\,\tilde{\omega}-i\,\omega}\,,
e^{\tilde{\omega}-i\,\omega}\,,e^{-\,\tilde{\omega}+i\,\omega}\,)$, 
$\tilde{\omega}$, $\omega$  real 
\begin{eqnarray}
&&\mon_{\mathrm{lox.}}\,=\,\left[\begin{array}{cc}
e^{\tilde{\omega}}\,\mathsf{R}_2(-\,\omega) \quad & \quad 0
\\
0 \quad & \quad e^{-\,\tilde{\omega}}\mathsf{R}_2(-\,\omega)
\end{array}\right]
\nonumber\\
&&\sym_{\mathrm{lox.}}\,=\,\left[\begin{array}{cc}
0 \quad & \quad -\id_2
\\
\id_2 \quad & \quad 0
\end{array}\right]
\label{Hamilton-periodic:loxodromic}
\end{eqnarray}
In a one parameter family of real linear Hamiltonian systems 
the creation of a loxodromic block requires the simultaneous 
exit of four eigenvalues from the real or imaginary axis. 
Such bifurcations will occur only at isolated values of the 
parameter corresponding to points where two pairs of hyperbolic 
or elliptic eigenvalues coalesce.
\item[5] {\em Parabolic blocks}
\\
A parabolic block comprises two degenerate unit eigenvalues paired with an 
eigenvector and a generalised eigenvector:
\begin{eqnarray}
\mon_{\mathrm{par.}}\,=\,\left[\begin{array}{cc}
1 \quad & \quad \kappa
\\
0 \quad & \quad 1
\end{array}\right]
\,,\quad\quad \sym_{\mathrm{par.}}\,=\,\left[\begin{array}{cc}
0 \quad & \quad -\,1
\\
1 \quad & \quad 0
\end{array}\right]
\label{Hamilton-periodic:parabolic}
\end{eqnarray}
The non diagonal entry $\kappa$ is fixed by the skew product of the 
elements of the basis yielding the normal form. \\
From the differential equations point of view, a parabolic block 
is the value over (a multiple of) the period of a marginally unstable
block of the linear flow $\fu$. Such block is spanned by linear combinations 
of periodic vector fields with coefficients polynomial in time 
\cite{Poincare,DeWittMoretteMaheshwariNelson}. 
A simple eigenvalue degeneration corresponds to the evolution in phase space 
of a periodic $x_1(t)$ and a non periodic $x_2(t)$ Jacobi fields  
\begin{eqnarray}
&&\fu(t,T')\,x_{1}(T')=x_{1}(t)
\nonumber\\
&&\fu(t,T')\,x_{2}(T')=x_{2}(t)=
\kappa\,\frac{t-T'}{T_{\cl}}x_{1}(t)
+y(t)
\label{Hamilton-periodic:invariant}
\end{eqnarray}
The conservation of the skew product
\begin{eqnarray}
1=(x_{2}^{\dagger}\,\sy\,x_{1})(t)\,,\qquad\qquad\forall\,t
\label{Hamilton-periodic:alltimes}
\end{eqnarray}
enforces the normalisation condition whence (\ref{Hamilton-periodic:parabolic})
follows. 
Parabolic blocks are a generic feature of systems with continuous symmetries.
Jordan blocks also appear in the presence of unstable degenerate 
eigenvalues different from unity \cite{Poincare,GelfandLidskii,Moser}. 
\end{enumerate}
Hyperbolic and loxodromic blocks characterise unstable directions of periodic 
orbits. Parabolic blocks are marginally unstable and exhibit a linear growth of 
the perturbation along the direction spanned by the generalised eigenvector. Finally
elliptic blocks describe under certain conditions stability \cite{GelfandLidskii,Moser}. 
If the characteristic frequencies $\{\omega_n\}_{n=1}^{N\,\leq\,d}$ of the 
elliptic blocks are mutually irrational the linear
flow is stable. Furthermore, Moser has proven in ref. \cite{Moser} that a 
periodic orbit  is {\em almost} stable versus nonlinear perturbations if 
the linearised flow is elliptic with mutually irrational frequencies.
Almost stable here means that there exists a formal power series (possibly divergent)
supplying a Ljapunov function for the periodic orbit. In consequence a parametric 
perturbation of the periodic orbit will remain for extremely
long times in the neighborhood of the periodic orbit.  

Finally it is worth stressing that a block contributing to the monodromy with a 
certain stability may change stability at generic times \cite{Ekeland}. 
As a matter of fact, the symplectic property requires the eigenvalues of a linear 
Hamiltonian flow to be continuous under time evolution but does not rule out 
discontinuities of the derivatives. Inverse hyperbolic blocks are generated by the 
time evolution from bifurcations of unstable elliptic blocks. 
The phenomenon is exemplified by the dynamics of a linear Hamiltonian system with 
Hamilton matrix periodic and positive definite. Rule out the trivial example 
of the harmonic oscillators and assume:
\begin{eqnarray}
\frac{d}{dt} [\delta x_{\cl}^{\dagger}\,\mathsf{H}\, \delta x_{\cl}]
= \delta x_{\cl}^{\dagger}\,\frac{d\mathsf{H}}{dt}\, \delta x_{\cl}\,>\,0\,,
\qquad\qquad \forall t
\label{Hamilton-periodic:convexity}
\end{eqnarray}
At initial time the eigenvalues are equal to one. Since the assumption
(\ref{Hamilton-periodic:convexity}) forbids the formation of hyperbolic blocks
at unity, as time increases the eigenvalues can only move on the unit circle.
In particular Krein positive eigenvalues 
(\ref{Hamilton-periodic:ellipticnormalization})
move counterclockwise on the upper half unit-circle and the Krein negative
ones on the lower half-circle. 
If two eigenvalues with opposite Krein signature meet at 
minus unity, they can satisfy the symplectic condition in two ways. 
They can cross each other and continue their motion on the unit circle or 
they can leave the unit circle and start moving on the negative semi-axis. 
Note that the second option is ruled out for eigenvalues with the same
Krein signature. A bifurcation to inverse hyperbolic would imply in that case
a change of the overall Krein signature. The eigenvalues will at some later time come 
back to minus unity and resume their motion on the unit circle, 
with Krein-negative ones now moving clockwise on the upper-half circle. 
All of that can happen for a stable $T_{\cl}$-periodic system: the eigenvalues must 
only be back on the unit circle at times multiple of the prime period. 
More information can be found in \cite{GelfandLidskii,Moser,Ekeland,YakubovichStarzhinskii}.

\subsubsection{Eigenvalue flow around a parabolic block}
\label{Hamilton:perturbative}

Consider a one parameter $\tau$ family of linear $T_{\cl}$-periodic Hamiltonian
systems
\begin{eqnarray}
&&\sy(t)\,\frac{d \fu_{\tau}}{dt\,}(t,t')\,=\,\mathsf{H}_{\tau}(t)\,\fu_{\tau}(t,t')\,,
\nonumber \\
&&\fu_{\tau}(t',t')\,=\,\id_{2\,d}
\label{Hamilton-perturbative:equation}
\end{eqnarray}
with analytic dependence in $\tau$. In what follows at $\tau$ equal zero 
the parametric dependence will be simply omitted.
At fixed times the linear flow evolves in $\tau$ according
to a linear Hamiltonian equation
\begin{eqnarray}
\sy(t)\,\frac{\de \fu_{\tau}}{\de \tau\,}(t,T)\,=\int_{T}^{t}\!\!dt'\,
\fu^{-1\,\dagger}_{\tau}(t,t')\,\delta\mathsf{H}(t')\,\fu^{-1}_{\tau}(t,t')\,
\fu_{\tau}(t,T)
\label{Hamilton-perturbative:curve}
\end{eqnarray}
with
\begin{eqnarray}
\delta\mathsf{H}(t)=\mathsf{H}_{\tau}(t)-\mathsf{H}(t)
\label{Hamilton-perturbative:Hamiltonian}
\end{eqnarray}
The parametric stability is determined by the $\tau$ dependence 
of the eigenvalues of the monodromy matrix. Up to first order 
accuracy in $\tau$, this latter is 
\begin{eqnarray}
\mon_\tau = \mon+\tau\,\,\sym^{\,\dagger}\!\!\int_{T}^{T\!+\!T_{\cl}}\!\!\!dt\,
\fu^{-1\,\dagger}(T\!+\!T_{\cl},t)\,\left.\frac{\de \mathsf{H}_{\tau}}{\de \tau}(t)
\right|_{\tau=0}\!\!\!\!\!\!\!\fu^{-1}(T\!+\!T_{\cl},t)\,\mon
\label{Hamilton-perturbative:monodromy}
\end{eqnarray}
with
\begin{eqnarray}
\sym=\Omega(T)=\Omega(T+T_{\cl})
\label{Hamilton-perturbative:hypothesis}
\end{eqnarray}
The leading correction to the eigenvalues is obtained projecting the
above equation on the unperturbed eigenvectors. 

An isolated zero mode of the periodic Sturm-Liouville fluctuation operator 
pairs up with an elementary parabolic block of the monodromy
as (\ref{Hamilton-periodic:parabolic}). Multiple zero modes brought about by Abelian 
symmetries also decouple to form an equal number of elementary parabolic blocks.
For generic $\delta \mathsf{H}$, the qualitative effect of the perturbation in $\tau$ 
in such cases is to shift away from unity the eigenvalue pair of the parabolic 
block. Since the symplectic structure is preserved, a parabolic block 
generically bifurcates into an elliptic or hyperbolic one. 
Groups of unit eigenvalues of the monodromy corresponding to non-Abelian 
symmetries or marginal degenerations can also be analysed in a similar way 
\cite{GelfandLidskii}.

The right periodic eigenvector is the phase space lift 
of a periodic Jacobi field, a zero mode of $\foP$ in $[T,T+T_{\cl}]$. 
If the zero mode stems from a continuous symmetry according to N\"other theorem 
in Darboux coordinates it will have the form  
\begin{eqnarray}
\mathfrak{r}_1\,=\,\left[\begin{array}{c}
\delta_{\mo}q_{\cl}(T) \\ \nabla \delta_{\mo}q_{\cl}(T)
\end{array}\right]
\label{Hamilton-perturbative:righteigenvector}
\end{eqnarray}
$\delta_{\mo}q_{\cl}$ is the vector field induced by the infinitesimal
generator of the symmetry transformation.
Together with a generalised eigenvector $\mathfrak{r}_2$,
(\ref{Hamilton-perturbative:righteigenvector}) specifies a dual basis:
\begin{eqnarray}
\begin{array}{l}
\mon\,\mathfrak{r}_1\,=\,\mathfrak{r}_1
\\
\mon\,\mathfrak{r}_2\,=\,\mathfrak{r}_2+\kappa\,\mathfrak{r}_1
\end{array}
\quad \Rightarrow \quad
\begin{array}{l}
\mon^{\dagger}\,\sym\,\mathfrak{r}_1\,=\,\sym\,\mathfrak{r}_1
\\
\mon^{\dagger}\,\sym\,\mathfrak{r}_2\,=\,\sym\,\mathfrak{r}_2-\kappa\,\sym\,\mathfrak{r}_1
\end{array}
\label{Hamilton-perturbative:lefteigenvectors}
\end{eqnarray}
whence it follows
\begin{eqnarray}
\mathfrak{l}_1\,=\,-\,\sym\,\mathfrak{r}_2\,
\quad\quad\,\mathfrak{l}_2\,=\,\sym\,\mathfrak{r}_1
\label{Hamilton-perturbative:righteigenvectors}
\end{eqnarray}
provided
\begin{eqnarray}
\mathfrak{l}_1^{\dagger}\,\mathfrak{r}_1\,\equiv\,
\mathfrak{r}_2^{\dagger}\,\sym\,\mathfrak{r}_1\,=\,1
\label{Hamilton-perturbative:skewproduct}
\end{eqnarray}
Here, the class of perturbations of interest comprises those ones which do not change 
the Morse index of the Sturm-Liouville operator associated to the second variation
of the Lagrangian around the periodic orbit.
Any strictly positive definite $T_{\cl}$-periodic term $\dU(t)$ added to 
the potential in the Sturm-Liouville operator
\begin{eqnarray}
\flu^{\dagger}\,\potau\,\flu=\,\flu^{\dagger}[\pot\,+\,\tau\,\dU]\,\flu
\end{eqnarray}
implements the condition: 
\begin{eqnarray}
\int_{T}^{T+T_{\cl}}\!\!\!\!\!dt\,
(\flu^{\dagger} \dU\,\flu)(t)\,>\,0\,,\qquad\qquad\qquad \forall\,\flu(t)
\label{Hamilton-perturbative:Lagrangianshift}
\end{eqnarray}
The perturbation of the Lagrangian introduces in Darboux $(q\,,p)$ coordinates 
the Hamiltonian perturbation
\begin{eqnarray}
\delta \mathsf{H}(t;\tau)\,=\,\tau\,\left[\begin{array}{cc} 
-\,\dU(t) \quad & \quad 0 \\
0\quad &\quad 0
\end{array}\right]
\label{Hamilton-perturbative:perturbation}
\end{eqnarray}
Up to leading order the eigenvalue equation for the perturbed parabolic block 
reduces to
\begin{eqnarray}
0\,=\,\mathrm{det}\,\left[\begin{array}{cc} 
1-\mathsf{m}(\tau) \quad & \quad \kappa
\\
\tau\, V 
\quad & \quad 1-\mathsf{m}(\tau)
\end{array}
\right]+O(\tau^{2})
\label{Hamilton-perturbative:characteristic}
\end{eqnarray}
with $V$ specified by
\begin{eqnarray}
V&:=&\mathfrak{l}_2^{\dagger}\,\left[
\sym^{\dagger}\!\int_{T}^{T\!+\!T_{\cl}}\!\!\!\!dt\,
\fu^{-1\,\dagger}(T\!+\!T_{\cl},t)\,\left.\frac{\de \mathsf{H}}{\de \tau}(t)\right|_{\tau=0}
\!\!\!\!\!\!\fu^{-1}(T\!+\!T_{\cl},t)\,\mon\right]\mathfrak{r}_1
\nonumber\\
&=&\int_{T}^{T\!+\!T_{\cl}}\!\!\!dt\,(\delta_{\mo}q_{\cl}^{\dagger}
\dU\,\delta_{\mo}q_{\cl})(t)
\label{Hamilton-perturbative:correction}
\end{eqnarray}
The eigenvalues are
\begin{eqnarray}
\mathsf{m}_{\pm}(\tau)&=&1\,\pm\,
\sqrt{|\tau\,\,\kappa\,V|}\,e^{i\,\pi\,\frac{1-\mathrm{sign}(V\,\kappa)}{4}}
+O(\tau)\,
\nonumber\\
&\sim&\exp\left\{\pm\,\sqrt{|\tau\,\kappa\,V|}
\,e^{i\,\pi\,\frac{1-\mathrm{sign}(V\,\kappa)}{4}}\right\}+O(\tau)
\label{Hamilton-perturbative:eigenvalues}
\end{eqnarray}
The exponentiation is legitimate because of the constraint imposed by the
symplectic structure. It determines the phase factor in the exponential 
modulo $2\,\pi$.

There are two main lessons to be drawn from (\ref{Hamilton-perturbative:eigenvalues}).
First, the characteristic frequency has a power expansion in $\sqrt{\tau}$ and not
in $\tau$ as the monodromy matrix. This is not surprising as the frequency of
an harmonic oscillator appears quadratically in the equations of the motion.
The eigenvalue split induced by an Hamiltonian perturbation on blocks of higher 
eigenvalue degeneration $N$ is amenable to the power series expansion \cite{GelfandLidskii,BenderOrzsag}
\begin{eqnarray}
\mathsf{m}_k(\tau)\,=\,\mathsf{m}_k+\Sigma_{n=1}^{\infty}\,\mathsf{m}_k^{(n)}\,
\tau^{\frac{n}{N}}\,,\quad\quad k\,=\,1,...,N
\label{Hamilton-perturbative:series}
\end{eqnarray}
The second lesson is that the nature of the block emerging from the bifurcation is invariantly
determined by the signature of the product $V\,\kappa$. According to the conventions 
adopted for the symplectic pseudo-metric, $V$ is positive definite and the signature of 
the non-diagonal element in the parabolic block directly specify the bifurcation. \\
The physical meaning of the non diagonal element in the parabolic block
is better realised by considering the ``default'' zero mode encountered
in the Gutzwiller trace formula.\\
By energy conservation classical periodic orbits occur on closed curves 
of level of the Hamiltonian $E=\ha(x)$.
Periodic orbits are then expected to appear in one parameter families the period 
$T_{\cl}(E)$ whereof smoothly depends on the energy.
In such a case the periodic eigenvector of the monodromy matrix 
is the phase space ``velocity'' of the classical trajectory
\begin{eqnarray}
\mathfrak{r}_1\,=\,\dot{x}_{\cl}(T\,;\,E,...)
\label{Hamilton-perturbative:velocitylift}
\end{eqnarray}
In the absence of further conservation laws $x_{\cl}(t)$ is generically 
the only periodic eigenvector in $[T,T+T_{\cl}]$.
The associated generalised eigenvector is found by observing that for all $t$:
\begin{eqnarray}
\dot{x}_{\cl}(t+T_{\cl}(E,...)\,;\,E,...)\,=\,\dot{x}_{\cl}(t\,;\,E,...)
\label{Hamilton-perturbative:periodicity}
\end{eqnarray}
The derivative with respect to the energy yields the phase space lift of 
a non periodic Jacobi field. One can identify in (\ref{Hamilton-periodic:invariant}) 
\begin{eqnarray}
x_{2}(t)=\frac{\de x_{\cl}}{\de E\,}(t)\,,
\qquad\qquad
y(t)=\left.\frac{\de x_{\cl}}{\de E\,}(t)\right|_{\frac{t-T'}{T_{\cl}}=\mathrm{constant}}
\label{Hamilton-periodic:identifications}
\end{eqnarray}
the second vector field being periodic by construction. Thus $y(T)$ 
provides the sought generalised eigenvector $\mathfrak{r}_2$.
Furthermore energy conservation along the trajectory 
\begin{eqnarray}
1\,=\,\left(\frac{\de \ha}{\de E}\right)(x_{\cl}(t))\,=\,
\left(\frac{\de x_{\cl}}{\de E}^{\dagger}\,\sym\,\dot{x}_{\cl}\right)(t)\,,\qquad\qquad \forall\,t
\label{Hamilton-perturbative:parabolicnormalization}
\end{eqnarray}
yields for the family of periodic orbits 
\begin{eqnarray}
\kappa(E)\,=\,-\,\frac{d T_{\cl}}{d E}(E) 
\label{Hamilton-perturbative:nondiagonal}
\end{eqnarray} 
Hence the non diagonal element of the parabolic block is equal to the variation
in period of two orbits, infinitesimally separated in energy.

%% file: Lie.tex
\section{Lie's first fundamental theorem}
\label{Lie}

A Lie group is a smooth manifold $G$ on which the group operations of 
\begin{eqnarray}
\begin{array}{lll}
\mbox{product}:\quad&\quad G\,\times \,G \,\rightarrow\, G \quad&\quad 
\mathsf{g}(\mo)\cdot\mathsf{g}(\mos)=\mathsf{g}(f(\mo,\mos))
\\
\mbox{inverse} : \quad&\quad G\,\rightarrow\, G \quad&\quad 
\mathsf{g}(\mo)\cdot\mathsf{g}(\mo^{(-1)})=\mathsf{g}(0)=\mbox{identity}
\end{array}
\end{eqnarray}
are defined.  
The  analytic mapping $f=(f^1,..,f^N)$ governs the composition law governing the
group operations
\begin{eqnarray}
&&\mo^{a}=f^a(\mo,0)=f^a(0,\mo)
\nonumber\\
&&f^a(\mo,f(\mos,\mov))=f^a(f(\mo,\mos),\mov)
\label{Lie:composition}
\end{eqnarray}
A Lie group acts on a $d$-dimensional configuration space $\mf$ through 
a smooth mapping $\varphi$
\begin{eqnarray}
\varphi: G\,\times\,\mf\,\rightarrow \mf
\end{eqnarray}
The mapping induces transformation laws of point coordinates of $\mf$ 
\begin{eqnarray}
 q^{\alpha}_{[\mo]}= \varphi^{\alpha}(q,\mo)
\label{Lie:transformation}
\end{eqnarray}
The origin in the $\{\mo\}_{a=1}^{N}$ space is chosen to correspond to the 
identity transformation on $\mf$
\begin{eqnarray}
\varphi^{\alpha}(q,0)=q^{\alpha}
\end{eqnarray}
Lie's first fundamental theorem \cite{Gilmore} relates the derivatives of (\ref{Lie:transformation})
at a generic point $\mo$ of $G$ to the vector fields induced on $\mf$ by the 
infinitesimal generators of the group transformations.
\begin{eqnarray}
\left. \frac{\de \varphi^{\alpha}}{\de \mo^a}(q,\mo)\right|_{\mo=0}:=v^{\alpha}_{a}(q)
\label{Lie:infinitesimal}
\end{eqnarray}
The group composition law (\ref{Lie:composition}) permits to use 
both left or right infinitesimal translations at $\mo$.
A left translation at $\mo$ is 
\begin{eqnarray}
\varphi^{\alpha}(\varphi(q,\mo),\mos)=
\varphi^{\alpha}(q,f(\mos,\mo))
\label{Lie:momos}
\end{eqnarray}
Differentiating both sides with respect to the $\mos$'s in zero yields
\begin{eqnarray}
v^{\alpha}_{a}(\varphi(q,\mo))=
\left.\frac{\de f^{b}}{\de \mos^a}(\mos,\mo)\right|_{\mos=0} \,
\frac{\de \varphi^{\alpha}}{\de \mo^b}(q,\mo)
\label{Lie:leftder}
\end{eqnarray}
The matrix 
\begin{eqnarray}
(\mathfrak{L}^{-1})^{b}_{a}(\mo):=\left.\frac{\de f^{b}}{\de \mos^a}(\mos,\mo)\right|_{\mos=0} 
\label{Lie:left}
\end{eqnarray}
characterises the infinitesimal left-translation.\\
Analogously, exchanging $\mo$ with $\mos$ in (\ref{Lie:momos}) it is possible to
express the derivatives of group transformations at $\mo$ in terms of 
an infinitesimal right translation
\begin{eqnarray}
\frac{\de \varphi^{\alpha}}{\de q^\beta}(q,\mo)\, v^{\beta}_{a}(q)=
\left.\frac{\de f^{b}}{\de \mos^a}(\mo,\mos)\right|_{\mos=0}\, 
\frac{\de \varphi^{\alpha}}{\de \mo^b}(q,\mo):=(\mathfrak{R}^{-1})^{b}_{a}(\mo)\,
\frac{\de \varphi^{\alpha}}{\de \mo^b}(q,\mo)
\label{Lie:rightder}
\end{eqnarray}
Comparing the two expressions (\ref{Lie:leftder}), (\ref{Lie:rightder}) one finds
\begin{eqnarray}
\frac{\de \varphi^{\alpha}}{\de \mo^a}(q,\mo)=\mathfrak{R}^{b}_{a}(\mo)
\frac{\de \varphi^{\alpha}}{\de q^\beta}(q,\mo)\, v^{\beta}_{b}(q)=
\mathfrak{L}^{b}_{a}(\mo)\,v^{\alpha}_{b}(\varphi(q,\mo))
\label{Lie:derivative}
\end{eqnarray}
and therefore
\begin{eqnarray}
(\mathfrak{R}^{-1})^{b}_{a}(\mo)\,\mathfrak{L}^{c}_{b}(\mo)\, v^{\alpha}_{c}(\varphi)=
\frac{\de \varphi^{\alpha}}{\de q^\beta}\, v^{\beta}_{a}(q)
\label{Lie:consistency}
\end{eqnarray}
This latter equality specifies the {\em adjoint} representation of the action of the group.
Namely a right translation can be represented in the guise of a left 
translation:
\begin{eqnarray}
\varphi^{\alpha}(\varphi(q,\mos),\mo)=\varphi^{\alpha}(\varphi(q,\mo),f(\mo,f(\mos,\mo^{(-1)}))
\label{Lie:leftright}
\end{eqnarray}
exploiting the expression of the identity
\begin{eqnarray}
q^{\alpha}=\varphi^{\alpha}(\varphi(q,\mo),\mo^{(-1)})
\end{eqnarray}
Differentiation at $\mos$ equal zero yields:
\begin{eqnarray}
\mathrm{Ad}^{b}_{a}(\mo)=\left.\frac{\de f^{b}}{\de \mo^{a}}(\mo,f(\mos,\mo^{(-1)})\right|_{\mos=0}:=
(\mathfrak{R}^{-1})^{c}_{a}(\mo)\,\mathfrak{L}^{b}_{c}(\mo)
\label{Lie:adjoint}
\end{eqnarray}
From the representation (\ref{Lie:derivative}) of the derivatives of group transformations
it is straightforward to derive the structure constants of the group.
To wit, the existence of a global parametrisation of group transformations in terms of
the variable $\mo$ requires
\begin{eqnarray}
\frac{\de^{2} \varphi^{\alpha}}{\de \mo^{a}\de \mo^b}=
\frac{\de^{2} \varphi^{\alpha}}{\de \mo^{b}\de \mo^a}
\end{eqnarray} 
The identity implies for infinitesimal left translations
\begin{eqnarray}
\mathfrak{L}^{a'}_{a}\,\mathfrak{L}^{b'}_{b}\,
\left[\,v^{\beta}_{b'}\,\frac{\de v^{\alpha}_{a'}}{\de \varphi^{\beta}}-
v^{\beta}_{a'}\,\frac{\de v^{\alpha}_{b'}}{\de \varphi^{\beta}}\right]=
-\left[\frac{\de \mathfrak{L}^{c}_{a}}{\de \mo^{b}}-
\frac{\de \mathfrak{L}^{c}_{b}}{\de \mo^{a}}\,\right]\,v_{c}^{\alpha}
\label{Lie:separation}
\end{eqnarray}
The vector fields $v^{\alpha}_{a}$ do not depend explicitly on the group coordinates 
$\mo$. Hence the condition (\ref{Lie:separation}) admits solution if and only if it is possible to
separate the variables. In other words there must exist some constants $C^{d}_{a\,b}$
such that
\begin{eqnarray}
\left[\frac{\de \mathfrak{L}^{c}_{a}}{\de \mo^{b}}-
\frac{\de \mathfrak{L}^{c}_{b}}{\de \mo^{a}}\right]=- C^{c}_{a'\,b'}
\mathfrak{L}^{a'}_{a}\,\mathfrak{L}^{b'}_{b}
\label{Lie:MaurerCartan}
\end{eqnarray}
is satisfied.
The constants $C^{c}_{a\,b}$ are the structure constant of the group while (\ref{Lie:MaurerCartan}) 
are the Maurer-Cartan structure equations.

Finally, in order to prove that (\ref{moduli-moduli:insertion}) is the invariant measure of 
the group one observes that if
\begin{eqnarray}
&&\varphi^{\alpha}(q,\mo)=\varphi^{\alpha}(\varphi(q,\mov),\mos)
\nonumber\\
&&\mo^a=f^a(\mos,\mov)
\end{eqnarray}
differentiating the first equation with respect to $\mos^b$ yields
\begin{eqnarray}
\frac{d f^{b}}{d \mos^a}\mathfrak{L}_{b}^{c}(f) v_{c}(\varphi^{\alpha}(q,\mo))
=\mathfrak{L}_{a}^{b}(\mos)v_{b}(\varphi(\varphi(q,\mov),\mos))
\end{eqnarray}
Therefore one gets into 
\begin{eqnarray}
\frac{d f^{c}}{d \mos^a}\mathfrak{L}_{c}^{b}(f)=\mathfrak{L}_{a}^{b}(\mos)
\end{eqnarray}
or equivalently
\begin{eqnarray}
d \mo^b \mathfrak{L}_{b}^{a}(\mo)=d \mos^{b} \mathfrak{L}_{b}^{a}(\mos)
\end{eqnarray}
whence it follows that 
\begin{eqnarray}
dG=\prod_{a=1}^{N}d\mo^{a}\,\det\mathfrak{L}(\mo)
\end{eqnarray}
is the right invariant measure. However for a compact connected group the
invariant measure is unique (up to a constant factor) and therefore also
\begin{eqnarray}
\det\mathfrak{L}(\mo)\propto\det\mathfrak{R}(\mo)
\end{eqnarray}
must hold.

More details can be found in \cite{Frankel,Gilmore,Nakahara,SattingerWeaver}.


%% file: fresnel.tex
\section{Fresnel Integrals}
\label{Fresnel}

The paradigm of Fresnel integrals is provided by the one dimensional case 
\begin{eqnarray}
\iota(z)=\int_\mathrm{R}\frac{dq}{\sqrt{2\,\pi}\,e^{i\frac{\pi}{4}}}
\,e^{i\,\frac{z\,q^2}{2}}
\label{Fresnel:paradigm}
\end{eqnarray}
with $z$ a real number.
The integral is not absolutely convergent since the integrand is
in modulo equal to one. Nevertheless intuitively one can hope that the 
integral converges on the real axis due to the increasingly fast oscillations 
of the integrand.\\
A quantitative analysis can be performed on the complex $q$-plane.
Since the integrand is even, it is enough to consider the first quadrant 
of the complex plane, $\mathrm{R}_{+}\,\times\,i\,\mathrm{R}_+$. 
The integral can be made absolutely convergent if
\begin{eqnarray}
\mbox{Re}\left\{i\,z\,q^2\right\}\,<\,0 \Rightarrow 
\cos\left(2\,\arg q +\arg z+\frac{\pi}{2}\right)\,<\,0
\label{Fresnel:absoluteconvergence}
\end{eqnarray}
A Gaussian integral is recovered each time 
\begin{eqnarray}
2\,\arg q +\arg z+\frac{\pi}{2}=\pi
\nonumber
\end{eqnarray}
The absence of poles in the domain of convergence in the complex plane
permits to enclose the Fresnel integral into a null circuit receiving 
the other non-vanishing contribution from a Gaussian integral:
\begin{eqnarray}
0=\oint dq\,e^{i\,\frac{z\,\,q^2}{2}}=\int_{\mathrm{R}_+}dq\,e^{i\,
\frac{z\,q^2}{2}}-e^{i\frac{\pi}{4}}
\,\int_{\mathrm{R}_+}d|q|\,e^{-\frac{z\,|q|^2}{2}}\,\quad &\mbox{if}&\quad \arg z\,=\,0
\nonumber\\
0=\oint dq\,e^{i\,\frac{\hbar\,q^2}{2}}=e^{-i \frac{\pi}{4}}\,\int_{\mathrm{R}_+}d|q|\,
e^{\frac{z\,|q|^2}{2}}-\int_{\mathrm{R}_+}dq\,e^{i\,\frac{z\,q^2}{2}}\,\quad &\mbox{if}&
\quad \arg z\,=\,\pi
\label{Fresnel:circuit}
\end{eqnarray}
The final result is
\begin{eqnarray}
\iota(z)=\int_\mathrm{R}\frac{dq}{\sqrt{2\,\pi}\,e^{i\frac{\pi}{4}}}
\,e^{i\,\frac{z\,q^2}{2}}=\sqrt{\frac{2\,\pi}{|z|}}\,e^{-i \,\pi
\frac{1-\mathrm{sign}z}{4}}
\label{Fresnel:result}
\end{eqnarray}
The multidimensional generalization is
\begin{eqnarray}
\int_{\mathrm{R}^N}\frac{d^Nq}{(2\,\pi)^{N/2}\,e^{i\,d\frac{\pi}{4}}}\, 
e^{i\,\frac {q^\dagger\,\,\mathsf{L}^{(N)}\,q}{2}}=\frac{e^{-i\,\frac{\pi}{2}
\ind\mathsf{L}^{(N)}}}{\sqrt{\left|\det \mathsf{L}^{(N)}\right|}}
\label{Fresnel:multidimensional}
\end{eqnarray}
$\ind\mathsf{L}^{(N)}$ being the number of negative eigenvalues of the symmetric 
matrix $\mathsf{L}^{(N)}$. 

Quadratic path integrals are the continuum limit of a lattice Fresnel integral
obtained from the discretisation of the action
\begin{eqnarray}
\ac^{(N)}=\sum_n \Delta t\, \la^{(N)}_n
\label{Fresnel:action}
\end{eqnarray}
defined by mid-point rule 
\begin{eqnarray}
\la^{(N)}_n&=&\frac{1}{2}\left\{[\flu(n)-\flu(n-1)]^{\dagger}\,\mass(n)\,
[\flu(n)-\flu(n-1)]
\right.\nonumber\\
&&+2\,[\flu(n)-\flu(n-1)]^{\dagger}\,\vpt(n)\,
\frac{\flu(n)+\flu(n-1)}{2}
\nonumber\\
&&\left.
+\frac{[\flu(n)+\flu(n-1)]^{\dagger}}{2}\,\pot(n)\,\frac{\flu(n)+\flu(n-1)}{2}
\right\}
\label{Fresnel:Lagrangian}
\end{eqnarray}
with
\begin{eqnarray}
\flu(n)\,\equiv\,\flu\left(T'+n\,\Delta t\right)\,,\quad\quad \Delta t\,=\,\frac{T-T'}{N}
\nonumber
\end{eqnarray}
The continuum limit does not depend on the discretisation of the potential term.
The form (\ref{Fresnel:multidimensional}) of the Fresnel integral is attained
by collecting the quantum fluctuation in a single $N d$-dimensional vector $\Delta Q$:
\begin{eqnarray}
&&\ac^{(N)}=\Delta Q^{\dagger}\,\mathsf{L}^{(N)}\,\Delta Q
\nonumber\\
&&\delta Q^{\dagger}=\left[ \flu^1(0) \,,...,\flu^d(0),..., \flu^1(N)\,, .\, . \, .\,, \flu^d(N)\right]
\label{Fresnel:vector}
\end{eqnarray}
The matrix $\mathsf{L}^{(N)}$ depends both on the discretisation and the lattice 
boundary conditions. For example, a one dimensional harmonic oscillator with periodic
boundary conditions yields
\begin{eqnarray} 
\mathsf{L}^{(N)}=\left[
\begin{array}{cccccc} 
2 +\omega^2 \, &\,-\,1 \,&\, 0\, &\, 0 \,& ... & \,-\,1 \\
-\,1 \,&\,  2 +\omega^2 \,&\,-\, 1 \,&\, 0 \,& ... &\, 0 \\
... \, &\, ...\, &\, ...\, &\, ... \,&\, ... \,&\, ... \, \\
... \,&\, ... \,&\, ... \,&\, ... \,&\, ... \,&\, ... \\
 -\,1 \,&\,  0  \,&\,  0  \,&\, ... \,&\,-\,1 \,&\,  2+\omega^2
\end{array}\right]
\label{Fresnel:periodic}
\end{eqnarray}
From (\ref{Fresnel:action}) it follows that the lattice path integral is finally
\begin{eqnarray}
\iota^{(N)}(\mathfrak{B})=
\int_{\mathbf{R}^d}\Pi_{n=1}^{N}
\left[\frac{d^d\flu(n)}{(2\,\pi\,\hbar\,\Delta t)^{d/2}\,e^{i\,d\,\frac{\pi}{4}}}\right]\,
e^{\frac{i\,\Delta t}{2\,\hbar} \delta Q^{\dagger}\,\mathsf{L}^{(N)}\,\delta Q}
\label{Fresnel:pathintegral}
\end{eqnarray}

\section{Some exact path integral formulae}
\label{formulae}

The stationary phase is exact for quadratic integrals. Namely its effect is to
decouple the classical from ``quantum'' contributions to the action functional.  
The propagator path integral becomes
\begin{eqnarray}
K(Q,T|Q',T')\,&=&\,e^{i\,\frac{\ac(q_{\cl})}{\hbar}}\,\int_{\flu(T')=\flu(T)=0}
\mes[\flu(t)]\,e^{i \frac{\ac(\sqrt{\hbar}\,\flu)}{\hbar}}
\nonumber\\
&=&e^{i\,\frac{\ac(q_{\cl})}{\hbar}}\,\iota(T,T')
\label{formulae:propagator}
\end{eqnarray}
All the spatial dependence is stored in the action function evaluated on the
classical trajectory $q_{\cl}$ matching the boundary conditions. The path integral 
$\iota(T,T')$ reduces to a pure function of the time interval. 
The observation allows to shortcut the analysis of the continuum limit by a self 
consistency argument. 

\subsubsection{Free particle propagator}
\label{formulae:free}

The free propagator is known to be
\begin{eqnarray}
K_{\mathrm{free}}(Q,T|Q',T')\,=\,
\frac{e^{\frac{i\,m\,(Q-Q')^{2}}{2\,\hbar\,(T-T')}-i\,\frac{d}{4}\,\pi}}
{(2\,\pi\,\hbar)^{\frac{d}{2}}}
\label{formulae-free:free}
\end{eqnarray}
Comparison with (\ref{formulae:propagator}) fixes the normalisation of the
path integral to
\begin{eqnarray}
K_{\mathrm{free}}(0,T|0,T')\,=\,\iota_{\mathrm{free}}(T,T')
\label{formulae-free:prefactor}
\end{eqnarray}
In consequence the general formula becomes  
\begin{eqnarray}
K(Q,T|Q',T')\,=\,e^{i\,\frac{\ac(q_{\cl})}{\hbar}}\,
K_{\mathrm{free}}(0,T|0,T')\,\frac{\iota(T,T')}{\iota_{\mathrm{free}}(T,T')}
\label{formulae-free:shortcut}
\end{eqnarray}

\subsubsection{One dimensional harmonic oscillator}
\label{formulae:oscar}

The evaluation of the action functional
\begin{eqnarray}
\ac=\int_{T'}^{T}dt\,\left[\frac{m}{2}\dot{q}^2-\frac{m\,\omega^{2}}{2}\,q^{2}\,\right] 
\label{formulae-oscar:harmonic}
\end{eqnarray}
on a classical trajectory connecting $Q'$ to $Q$ in $\Delta T=T-T'$ (open extremals)
yields 
\begin{eqnarray}
\ac_{\cl}(Q,T|Q',T')\,=\,\frac{m\,\omega}{2\,\sin(\omega\,\Delta T)}\left[(Q^2+Q^{\prime\,2})
\cos(\omega\,\Delta T)-2\,Q\,Q'\right]
\label{formulae-oscar:Dirichletaction}
\end{eqnarray}
Divergences are encountered at 
\begin{eqnarray}
\omega\,\Delta T\,=\,n\,\pi
\label{formulae-oscar:times}
\end{eqnarray}
The divergences signal the existence of periodic orbits. The quantum propagator
in such cases reduces to a Dirac $\delta$ function. \\
Quantum fluctuations around open extremals are governed by the Sturm-Liouville 
operator
\begin{eqnarray}
&&L\,=\,-\,\frac{d^{2}\,}{d t^{2}}-\omega^2
\nonumber\\
&&\flu(T')\,=\,\flu(T)=0
\label{formulae-oscar:Dirichlet}
\end{eqnarray}
The operator is diagonal in the complete, for Dirichlet boundary conditions, basis of odd 
Fourier harmonics
\begin{eqnarray}
\chi_n(t)\,=\,\sqrt{\frac{2}{\Delta T}}\,\sin\left(\frac{n\,t}{\Delta T}\right)\,,
\quad\quad n\,=\,1,2,...
\label{formulae-oscar:Dirichletbasis}
\end{eqnarray}
On a time lattice an infinite dimensional orthogonal matrix changes the variables of 
integration in (\ref{formulae:propagator}) from the quantum fluctuation $\delta q$ 
to the amplitudes of the Fourier decomposition. The effect of the quantum fluctuations 
is enclosed in the eigenvalue ratio
\begin{eqnarray}
\frac{\iota(T,T')}{\iota_{\mathrm{free}}(T,T')}&=&e^{i\,\frac{\pi}{2}\ind \foD([T',T])}
\Pi_{n=1}^{\infty}\left|1-\left(\frac{\omega\,\Delta T}{n\,\pi}\right)^2\right|
\nonumber\\
&=&e^{i\,\frac{\pi}{2}\ind \foD([T',T])}
\left|\frac{\sin \left(\omega\,\Delta T\right)}{\omega\,\Delta T}\right|
\label{formulae-oscar:Dirichletratio}
\end{eqnarray}
The last equality follows from the analytical continuation of the $\zeta$-function
\begin{eqnarray}
\zeta(s)=\Sigma_{j=0}^{\infty}\,\frac{1}{\left(n^2+\omega^2\right)^{s}}\,\quad\quad 
\rea\,s>0
\end{eqnarray}
The Morse index is the number of negative eigenvalues of the Sturm-Liouville operator
(\ref{formulae-oscar:Dirichlet})
\begin{eqnarray}
\ind \foD([T',T])\,=\,\#\left\{n\,|\,1-\left(\frac{\omega\,\Delta T}{n\,\pi}\right)^2
\,<\,0\right\}=\mathrm{Int}\left[\frac{\omega\,\Delta T}{\pi}\right]
\label{formulae-oscar:Dirichletindex}
\end{eqnarray}
The trace of the quantum harmonic oscillator propagator corresponds to the integral 
\begin{eqnarray}
\Tr&&\!\!\!\!K(T,T')
\nonumber\\
&=&\int_{\rn}\!\!dQ \left[\frac{m\,\omega}
{2\,\pi\,\hbar\,\left|\sin \left(\omega\,\Delta T\right)\right|}\right]^{\frac{1}{2}}
e^{\frac{i\,\ac_{\cl}(Q,T|Q',T')}{\hbar}-i\,\frac{\pi}{2}\,
\left[\ind \foD([T',T])+\frac{1}{2}\right]}
\nonumber\\
&=&\frac{1}{\left|2\,\sin\frac{\omega\,\Delta T}{2}\right|}\,
\,e^{-\frac{i\,\pi}{2}\,\left[\frac{\mathrm{sign}\sin \left(\omega\,\Delta T\right)+1}{2}+
\ind \foD([T',T])\right]}
\label{formulae-oscar:trace}
\end{eqnarray}
The trace is also the inverse square root of the determinant of the self-adjoint 
operator 
\begin{eqnarray}
&& L\,=\,-\,\frac{d^{2}\,}{d t^{2}}-\omega^2
\nonumber\\
&&\flu(T')\,=\,\flu(T)\,,\qquad\qquad \dflu(T')\,=\,\dflu(T)
\label{formulae-oscar:periodic}
\end{eqnarray}
The latter admits as eigenfunctions both odd and even harmonics of the Fourier basis 
in the interval $[T',T]$. The path integral measure becomes
\begin{eqnarray}  
\left.\mes[\flu(t)\right|_{\mathrm{Per.}}
=\frac{dc_{0}\,e^{-\frac{i\,\pi}{4}}}{\sqrt{2\,\hbar\,\pi}}\,\Pi_{n\,>\,0}
\frac{dc_{n}^{\mathrm{even}}\,e^{-\frac{i\,\pi}{4}}}{\sqrt{2\,\hbar\,n\,\pi}}
\frac{dc_{n}^{\mathrm{odd}}\,e^{-\frac{i\,\pi}{4}}}{\sqrt{2\,\hbar\,n\,\pi}}
\label{formulae-oscar:periodicmeasure}
\end{eqnarray}
Thus the eigenvalue spectrum of (\ref{formulae-oscar:periodic})
\begin{eqnarray}
\ell_n\,=\, \left(\frac{2\,\pi\,n}{\Delta T}\right)^{2}-\omega^2\,,\quad\quad n=0,
\pm 1\,\pm 2\,,...
\label{formulae-oscar:periodiceigenvalues}
\end{eqnarray}
yields immediately the Morse index
\begin{eqnarray}
\ind \foP([T',T])&=&1+2\,\mathrm{Int}
\left[\frac{\omega\,\Delta T}{2\,\pi}\right]
\nonumber\\
&=&\frac{\mathrm{sign}\sin \left(\omega\,\Delta T\right)+1}{2}
+\mathrm{Int}\left[\frac{\omega\,\Delta T}{\pi}\right]
\label{formulae-oscar:periodicindex}
\end{eqnarray}
while the absolute value of the determinant can be extracted as above from the 
$\zeta$-function.